\documentclass
[aps,preprint,12pt,noshowpacs,showkeys,nofootinbib,eqsecnum,titlepage,hyperref]{revtex4}%
\usepackage{amssymb}
\usepackage{amsfonts}
\usepackage{amsmath}
\usepackage{graphicx}%
\providecommand{\U}[1]{\protect\rule{.1in}{.1in}}
\providecommand{\U}[1]{\protect\rule{.1in}{.1in}}
\providecommand{\U}[1]{\protect\rule{.1in}{.1in}}
\providecommand{\U}[1]{\protect\rule{.1in}{.1in}}
\providecommand{\U}[1]{\protect\rule{.1in}{.1in}}
\providecommand{\U}[1]{\protect\rule{.1in}{.1in}}
\providecommand{\U}[1]{\protect\rule{.1in}{.1in}}
\providecommand{\U}[1]{\protect\rule{.1in}{.1in}}
\providecommand{\U}[1]{\protect\rule{.1in}{.1in}}
\providecommand{\U}[1]{\protect\rule{.1in}{.1in}}

\ifx\pdfoutput\relax\let\pdfoutput=\undefined\fi
\newcount\msipdfoutput
\ifx\pdfoutput\undefined\else
\ifcase\pdfoutput\else
\ifx\paperwidth\undefined\else
\ifdim\paperheight=0pt\relax\else\pdfpageheight\paperheight\fi
\ifdim\paperwidth=0pt\relax\else\pdfpagewidth\paperwidth\fi
\fi\fi\fi
\begin{document}
\title{Background Independent String Field Theory }
\author{Itzhak Bars and Dmitry Rychkov}
\author{}
\affiliation{Department of Physics and Astronomy, University of Southern California, Los
Angeles, CA 90089-0484, USA}

\begin{abstract}
We develop a new background independent Moyal star formalism in bosonic open
string field theory, rendering it into a more transparent and computationally
more efficient theory. The new star product is formulated in a
half-phase-space, and because phase space is independent of any background
fields that the string propagates in, the interactions expressed with the new
star product are background independent. The interaction written in this basis
has a large amount of symmetry, including a supersymmetry OSp(d%
$\vert$%
2) that acts on matter and ghost degrees of freedom, and simplifies
computations. The BRST operator that defines the quadratic kinetic term of
string field theory may be regarded as the solution of the equation of motion
$\bar{A}\star\bar{A}=0$ of a purely cubic background independent string field
theory. We find an infinite number of non-perturbative solutions to this
equation, and are able to associate them to the BRST operator of conformal
field theories on the worldsheet. Thus, the background emerges from a
spontaneous-type breaking of a purely cubic highly symmetric theory. The form
of the BRST field breaks the symmetry in a tractable way such that the
symmetry continues to be useful in practical perturbative computations as an
expansion around some background. The new Moyal basis is called the $\sigma
$-basis, where $\sigma$ is the worldsheet parameter of an open string. A vital
part of the new star product is a natural and crucially needed mid-point
regulator in this continuous basis, so that all computations are finite. The
regulator is removed after renormalization and then the theory is finite only
in the critical dimension. Boundary conditions for D-branes at the endpoints
of the string are naturally introduced and made part of the theory as simple
rules in algebraic computations. The formalism is tested by computing some
perturbative quantities and finding agreement with previous methods. We are
now prepared for new non-perturbative computations. A byproduct of our
approach is an astonishing suggestion of the formalism: the roots of ordinary
quantum mechanics may originate in the rules of non-commutative interactions
in string theory.

\end{abstract}
\maketitle
\tableofcontents

\bigskip\newpage

\section{Introduction}

In this paper we develop string field theory (SFT)\ that is independent of
backgrounds. The progress is due to a new new background independent star
product to describe string-string interactions. This $\star$ product improves
the mathematical structure and the computational framework of open string
field theory (SFT) \cite{witten} under general conditions, including curved
spacetimes or more general string backgrounds consistent with conformal
symmetry on the worldsheet.

The basic approach in the current paper is similar to the Moyal star
formulation of string field theory (MSFT) previously constructed in flat
space-time \cite{B}-\cite{BP}. However, now we develop the Moyal star product
$\star$ in a new basis which is independent of any string backgrounds; this is
the $\sigma$-basis for the open string degrees of freedom $X^{M}\left(
\sigma\right)  $ as parametrized by the worldsheet parameter $\sigma$, with
$0\leq\sigma\leq\pi,~$at a fixed value of $\tau$. In this approach, rather
than expressing the string field as a functional $\psi\left(  X\right)  $ of
the full string coordinate $X^{M}\left(  \sigma\right)  ,$ the string field is
taken to be a functional $A\left(  x_{+},p_{-}\right)  $ of \textit{half of
the phase space} of the string, where $x_{+}^{M}\left(  \sigma\right)  $ is
the symmetric part of $X^{M}\left(  \sigma\right)  $ under reflections
relative to the midpoint, $x_{+}^{M}\left(  \sigma\right)  =\frac{1}{2}\left(
X^{M}\left(  \sigma\right)  +X^{M}\left(  \pi-\sigma\right)  \right)  ,$ while
$p_{-M}\left(  \sigma\right)  =\frac{1}{2}\left(  P_{M}\left(  \sigma\right)
-P_{M}\left(  \pi-\sigma\right)  \right)  $ is the antisymmetric part of the
momentum density $P_{M}=\partial S_{string}/\left(  \partial_{\tau}%
X^{M}\right)  ,$ where the string action $S_{string}$ corresponds to a
conformal field theory (CFT) with any set of background fields consistent with
the conformal symmetry of the worldsheet. Note that $p_{-}\left(
\sigma\right)  $ is the canonical conjugate to $x_{-}\left(  \sigma\right)  $
and commutes with $x_{+}\left(  \sigma\right)  $ in the first quantization of
the string. Thus the field $A\left(  x_{+},p_{-}\right)  $ is related to the
field $\psi\left(  X\right)  =\psi\left(  x_{+},x_{-}\right)  $ by a Fourier
transform from $x_{-}~$to $p_{-}.$ String joining in position space
$\psi\left(  X\left(  \sigma\right)  \right)  $ is represented by the new
Moyal product in the basis $A\left(  x_{+}\left(  \sigma\right)  ,p_{-}\left(
\sigma\right)  \right)  $. This $\star$ product is expressed in the half phase
space without any reference to the details of the CFT, thus being independent
of any backgrounds.

A second new feature is that, the BRST operator $\hat{Q}$ for any conformal
field theory is now represented as an anticommutator in MSFT, $\hat{Q}A\left(
x,p\right)  =\left\{  Q\left(  x,p\right)  ,A\left(  x,p\right)  \right\}
_{\star},$ purely in terms of only the new star product $\star$, involving the
string field $A\left(  x,p\right)  $ with another special string field
$Q\left(  x,p\right)  $ that represents the operation of $\hat{Q}$ on $A.$ We
give an explicit expression for the \textit{string field} $Q\left(
x,p\right)  $ for any corresponding conformal field theory (CFT). This field
$Q\left(  x,p\right)  $ satisfies (not as an operator $\hat{Q}^{2}=0,$ but as
a string field)
\begin{equation}
Q\left(  x,p\right)  \star Q\left(  x,p\right)  =0. \label{Q2zero}%
\end{equation}
In this formulation of string field theory, computations can be carried out
purely algebraically by starting from the MSFT action, without ever referring
to conformal field theory, thus avoiding the complexities of conformal maps
which is the difficult part of computations in other approaches to string
field theory \cite{witten}\cite{GJ}.

A third new feature is the introduction of D-brane boundary conditions at the
end points of the string in the SFT context. In our formalism part of the
information about D-branes can be introduced in a natural way through simple
alterations in the algebraic computational procedure as we will indicate.

A major simplification in structural clarity and computational technique in
MSFT is obtained in the current paper. The new formulation is not only more
transparent but it also displays larger symmetries that mix the matter and the
ghost degrees of freedom, such as the supersymmetry OSp$\left(  d|2\right)  .$
The quadratic $A\star A,$ cubic $A\star A\star A,$ or any higher products of
the string field all have the higher symmetry of the star product for any CFT.
However, the BRST operator is less symmetric due to the structure of ghosts
versus matter. For this reason, the quadratic kinetic term of string field
theory breaks the higher symmetry of the cubic interactions. Nevertheless, in
the Siegel gauge, in the case of flat $d=26$ background, there is an
accidental OSp$\left(  26|2\right)  $ supersymmetry which greatly simplifies
computations involving ghosts.

Moreover, it is possible to rearrange the \textit{effective quantum SFT}
action (where $\bar{A}$ includes all ghost numbers) into a purely cubic term
$S_{eff}=\frac{1}{3g_{0}^{2}}\int\left(  \bar{A}\star\bar{A}\star\bar
{A}\right)  ,$ with $\bar{A}=\left(  g_{0}A+Q\right)  $, and $Q\star Q=0$ as
in (\ref{Q2zero}). This form of the effective action displays the full
symmetry, and is \textit{background independent thanks to the background
independence of the new star product}. Then the emergent quadratic term,
$\int\left(  A\star Q\star A\right)  $ in the expansion in powers of $g_{0},$
is viewed as coming from a spontaneous-type breaking that rearranges the
purely cubic theory $\bar{A}^{3}$ into a perturbative expansion around a
background-dependent perturbative vacuum defined by the string field $Q\left(
x,p\right)  $. The purely cubic form of the action is the best way to see the
background independence of MSFT which is possible because of the background
independent new $\star$ product.

Although the motivation for developing the new formalism is to deal with the
issues of strings in curved spaces (in particular cosmological backgrounds)
and generalizations such as supersymmetry (which we are working on), it should
not escape the reader that the formalism sheds new light and develops new
tools of computation that apply also in the flat perturbative theory. Indeed,
the new Moyal product in the $\sigma$-basis, if taken with only trivial
flat-space string background, reproduces more efficiently the same
computational results as the Moyal $\star$ product in previous bases
\cite{B},\cite{DLMZ}, or the original star product that relies on the CFT
formalism on the worldsheet \cite{witten}.

MSFT is already known to successfully describe string theory in flat
space-time by using the previous Moyal $\star$ product \cite{B}. For example,
it yielded the 4-tachyon perturbative off-shell string scattering amplitudes
(beyond the on-shell Veneziano model) in copious detail not available before
the computation in \cite{BP}. Furthermore, MSFT led to the development of
analytic techniques \cite{BM}\cite{BKM}\cite{BP} for computing the
nonperturbative vacuum of SFT \cite{BKM}. Although the non-perturbative
program remained incomplete at that time, we will indicate how similar
analytic methods will work much better with the new Moyal star product. In
both perturbative and non-perturbative cases analytic results in string theory
that were not obtained before with other approaches were presented, thus
demonstrating the usefulness of MSFT as a tool that produces new analytic
results in string theory. In this paper the good features of MSFT that led to
successes are preserved while MSFT is generalized in several directions thanks
to the simplifications introduced by the $\sigma$-basis for the star product
and the BRST operator. The rest of this paper is organized as follows.

\begin{itemize}
\item In section (\ref{moyalMSFT}) we explain the new Moyal product that
represents string joining or splitting. This is an improvement over the old
discrete Moyal star product \cite{B}. The new product operates on string
fields $A\left(  x,p\right)  $ which are labeled by half of the phase space of
the string $\left(  x^{M}\left(  \sigma\right)  ,p_{M}\left(  \sigma\right)
\right)  $ which is denoted in low case letters $\left(  x,p\right)  ,$ in
contrast to the full phase space $\left(  X^{M}\left(  \sigma\right)
,P_{M}\left(  \sigma\right)  \right)  $ which is denoted in capital letters
$\left(  X,P\right)  .$

\item In section (\ref{iQMSection}) we discuss the similarities and
differences of the Moyal product \cite{Moyal} in ordinary quantum mechanics
(QM) versus the Moyal product in string field theory which represents string
joining. Although $\left(  \hat{x}\left(  \sigma\right)  ,\hat{p}\left(
\sigma\right)  \right)  $ is part of the phase space that consists of only
commuting operators in QM, this same half phase space becomes non-commutative
under the Moyal star $\star$ product that represents string joining. The
non-commutativity which is due to string interactions induces a QM-like system
which we call induced quantum mechanics (iQM). Based on the similarities
between QM and iQM Moyal products, we identify an algebraic system in the half
phase space $\left(  x,p\right)  $ whose properties are just like quantum
mechanics. The rules of iQM become the guiding principle for the rest of the
paper for constructing the interacting string field theory as well as for
performing practical computations.

\item In section (\ref{Regulator_section}) we discuss the regularization that
is essential to resolve midpoint singularities in explicit computations in
string field theory. We give the regulated and final version of the
\textit{background independent} star product, and provide an example of how
the regulated $\star$ product is used for computations. The regulator is
removed after renormalization of the cubic coupling constant $g_{0}$ as shown
in \cite{BP}. In this section we also indicate how D-branes with non-trivial
boundary conditions at the ends of the string are introduced and made part of
the new formalism. In Appendix-A it is shown that if the background CFT is the
flat background then the Moyal star product in the new $\sigma$-basis can be
explicitly related to the old discrete Moyal basis \cite{B} in which
computations of interacting strings were performed in the past \cite{BM}%
\cite{BKM}\cite{BP}, thus showing that the new star product reproduces all of
the previously successful computations.

\item In section (\ref{repr_msft}) we show how the elementary quantum
operators of the first quantized string, including the full phase space for
matter and $b,c$ ghosts, $\left(  X^{M}\left(  \sigma\right)  ,P_{M}\left(
\sigma\right)  \right)  ,$ are represented in terms of only the new
string-joining Moyal product in terms of the half phase space $\left(
x^{M}\left(  \sigma\right)  ,p_{M}\left(  \sigma\right)  \right)  $ of the
iQM. Using these representation rules we map the quantum \textit{operators}
$O\left(  X^{M}\left(  \sigma\right)  ,P_{M}\left(  \sigma\right)  \right)  ,$
associated with \textit{any conformal field theory} (CFT) on the worldsheet,
to the iQM space \textit{string field} $O\left(  x^{M}\left(  \sigma\right)
,p_{M}\left(  \sigma\right)  \right)  .$ In particular we construct the stress
tensor $T$, BRST current $J_{B}$ and BRST operator $Q$ as \textit{string
fields} in the half phase space, $T\left(  x,p\right)  ,$ $J_{B}\left(
x,p\right)  ,$ $Q\left(  x,p\right)  ,$ acting on general string fields
$A\left(  x,p\right)  $ only with the new Moyal star product within iQM. Then
we construct the action for MSFT in this iQM formalism. The proof that the new
MSFT action has a BRST gauge symmetry is given in Appendix-B.

\item In section (\ref{npSection}) we discuss the quantum effective action for
string field theory and show that it can be brought to a highly symmetric
purely cubic form. The equation of motion of the purely cubic SFT is $\bar
{A}\left(  x,p\right)  \star\bar{A}\left(  x,p\right)  =0.$ The BRST field
$Q\left(  x,p\right)  $ in Eq.(\ref{Q2zero}) is clearly a solution, $\bar
{A}_{sol}\left(  x,p\right)  =Q\left(  x,p\right)  $, and since with our
methods we can construct a $Q\left(  x,p\right)  $ for all CFTs on the
worldsheet, we clearly have an inifite number of solutions. Then, including
fluctuations, the general field is $\bar{A}\left(  x,p\right)  =Q\left(
x,p\right)  +g_{0}A\left(  x,p\right)  .$ This allows us to regard the
perturbative setup of string field theory in powers of $g_{0}$ as the analog
of a spontaneously broken version of a highly symmetric cubic theory. The last
paragraphs in this section show how we can easily construct a large class of
non-perturbative solutions in string field by using our methods.

\item In section (\ref{flatCFT}) we give an explicit expression for the
perturbative vacuum in this formalism and outline how to perform perturbative
computations. We give the result of a perturbative computation for the
\textit{off-shell 3-tachyon amplitude} for the case of the CFT with a flat
background. This is a test that our new methods reproduce previously computed
non-trivial quantities. As already mentioned, the discussion in Appendix-A
guarantees that for the CFT with a flat background there will always be
agreement between computations performed with the new Moyal star product and
the corresponding computations, such as amplitudes \cite{BP}, performed in
terms of the discrete Moyal product.

\item Finally in section (\ref{outlook}) we give an outline of where we are
heading with non-perturbative computations and physical applications of this formalism.
\end{itemize}

\section{What is the Moyal $\star$ product in MSFT? \label{moyalMSFT}}

The well known Moyal product \cite{Moyal} is a non-commutative associative
product between any two classical functions of phase space, $A_{1}\left(
x,p\right)  \star$ $A_{2}\left(  x,p\right)  =A_{12}\left(  x,p\right)  ,$
that yields a new function in phase space. The property of this product is
that it reproduces one to one all products of corresponding quantum operators
$\hat{A}_{1}(\hat{x},\hat{p})\hat{A}_{2}(\hat{x},\hat{p})=A_{12}(\hat{x}%
,\hat{p})$ in standard quantum mechanics. The Moyal product is given by
\begin{equation}
\left(  A_{1}\star A_{2}\right)  \left(  x,p\right)  =A_{1}\left(  x,p\right)
\exp\left[  \frac{i\hbar}{2}\left(  \overleftarrow{\partial_{x}}%
\cdot\overrightarrow{\partial_{p}}-\overrightarrow{\partial_{x}}%
\cdot\overleftarrow{\partial_{p}}\right)  \right]  A_{2}\left(  x,p\right)  ~.
\end{equation}
where the left/right arrows mean differentiation of the function on the
left/right sides. For practical computations it is also convenient to write it
in the forms
\begin{align}
\left(  A_{1}\star A_{2}\right)  \left(  x,p\right)   &  =A_{1}\left(  \left(
x^{\prime}+\frac{i\hbar}{2}\frac{\overrightarrow{\partial}}{\partial
p}\right)  ,\left(  p^{\prime}-\frac{i\hbar}{2}\frac{\overrightarrow{\partial
}}{\partial x}\right)  \right)  A_{2}\left(  x,p\right)
,\label{orderedMoyal1}\\
&  =A_{1}\left(  x^{\prime},p^{\prime}\right)  A_{2}\left(  \left(
x-\frac{i\hbar}{2}\frac{\overleftarrow{\partial}}{\partial p^{\prime}}\right)
,\left(  p+\frac{i\hbar}{2}\frac{\overleftarrow{\partial}}{\partial x^{\prime
}}\right)  \right)  , \label{orderedMoyal2}%
\end{align}
where $\left(  x^{\prime},p^{\prime}\right)  $ is set to $\left(  x,p\right)
$ after differentiation. All results of quantum mechanics in the standard
operator Hamiltonian formalism can be replicated in the Moyal formalism. For
example, the basic phase space variables satisfy, $x\star p=xp+\frac{i\hbar
}{2}$ and $p\star x=xp-\frac{i\hbar}{2},$ while their star commutator is
\begin{equation}
\left[  x^{\mu},p_{\nu}\right]  _{\star}=x^{\mu}\star p_{\nu}-p_{\nu}\star
x^{\mu}=i\hbar\delta_{\nu}^{\mu}, \label{commxp}%
\end{equation}
which is the expected result for the basic canonical quantization rules of the
operators $\left[  \hat{x},\hat{p}\right]  =i\hbar$ in the Hamiltonian formalism.

In the form of Eq.(\ref{moyal*}) the Moyal product is valid for fermions
provided the orders of fermionic $x^{M},p_{M}$ or fermionic $A_{1},A_{2}$ are
respected. To include both bosons and fermions $\left(  x^{M},p_{M}\right)  $
we denote $M=\left(  \mu,m\right)  $ where $\mu$ labels bosons and $m$ labels
fermions; we also permit functions of phase space $A\left(  x,p\right)  $ that
could be either bosonic or fermionic. We will define $\left(  -1\right)  ^{M}$
or $\left(  -1\right)  ^{A}$ where the exponent $M$ or $A$ means the grade,
$M,A=0$ for bosons and $M,A=1$ for fermions. Thus $\left(  -1\right)
^{M\text{ or }A}$ $=\left(  -1\right)  ^{0}=1$ for bosons, and $\left(
-1\right)  ^{M\text{ or }A}$ $=\left(  -1\right)  ^{1}=-1$ for fermions.
Changing the order of factors would cost an extra minus sign when they are
both fermions, such as $~\overrightarrow{\partial}_{p_{M}}\overleftarrow
{\partial}_{x^{M}}=\overleftarrow{\partial}_{x^{M}}\overrightarrow{\partial
}_{p_{M}}\left(  -1\right)  ^{MN}~$. Then, the Moyal product that works for
both bosons and fermions is
\begin{equation}
\left(  A_{1}\star A_{2}\right)  \left(  x,p\right)  =A_{1}\left(  x,p\right)
\exp\left[  \frac{i\hbar}{2}\left(  \overrightarrow{\partial}_{p_{M}%
}\overleftarrow{\partial}_{x^{M}}~-~\overleftarrow{\partial}_{p_{M}%
}\overrightarrow{\partial}_{x^{M}}\right)  \right]  A_{2}\left(  x,p\right)  .
\label{moyal*}%
\end{equation}
where the order of the derivatives and their order relative to $A_{1},A_{2}$
need to be respected. This gives the commutation rules (\ref{commxp}) for
boson as above, and also gives the anticommutation rule for fermions
\begin{equation}
\{x^{m},p_{n}\}_{\star}=x^{m}\star p_{n}+p_{n}\star x^{m}=i\hbar\delta_{n}%
^{m}, \label{commxpFermi}%
\end{equation}
where we watch the orders when taking derivatives as follows,
\begin{equation}
A\overleftarrow{\partial}_{x^{N}}=\left(  -1\right)  ^{AN}\partial_{x^{N}%
}A,\text{\ or}\;\;x^{M}\overleftarrow{\partial}_{x^{N}}=\left(  -1\right)
^{MN}\partial_{x^{N}}x^{M}=\left(  -1\right)  ^{M}\delta_{N}^{M}.
\label{diffRulesBF}%
\end{equation}
Then the general star commutator for any $M=\left(  \mu,m\right)  $ gives
\begin{equation}
\lbrack x^{M},p_{N}\}_{\star}=i\hbar\delta_{N}^{M}, \label{superxp}%
\end{equation}
for both bosons and fermions, where the symbol $[\cdot,\cdot\}_{\star}$ means
either commutator or anticommutator as needed. The expressions in
(\ref{orderedMoyal1},\ref{orderedMoyal2}) could be used also for both bosons
and fermions.

We emphasize that the usual anticommutation rules for fermions, with both
$x^{m}$ and $p_{m}$ Hermitian must contain $\hbar$ rather than $i\hbar$ on the
right hand side of Eq.(\ref{commxpFermi}). In this paper, for notational
uniformity of the Moyal product for both bosons and fermions, we wish to use
$i\hbar,$ so this implies that in this paper $x^{m}$ for fermions is defined
to be antihermitian while $p_{m}$ is Hermitian.

\subsection{The new $\star$ product \label{new*}}

In the Moyal formulation of SFT (MSFT), string joining, not ordinary quantum
mechanics, is also represented by a stringy Moyal $\star$ product as
discovered in \cite{B}. In this paper we suggest a new formulation of MSFT
with the following new version of the $\star$ in the $\sigma$-basis
\begin{equation}
\star~=\exp\left[  \frac{i}{4}\int_{0}^{\pi}d\sigma~\text{sign}\left(
\frac{\pi}{2}-\sigma\right)  \left(  \overrightarrow{\partial}_{p_{M}\left(
\sigma\right)  }\overleftarrow{\partial}_{x^{M}\left(  \sigma,\varepsilon
\right)  }-\overleftarrow{\partial}_{p_{M}\left(  \sigma\right)
}\overrightarrow{\partial}_{x^{M}\left(  \sigma,\varepsilon\right)  }\right)
\right]  . \label{msftstar}%
\end{equation}
One of the most important properties of this expression is that it is
independent of any background conformal field theory (CFT) that describes the
free string. The reason for background independence independence is the fact
that it is defined in phase space. The notion of phase space is quite
independent of any Lagrangian formalism on the worldsheet that describes the
free string propagating in some background fields. Although the Lagrangian
provides a background dependent relation between velocities and momenta
through the standard definition $P_{M}=\partial S/\left(  \partial_{\tau}%
X^{M}\right)  $ , the properties of phase space $\left(  X^{M},P_{M}\right)  $
are completely independent of the Lagrangian. Note in particular the natural
upper and lower indices: no background metric is involved in lowering the
index of $P_{M}$. This is the key to background independence. Since it will
require quite a few details to explain how this $\star$ product is derived, we
highlight here the crucial conceptual features of this formalism to focus the
reader on aspects that are important for SFT.

\begin{enumerate}
\item The $\star$ in (\ref{msftstar}) contains the small parameter
$\varepsilon$ in the $x\left(  \sigma,\varepsilon\right)  $ which is a
regulator to separate the midpoint clearly in the process of string joining.
As explained later, a regulator is crucially needed to resolve ambiguities and
related associativity anomalies that were identified in the past
\cite{BM}\cite{Erler}. This avoids the danger of formal manipulations that
could lead to wrong computations, thus making the new MSFT a finite theory in
which every step of a computation is well defined.

\item We define the basis on which the derivatives $\partial_{p\left(
\sigma\right)  },\partial_{x\left(  \sigma,\varepsilon\right)  }$ in the star
product act. The string field $A\left(  x\left(  \sigma,\varepsilon\right)
,p\left(  \sigma\right)  \right)  $ is a functional of \textit{half of the
phase space} of the string as follows. This half-phase-space $\left(  x\left(
\sigma,\varepsilon\right)  ,p\left(  \sigma\right)  \right)  $ is the
regulated $\sigma$-basis which is closely related to the position basis
$\psi\left(  X\left(  \sigma,\varepsilon\right)  \right)  $ by a Fourier
transform. The regulated string position $X\left(  \sigma,\varepsilon\right)
,$ which contains a small parameter $\varepsilon,$ is related to the
unregulated one $X\left(  \sigma\right)  $ by a simple redefinition of the
choice of independent position degrees of freedom to be used to label the
string field $\psi\left(  X\left(  \sigma,\varepsilon\right)  \right)  .$ The
relation is $X\left(  \sigma,\varepsilon\right)  =\exp\left(  -\varepsilon
\sqrt{-\partial_{\sigma}^{2}}\right)  X\left(  \sigma\right)  $ as will be
explained below in some detail. The unregulated momentum $P\left(
\sigma\right)  $ is the canonical conjugate to the unregulated $X\left(
\sigma\right)  $. The position and momentum degrees of freedom are split into
the symmetric/antisymmetric parts relative to the midpoint at $\sigma=\pi/2$
\begin{align}
x_{\pm}\left(  \sigma,\varepsilon\right)   &  =\frac{1}{2}\left(  X\left(
\sigma,\varepsilon\right)  \pm X\left(  \pi-\sigma,\varepsilon\right)
\right)  ,\;\label{x+-}\\
p_{\pm}\left(  \sigma\right)   &  =\frac{1}{2}\left(  P\left(  \sigma\right)
\pm P\left(  \pi-\sigma\right)  \right)  . \label{p+-}%
\end{align}
For simplicity of notation in most of the paper we will omit the $\pm$ on
$x_{+}$ and $p_{-}$ (but keep the $\pm$ for $p_{+}$ and $x_{-}$), thus we will
use interchangeably the following notation
\begin{equation}
x_{+}\left(  \sigma,\varepsilon\right)  \equiv x\left(  \sigma,\varepsilon
\right)  \;\text{and\ }p_{-}\left(  \sigma\right)  \equiv p\left(
\sigma\right)  . \label{x+=x}%
\end{equation}
Thus, the position space string field $\psi\left(  X\left(  \sigma
,\varepsilon\right)  \right)  $ is a functional of the symmetric and
antisymmetric parts of the string position $\psi\left(  X\right)  =\psi\left(
x_{+},x_{-}\right)  .$ The field $A\left(  x_{+}\left(  \sigma,\varepsilon
\right)  ,p_{-}\left(  \sigma\right)  \right)  $ is a half-fourier transform
of the field $\psi\left(  X\right)  =\psi\left(  x_{+},x_{-}\right)  $ in the
antisymmetric variable only%
\begin{equation}
\psi\left(  x_{+},x_{-}\right)  \underset{x_{-}\rightarrow p_{-}}%
{\overset{Fourier}{\rightarrow}}~A\left(  x_{+},p_{-}\right)
\label{FourierPsiA}%
\end{equation}
Hence the antisymmetric $p_{-}\left(  \sigma\right)  =-p_{-}\left(  \pi
-\sigma\right)  $ is the canonical conjugate to the antisymmetric part of the
string $x_{-}\left(  \sigma,\varepsilon\right)  $ in usual first quantization
if the regulator $\varepsilon$ is ignored. Similarly, the even label
$x_{+}\left(  \sigma,\varepsilon\right)  =x_{+}\left(  \pi-\sigma
,\varepsilon\right)  $ is the symmetric part of the position, and it includes
a regulated midpoint $\bar{x}\left(  \varepsilon\right)  =X\left(
\pi/2,\varepsilon\right)  =x_{+}\left(  \pi/2,\varepsilon\right)  $. Due to
the symmetry/antisymmetry properties, it is evident that $x_{+}\left(
\sigma,\varepsilon\right)  ,$ $p_{-}\left(  \sigma\right)  $ are compatible
observables: namely, their operator counterparts $\hat{x}_{+}\left(
\sigma,\varepsilon\right)  ,$ $\hat{p}_{-}\left(  \sigma\right)  $ commute
with each other in the first quantization of the string, that is, although
$\left[  \hat{X}\left(  \sigma,\varepsilon\right)  ,\hat{P}\left(
\sigma^{\prime}\right)  \right]  \neq0$ does not vanish at $\sigma
=\sigma^{\prime}$, the commutator $\left[  \hat{x}_{+}\left(  \sigma
,\varepsilon\right)  ,\hat{p}_{-}\left(  \sigma^{\prime}\right)  \right]  =0$
does vanish for all values of $\sigma,\sigma^{\prime}$. Hence as eigenvalues
of commuting operators, $\left(  x_{+}\left(  \sigma,\varepsilon\right)
,p_{-}\left(  \sigma\right)  \right)  $ is a set of complete labels for the
string field $A\left(  x_{+}\left(  \sigma,\varepsilon\right)  ,p_{-}\left(
\sigma\right)  \right)  $ which is nothing but the wavefunction in first
quantization taken in an appropriate basis $A\left(  x_{+}\left(
\sigma,\varepsilon\right)  ,p_{-}\left(  \sigma\right)  \right)  =\langle
x_{+}\left(  \sigma,\varepsilon\right)  ,p_{-}\left(  \sigma\right)
|A\rangle.$

\item The dot products that appear in the $\star,$ such as $\overleftarrow
{\partial}_{x\left(  \sigma,\varepsilon\right)  }\cdot\overrightarrow
{\partial}_{p\left(  \sigma\right)  },$ mean
\begin{equation}
\overrightarrow{\partial}_{p\left(  \sigma\right)  }\cdot\overleftarrow
{\partial}_{x\left(  \sigma,\varepsilon\right)  }=\frac{\overrightarrow
{\partial}}{\partial p_{M}\left(  \sigma\right)  }\frac{\overleftarrow
{\partial}}{\partial x^{M}\left(  \sigma,\varepsilon\right)  },
\label{superdot}%
\end{equation}
where the sum over $M=\left(  \mu,m\right)  $ includes matter and ghosts
$\left(  x^{\mu},x^{m}\right)  ,$ where the ghosts $x^{m}$ with $m=c,b$
account for the fermionic ghost degrees of freedom $b_{\pm\pm}\left(
\sigma,\tau\right)  $ and $c^{\pm}\left(  \sigma,\tau\right)  $ as given
explicitly in Eq.(\ref{BCXP}). This $\star$ product applies to both bosons
(label $\mu$) and fermions (label $m$). Changing the order of factors would
cost a minus sign in the fermion-like ghost directions $M=m$ as explained
after Eq.(\ref{commxp}).

\item Because no metric appears in the contraction of covariant and
contravariant indices of canonical variables, the dot product in
Eq.(\ref{superdot}) is background independent for any set of background fields
in a conformal field theory that describes the string action and the BRST operator.

\item The matter and ghosts in each supervector $x^{M},p_{M}$ in
Eq.(\ref{commxp}) can be rotated into each other under OSp$\left(  d|2\right)
$ supertransformations, $x^{M}\rightarrow x^{N}\left(  S^{-1}\right)
_{N}^{~M}$ and $p_{M}\rightarrow S_{M}^{~N}p_{N}$ where $S\in$OSp$\left(
d|2\right)  $ mixes matter and ghost degrees of freedom. The new star product
in Eq.(\ref{msftstar}), and therefore the string interaction terms in the new
MSFT are evidently symmetric under this OSp$\left(  d|2\right)  $. Although
this symmetry is broken by the BRST operator, keeping track of this symmetry
greatly simplifies computations.

\item The sum over the degrees of freedom in the integral in (\ref{msftstar}),
$\frac{i}{4}\int_{0}^{\pi}d\sigma~$sign$\left(  \frac{\pi}{2}-\sigma\right)
\left(  \cdots\right)  ,$ may be rewritten as a half-range integral $\frac
{i}{2}\int_{0}^{\pi/2}d\sigma~\left(  \cdots\right)  $ after taking into
account the antisymmetric properties of $\left(  \cdots\right)  .$ This shows
clearly that only half of the string phase space degrees of freedom are
relevant in our formulation. We prefer the version with the full range
integral because the $sign\left(  \frac{\pi}{2}-\sigma\right)  $ factor will
clarify several interesting features.

\item The antisymmetry of the integrand $\left(  \cdots\right)  $ in
(\ref{msftstar}) shows that the midpoint $\bar{x}\left(  \varepsilon\right)
=x\left(  \pi/2,\varepsilon\right)  $ has no contribution to the non-trivial
properties of the $\star$ since, at the midpoint, the quantity $\left(
\cdots\right)  $ vanishes due to the antisymmetry of $\partial/\partial
p_{-}\left(  \sigma\right)  ,$ namely $\partial/\partial p_{-}\left(
\pi/2\right)  =0.$ Accordingly, the $\star$ product (\ref{msftstar}) is local
at the midpoint since $\partial/\partial\bar{x}\left(  \varepsilon\right)  $
does not occur in it. That is, in the product $A_{1}\star A_{2}=A_{12}$ the
three fields $A_{1},A_{2},A_{12}$ are all functions of the same midpoint
$\bar{x}\left(  \varepsilon\right)  ,$ showing that the joining of strings is
implemented \textit{locally at the same midpoints} of the first, second and
final strings. This desired property of string joining is automatically
implemented in our formalism using, $x_{+}^{M}\left(  \sigma,\varepsilon
\right)  $ for all $\sigma,$ without the need of a special treatment of the
midpoint. This is a very important key feature that greatly simplified our formalism.
\end{enumerate}

In the next section we will first review some of the properties of the
standard Moyal product in quantum mechanics for a self sufficient
presentation, and then indicate how to deduce from those properties that
string joining is also conveniently expressed as the $\star$ product given above.

\section{Moyal $\star$ in QM as inspiration for string joining
\label{iQMSection}}

It is useful to recall the essential ingredients of how the Moyal product
works in quantum mechanics (QM) because these same mathematical ingredients
are behind the Moyal star product that describes string joining or splitting
in the context of SFT \cite{witten} as discovered in \cite{B}. We will use the
same method as \cite{B} again in this paper to construct the new $\star$
product in the $\sigma$-basis and show that in this formalism MSFT may be
viewed as a quantum mechanics type system which we call henceforth
\textit{induced quantum mechanics} (iQM) to distinguish it from ordinary QM.

In QM each quantum operator $\hat{A}(\hat{x},\hat{p})$ has a matrix
representation in position space, $\langle x_{L}|\hat{A}|x_{R}\rangle
=\psi\left(  x_{L},x_{R}\right)  ,$ where $x_{L},x_{R}$ are eigenvalues of the
position operator $\hat{x}.$ All properties of the product of two operators
$\hat{A}_{12}=\hat{A}_{1}\hat{A}_{2}$ is captured by the matrix product,
$\langle x_{L}|\hat{A}_{12}|x_{R}\rangle=\langle x_{L}|\hat{A}_{1}\hat{A}%
_{2}|x_{R}\rangle=\int dz\langle x_{L}|\hat{A}_{1}|z\rangle\langle z|\hat
{A}_{2}|x_{R}\rangle,$ which we may write in terms of the corresponding
functions
\begin{equation}
\psi_{12}\left(  x_{L},x_{R}\right)  =\int dz\psi_{1}\left(  x_{L},z\right)
\psi_{2}\left(  z,x_{R}\right)  . \label{matrixProdQM}%
\end{equation}
This QM notation invites the reader to think of the function $\psi\left(
x_{L},x_{R}\right)  $ as an infinite dimensional matrix with continuous
labels, that is associated to a quantum operator $\hat{A}$. Applying this
notion to SFT \cite{B}, the function $\psi\left(  x_{L},x_{R}\right)  $ will
be replaced by the string field in position space $\psi\left(  X\left(
\sigma\right)  \right)  =\psi\left(  x_{L}\left(  \sigma\right)  ,\bar
{x},x_{R}\left(  \sigma\right)  \right)  $, where $\bar{x}\equiv X\left(
\pi/2\right)  $ is the midpoint of the string and $x_{L,R}\left(
\sigma\right)  $ are half-strings that correspond to the left/right portions
of the full string relative to the midpoint,
\begin{equation}
x_{L}\left(  \sigma\right)  \equiv\left\{  X\left(  \sigma\right)
,~0\leq\sigma<\pi/2\right\}  ,\;\;x_{R}\left(  \sigma\right)  \equiv\left\{
X\left(  \pi-\sigma\right)  ,~0\leq\sigma<\pi/2\right\}  .
\label{leftMiddleRight}%
\end{equation}
The midpoint should be subtracted from $\left(  x_{L}\left(  \sigma\right)
,x_{R}\left(  \sigma\right)  \right)  ;$ we will proceed as if this has been
taken into account in order not to cloud the main idea and will return to this
detail later. Then, as suggested by Witten \cite{witten}, the string field
will be treated like a matrix, so that the matrix product of string fields of
the form%
\begin{equation}
\psi_{12}\left(  x_{L}\left(  \sigma\right)  ,\bar{x},x_{R}\left(
\sigma\right)  \right)  =\int Dz\left(  \sigma\right)  ~\psi_{1}\left(
x_{L}\left(  \sigma\right)  ,\bar{x},z\left(  \sigma\right)  \right)
~\psi_{2}\left(  z\left(  \sigma\right)  ,\bar{x},x_{R}\left(  \sigma\right)
\right)  , \label{Wproduct}%
\end{equation}
corresponds to Witten's star product for computing the probability amplitude
for two strings that join at their midpoints to create a new string. Note that
the midpoint $\bar{x}$ is fixed, it is a common point of the first, second and
joined strings, and must be excluded in the integral $Dz\left(  \sigma\right)
.$ This relates to the remarks in item $7$ above about the good properties of
our new $\star$ product in Eq.(\ref{msftstar}) that automatically accomplishes
the exclusion of the midpoint in the star product without separating it from
the rest of the string as a special point.

This definition of the product among fields in SFT conveys the general idea
formally for the joining of strings, but the implementation of how the matrix
product (\ref{Wproduct}) is to be carried out requires careful definition and
considerable technical detail. Most current practitioners in SFT
\cite{giddings}-\cite{schnabl} implement the star product by performing
computations in conformal field theory, as it was done historically in the
initial computations \cite{witten}\cite{giddings}. By contrast, in MSFT
computations are performed using only the Moyal $\star$ where it becomes a
straightforward algebraic computation without ever needing the complicated
gymnastics of conformal maps. This was demonstrated in the past \cite{BM}%
\cite{BKM}\cite{BP} but it becomes considerably simpler and transparent in the
new formalism.

To see how to convert the matrix product to the Moyal product we go back to
QM. Each operator constructed from the basic canonical conjugate operators
$(\hat{x},\hat{p})$ has a classical image $A\left(  x,p\right)  $ assembled as
follows. First consider the matrix elements $\langle x_{L}|\hat{A}%
|x_{R}\rangle=\psi\left(  x_{L},x_{R}\right)  $ as above. Then define the
classical phase space image $A\left(  x,p\right)  $ of the operator $\hat{A}$
by taking half Fourier transform as follows%
\begin{equation}
A\left(  x,p\right)  =\frac{1}{2\pi}\int_{-\infty}^{\infty}dy~e^{2ipy}%
~\psi\left(  x,y\right)  =\frac{1}{2\pi}\int_{-\infty}^{\infty}dy~e^{2ipy}%
~\langle x+y|\hat{A}|x-y\rangle\label{prescription1}%
\end{equation}
where we have rewritten the function $\psi\left(  x_{L},x_{R}\right)  $ in
terms of $\left(  x,y\right)  $ which are the symmetric and antisymmetric
combinations of $x_{L},x_{R},$
\begin{equation}
\psi\left(  x,y\right)  \equiv\psi\left(  x_{L},x_{R}\right)  =\langle
x_{L}|\hat{A}|x_{R}\rangle,\;x=\frac{x_{L}+x_{R}}{2},\;y=\frac{x_{L}-x_{R}}%
{2}. \label{prescription2}%
\end{equation}
The operator $\hat{A}$ that appears in Eqs.(\ref{prescription1}%
,\ref{prescription2}) can be reconstructed from its classical image $A\left(
x,p\right)  $ by substituting operators $\left(  \hat{x},\hat{p}\right)  $
instead of the classical phase space $\left(  x,p\right)  $ as follows,
$\hat{A}(\hat{x},\hat{p})=\int dxdpA\left(  x,p\right)  \left\{  \delta\left(
x-\hat{x}\right)  \delta\left(  p-\hat{p}\right)  \right\}  ,$ provided some
operator ordering prescription for the delta functions is given. The
prescription given by Weyl \cite{Moyal}, which insures that $\hat{A}(\hat
{x},\hat{p})$ is a Hermitian operator, is to replace the delta functions by
their integral representations such that the operators $(\hat{x},\hat{p})$
appear in the same exponential as follows%
\begin{equation}
\hat{A}(\hat{x},\hat{p})=\frac{1}{\left(  2\pi\right)  ^{2}}\int dx^{\prime
}dp^{\prime}dxdp~A\left(  x,p\right)  ~\exp\left(  ip^{\prime}\left(
x-\hat{x}\right)  -ix^{\prime}\left(  p-\hat{p}\right)  \right)  .
\label{moyal2}%
\end{equation}
It can be checked that Eqs.(\ref{prescription1},\ref{moyal2}) are consistent
with each other by inserting the operator (\ref{moyal2}) back into
Eq.(\ref{prescription1}), computing the matrix element in position space by
using $\langle x_{L}|e^{ix^{\prime}\hat{p}-ip^{\prime}\hat{x}}|x_{R}\rangle=$
$e^{ix^{\prime}p^{\prime}/2}e^{-ip^{\prime}x_{R}}\delta\left(  x^{\prime
}-x_{L}+x_{R}\right)  ,$ and performing the integrals. This shows that the
classical function $A\left(  x,p\right)  $ is the same in both
Eqs.(\ref{prescription1},\ref{moyal2}).

Now that we have a one to one correspondence between quantum operators and
their classical images, we can ask the following question: if we compute the
product of two operators in QM to obtain a new one $\hat{A}_{12}=\hat{A}%
_{1}\hat{A}_{2}$, what is the rule for computing the phase space function
$A_{12}\left(  x,p\right)  $ which is the image of $\hat{A}_{12}$ from the
images $A_{1}\left(  x,p\right)  $ and $A_{2}\left(  x,p\right)  ?$ The answer
to this question is the Moyal product, namely $A_{12}\left(  x,p\right)
=A_{1}\left(  x,p\right)  \star A_{2}\left(  x,p\right)  $ where the $\star$
product is defined in Eq.(\ref{moyal*}). Recalling that the matrix elements in
position space $\left(  \psi_{1},\psi_{2},\psi_{12}\right)  \left(
x_{L},x_{R}\right)  $ also reproduce the operator product, this means that the
matrix product in Eq.(\ref{matrixProdQM}) is also equivalent to the Moyal
product, provided each function $\left(  \psi_{1},\psi_{2},\psi_{12}\right)
\left(  x_{L},x_{R}\right)  $ is related to its half-Fourier transform
$\left(  A_{1},A_{2},A_{12}\right)  \left(  x,p\right)  $ according to the
prescription in Eqs.(\ref{prescription1},\ref{prescription2}).

Coming back to SFT, the content of the previous paragraph was the basic
inspiration in \cite{B} that led to the rewriting of the matrix-like product
in Eq.(\ref{Wproduct}) as a Moyal product. In \cite{B} this was done for the
modes of the string in a flat target spacetime. Using Eq.(\ref{msftstar}) we
now do it in the $\sigma$-basis which can be used in the presence of any set
of background fields in a conformal field theory that describes a string. So,
imitating Eq.(\ref{moyal*}) we are led to its stringy analog in
(\ref{msftstar}) in order to reproduce the \textit{matrix product in string
joining}
\begin{equation}
A_{12}\left(  x_{+}\left(  \sigma,\varepsilon\right)  ,p_{-}\left(
\sigma\right)  \right)  =A_{1}\left(  x_{+}\left(  \sigma,\varepsilon\right)
,p_{-}\left(  \sigma\right)  \right)  \star A_{2}\left(  x_{+}\left(
\sigma,\varepsilon\right)  ,p_{-}\left(  \sigma\right)  \right)  ,
\label{A12-1-2}%
\end{equation}
where we replace the pointlike $\left(  x,p\right)  $ in Eq.(\ref{moyal*}) by
the stringlike $\left(  x_{+}\left(  \sigma,\varepsilon\right)  ,p_{-}\left(
\sigma\right)  \right)  ,$ including the regulator $\varepsilon.$ For now
ignore the complication of the midpoint in Eq.(\ref{Wproduct}) which is what
the $\varepsilon$ is for; we will explain this below. The basis $\left(
x_{+}\left(  \sigma,\varepsilon\right)  ,p_{-}\left(  \sigma\right)  \right)
$ emerges from taking into account the map between $\psi$ and $A$ in
Eq.(\ref{prescription1},\ref{prescription2}). In this map we must take
$x_{+}\left(  \sigma,\varepsilon\right)  =\frac{1}{2}\left(  x_{L}\left(
\sigma,\varepsilon\right)  +x_{R}\left(  \sigma,\varepsilon\right)  \right)
,$ while $p_{-}\left(  \sigma\right)  $ should be the Fourier transform
parameter for $x_{-}\left(  \sigma,\varepsilon\right)  =\frac{1}{2}\left(
x_{L}\left(  \sigma,\varepsilon\right)  -x_{R}\left(  \sigma,\varepsilon
\right)  \right)  $. This means $x_{+}\left(  \sigma,\varepsilon\right)  $ is
the symmetric part of the string $X\left(  \sigma,\varepsilon\right)  $, while
$p_{-}\left(  \sigma\right)  $ is the antisymmetric part of the momentum
$P\left(  \sigma\right)  .$ So, in terms of the full string degrees of freedom
$\left(  X\left(  \sigma,\varepsilon\right)  ,P\left(  \sigma\right)  \right)
$ the relevant symmetric/antisymmetric parts are given precisely by
Eqs.(\ref{x+-}-\ref{x+=x})$.$ This explains the logic why, except for some
midpoint details, string joining is represented by the Moyal star product. We
will discuss the details of the midpoint, but for now note that $x_{+}\left(
\sigma,\varepsilon\right)  $ includes the midpoint but the string joining
Moyal star excludes it as outlined in item (7) in section (\ref{new*}). So no
special treatment of the midpoint is needed.

It is important to emphasize that the new formalism applies now to string
theory for any set of conformally consistent background fields contained in
the worldsheet CFT. This is because the $\star$ product is expressed in terms
of the phase space degrees of freedom which is a notation that is independent
of the geometrical details of the background fields. The information about the
background geometry is buried in the canonical formalism that includes the
relation between the velocity and momentum as well as in the stress tensor or
BRST operator of the CFT. But none of this geometrical information enters in
the star product (\ref{msftstar}). Hence the star product for string joining
defined in this way provides a background independent method of expressing
string-string interactions via the joining of strings.

\subsection{String Joining iQM versus QM \label{QM=iQM}}

Although the setup presented above for the Moyal $\star$ in MSFT resembles
quantum mechanics (QM), the stringy $\star$ in Eq.(\ref{msftstar}) does not
arise because of the first quantization of the string as was the case of the
particle in Eq.(\ref{moyal*}). Instead, the star product (\ref{msftstar}) is
designed to formulate string joining or splitting. The resemblance to QM is
intriguing and it even invites the thought that string joining could be
considered as an origin for QM as we comment at the end of this section.

Below we will use the induced quantum mechanics (iQM) basis $\left(
x_{+}\left(  \sigma,\varepsilon\right)  ,p_{-}\left(  \sigma\right)  \right)
$ to construct a representation of the first quantized QM operators of the
string $\left(  \hat{X}\left(  \sigma,\varepsilon\right)  ,\hat{P}\left(
\sigma\right)  \right)  .$ That is, QM will be built from iQM. This
representation will clarify the relation and the difference between QM and iQM
while also show how to represent all string theory operators, such as the
stress tensor, BRST current $J_{B}\left(  \sigma\right)  $ and the BRST
operator, in the iQM basis $\left(  x_{+}\left(  \sigma,\varepsilon\right)
,p_{-}\left(  \sigma\right)  \right)  $ for any conformal field theory on the
worldsheet (CFT) that describes the string moving in a conformally consistent
set of background fields.

Starting from the definitions of $x_{\pm},p_{\pm}$ in Eq.(\ref{x+-}), we write
the full string first quantized QM operators $\hat{X}\left(  \sigma
,\varepsilon\right)  ,\hat{P}\left(  \sigma\right)  $ in terms of the
antisymmetric and symmetric parts%
\begin{equation}
\hat{X}^{M}\left(  \sigma,\varepsilon\right)  =\hat{x}_{+}^{M}\left(
\sigma,\varepsilon\right)  +\hat{x}_{-}^{M}\left(  \sigma,\varepsilon\right)
,\;\;\hat{P}_{M}\left(  \sigma\right)  =\hat{p}_{+M}\left(  \sigma\right)
+\hat{p}_{-M}\left(  \sigma\right)  ,
\end{equation}
The regulator $\varepsilon$ will be carefully discussed in section
\ref{Regulator_section} but it is not needed to explain the concepts in this
section. Although we will carry on the regulator $\varepsilon$ for consistency
with the rest of the paper, the reader would probably understand this section
more easily by first setting $\varepsilon=0$ everywhere, and then reviewing it
again by recalling the definition of the regulated position operator, $\hat
{X}^{M}\left(  \sigma,\varepsilon\right)  =\exp(-\varepsilon\sqrt
{-\partial_{\sigma}^{2}})\hat{X}^{M}\left(  \sigma\right)  ,$ and that
$M=\left(  \mu,m\right)  $ denotes $\mu$ for spacetime and $m$ for ghosts.

The QM rules for the first quantization of the string are (anticommutator for
the ghosts)
\begin{equation}
\lbrack\hat{X}^{M}\left(  \sigma,\varepsilon\right)  ,\hat{P}_{N}\left(
\sigma^{\prime}\right)  \}=[e^{-\varepsilon\left\vert \partial_{\sigma
}\right\vert }\hat{X}^{M}\left(  \sigma\right)  ,\hat{P}_{N}\left(
\sigma^{\prime}\right)  \}=i\delta_{N}^{M}\delta_{\varepsilon}\left(
\sigma,\sigma^{\prime}\right)  , \label{basic[XP]}%
\end{equation}
where $\left\vert \partial_{\sigma}\right\vert \equiv\sqrt{-\partial_{\sigma
}^{2}}$ and
\begin{equation}
\delta_{\varepsilon}\left(  \sigma,\sigma^{\prime}\right)  \equiv
e^{-\varepsilon\left\vert \partial_{\sigma}\right\vert }\delta\left(
\sigma,\sigma^{\prime}\right)
\end{equation}
is a regulated delta function defined below. Here $P_{M}\left(  \sigma\right)
$ is defined for any conformal field theory, with arbitrary string background
fields, in the canonical way, from the string Lagrangian, $P_{M}\left(
\sigma\right)  =\partial L_{string}/\partial(\partial_{\tau}X^{M}\left(
\sigma\right)  $). Note that, for any set of background fields in the CFT
$L_{string}$, the position $X^{M}\left(  \sigma\right)  $ is a contravariant
vector while the momentum $P_{M}\left(  \sigma\right)  $ is a covariant
vector, so that the commutation rules (\ref{basic[XP]}), with $\delta_{N}%
^{M}\delta_{\varepsilon}\left(  \sigma,\sigma^{\prime}\right)  $ on the right
hand side, are \textit{background independent}. Furthermore, the
$\delta\left(  \sigma,\sigma^{\prime}\right)  $ is a Dirac delta function
which must be consistent with Neumann or Dirichlet boundary conditions at the
end points $\sigma=0,\pi$ and $\sigma^{\prime}=0,\pi$ (D-branes). There are
two types: $\delta_{\varepsilon}^{nn}\left(  \sigma,\sigma^{\prime}\right)  $
for Neumann-Neumann, and $\delta_{\varepsilon}^{dd}\left(  \sigma
,\sigma^{\prime}\right)  $ for Dirichlet-Dirichlet, as given in
Eqs.(\ref{deltann}-\ref{deltadd2}). Mostly we simply write $\delta
_{\varepsilon}\left(  \sigma,\sigma^{\prime}\right)  $ and use the appropriate
version when necessary. When it becomes important to keep track of boundary
condition properties in different directions $M,$ more carefully we will write
$\delta_{N}^{M}\delta_{\varepsilon M}\left(  \sigma,\sigma^{\prime}\right)  ,$
where $\delta_{\varepsilon M}\left(  \sigma,\sigma^{\prime}\right)  $ stands
for $\delta_{\varepsilon}^{nn}\left(  \sigma,\sigma^{\prime}\right)  $ or
$\delta_{\varepsilon}^{dd}\left(  \sigma,\sigma^{\prime}\right)  .$ There will
be more refinements introduced for the $\delta\left(  \sigma,\sigma^{\prime
}\right)  $'s$,$ including \textquotedblleft even\textquotedblright%
\ $\delta^{+}\left(  \sigma,\sigma^{\prime}\right)  $, \textquotedblleft
odd\textquotedblright\ $\delta^{-}\left(  \sigma,\sigma^{\prime}\right)  ,$
and \textquotedblleft midpoint\textquotedblright\ $\hat{\delta}\left(
\sigma,\sigma^{\prime}\right)  $\ versions, that include also a regulator
$\varepsilon$, as will be shown explicitly below. Thus, D-branes enter our
formalism in this way through the details of these delta functions that are
sensitive to the boundary conditions applied to the ends of the string.

From (\ref{basic[XP]}) the QM rules for the operators $\hat{x}_{\pm},\hat
{p}_{\pm}$ are derived. First note that the even degrees of freedom commute
with the odd ones%
\begin{equation}
\lbrack\hat{x}_{\pm}\left(  \sigma,\varepsilon\right)  ,\hat{p}_{\mp}\left(
\sigma^{\prime}\right)  ]=0, \label{xpCommute}%
\end{equation}
while the non-trivial QM rules are
\begin{align}
\lbrack\hat{x}_{\pm}\left(  \sigma,\varepsilon\right)  ,\hat{p}_{\pm}\left(
\sigma^{\prime}\right)  ]  &  =\frac{1}{4}\left[  e^{-\varepsilon\left\vert
\partial_{\sigma}\right\vert }\left(  \hat{X}\left(  \sigma\right)  \pm\hat
{X}\left(  \pi-\sigma\right)  \right)  ,\left(  P\left(  \sigma^{\prime
}\right)  \pm\hat{P}\left(  \pi-\sigma^{\prime}\right)  \right)  \right]
\label{comm+-}\\
&  =\frac{i}{2}e^{-\varepsilon\left\vert \partial_{\sigma}\right\vert }\left(
\delta\left(  \sigma,\sigma^{\prime}\right)  \pm\delta\left(  \pi
-\sigma,\sigma^{\prime}\right)  \right)  \equiv\frac{i}{2}\delta_{\varepsilon
}^{\pm}\left(  \sigma,\sigma^{\prime}\right)  . \label{delta+-}%
\end{align}
The last equation defines the even and odd regulated delta functions
$\delta_{\varepsilon}^{\pm}\left(  \sigma,\sigma^{\prime}\right)  $ that will
appear again later. Their explicit formulas are given in Eqs.(\ref{deltann+}%
-\ref{deltadd-}) and their properties are illustrated for a small
$\varepsilon$ in Figs.(1,2).

The eigenvalues of the operator $\hat{X}^{M}\left(  \sigma,\varepsilon\right)
$ form a complete set of labels for the wavefunction (string field) in
position space $\psi\left(  X\left(  \sigma,\varepsilon\right)  \right)
\equiv\psi\left(  x_{+}\left(  \sigma,\varepsilon\right)  ,x_{-}\left(
\sigma,\varepsilon\right)  \right)  .$ This labelling is equivalent to the
position space labelling without a regulator, namely $\Psi\left(  X\left(
\sigma\right)  \right)  =\psi\left(  X\left(  \sigma,\varepsilon\right)
\right)  ,$ but as independent labels we prefer to use the eigenvalues of the
regulated $X\left(  \sigma,\varepsilon\right)  $ which amounts to a
reshuffling of the eigenvalues of the unregulated $X\left(  \sigma\right)  .$
Either way, the unregulated operator $\hat{P}\left(  \sigma\right)  $ is
represented in position space by the unregulated derivative $\hat{P}\left(
\sigma\right)  \rightarrow-i\partial/\partial X\left(  \sigma\right)  .$
However, since we insist on the regulated labeling we must use the chain rule
to compute it on $\psi\left(  X\left(  \sigma,\varepsilon\right)  \right)  ,$
that is%
\begin{align}
i\hat{P}_{M}\left(  \sigma\right)  \psi\left(  X\left(  \cdot,\varepsilon
\right)  \right)   &  =\frac{\partial\psi\left(  X\left(  \cdot,\varepsilon
\right)  \right)  }{\partial X^{M}\left(  \sigma\right)  }=\int d\sigma
^{\prime}\frac{\partial\psi\left(  X\left(  \cdot,\varepsilon\right)  \right)
}{\partial X^{N}\left(  \sigma^{\prime},\varepsilon\right)  }\frac{\partial
X^{N}\left(  \sigma^{\prime},\varepsilon\right)  }{\partial X^{M}\left(
\sigma\right)  }\nonumber\\
&  =\int d\sigma^{\prime}\frac{\partial\psi\left(  X\left(  \cdot
,\varepsilon\right)  \right)  }{\partial X^{M}\left(  \sigma^{\prime
},\varepsilon\right)  }e^{-\varepsilon\left\vert \partial_{\sigma^{\prime}%
}\right\vert }\delta\left(  \sigma,\sigma^{\prime}\right) \nonumber\\
&  =e^{-\varepsilon\left\vert \partial_{\sigma}\right\vert }\frac{\partial
\psi\left(  X\left(  \cdot,\varepsilon\right)  \right)  }{\partial
X^{M}\left(  \sigma,\varepsilon\right)  }. \label{Pdiff}%
\end{align}
Here we have used the definition of $X\left(  \sigma^{\prime},\varepsilon
\right)  $, and the natural differential rule for the unregulated symbols
$\partial X\left(  \sigma^{\prime}\right)  /\partial X\left(  \sigma\right)
=\delta\left(  \sigma,\sigma^{\prime}\right)  ,$ to obtain the following rules
of computation.
\begin{equation}%
\begin{array}
[c]{l}%
\text{regulated delta:\ \ }\frac{\partial X\left(  \sigma^{\prime}%
,\varepsilon\right)  }{\partial X\left(  \sigma\right)  }=e^{-\varepsilon
\left\vert \partial_{\sigma}\right\vert }\delta\left(  \sigma,\sigma^{\prime
}\right)  \equiv\delta_{\varepsilon}\left(  \sigma,\sigma^{\prime}\right)  ,\\
\text{unregulated delta:\ }\frac{\partial X\left(  \sigma^{\prime}%
,\varepsilon\right)  }{\partial X\left(  \sigma,\varepsilon\right)  }%
=\delta\left(  \sigma,\sigma^{\prime}\right)  ,\\
\text{regulated delta:\ }e^{-\varepsilon\left\vert \partial_{\sigma
}\right\vert }\frac{\partial X\left(  \sigma^{\prime},\varepsilon\right)
}{\partial X\left(  \sigma,\varepsilon\right)  }=e^{-\varepsilon\left\vert
\partial_{\sigma}\right\vert }\delta\left(  \sigma,\sigma^{\prime}\right)
\equiv\delta_{\varepsilon}\left(  \sigma,\sigma^{\prime}\right)  ,
\end{array}
\label{DiffruleRegX}%
\end{equation}
The last one, which is the only rule needed to represent the momentum
according to Eq.(\ref{Pdiff}), always yields well defined regulated results.
So, when $\psi$ is just $X\left(  \sigma^{\prime},\varepsilon\right)  ,$ we
get $\hat{P}\left(  \sigma\right)  \psi=-ie^{-\varepsilon\left\vert
\partial_{\sigma}\right\vert }\frac{\partial X\left(  \sigma^{\prime
},\varepsilon\right)  }{\partial X\left(  \sigma,\varepsilon\right)
}=-i\delta_{\varepsilon}\left(  \sigma,\sigma^{\prime}\right)  ,$ which is
consistent with the commutation rules (\ref{basic[XP]}). Rewriting this in
terms of the symmetric/antisymmetric labels we get the representation for
$\hat{p}_{\pm}$
\begin{equation}
i\hat{p}_{\pm M}\left(  \sigma\right)  \psi\left(  x_{+}\left(  \cdot
,\varepsilon\right)  ,x_{-}\left(  \cdot,\varepsilon\right)  \right)
=\frac{1}{2}e^{-\varepsilon\left\vert \partial_{\sigma}\right\vert }\left(
\frac{\partial\psi\left(  x_{+},x_{-}\right)  }{\partial x_{\pm}\left(
\sigma,\varepsilon\right)  }\right)  \label{p+-Psi}%
\end{equation}
and the rules for computation are%
\begin{equation}
\frac{\partial x_{\pm}\left(  \sigma^{\prime},\varepsilon\right)  }{\partial
x_{\pm}\left(  \sigma,\varepsilon\right)  }=\left(  \delta\left(
\sigma,\sigma^{\prime}\right)  \pm\delta\left(  \pi-\sigma,\sigma^{\prime
}\right)  \right)  \equiv\delta^{\pm}\left(  \sigma,\sigma^{\prime}\right)
,\text{ unregulated delta,} \label{diffrulesxx}%
\end{equation}
so that when $\psi$ is just $x_{\pm}\left(  \sigma^{\prime},\varepsilon
\right)  ,$ we get $\hat{p}_{\pm}\left(  \sigma\right)  \psi=-\frac{i}%
{2}e^{-\varepsilon\left\vert \partial_{\sigma}\right\vert }\frac{\partial
x_{\pm}\left(  \sigma^{\prime},\varepsilon\right)  }{\partial x_{\pm}\left(
\sigma,\varepsilon\right)  }=-i\delta_{\varepsilon}^{\pm}\left(  \sigma
,\sigma^{\prime}\right)  ,$ a regulated delta, which is consistent with the
commutation rules (\ref{comm+-}).

In summary, in position space we have the following representation of the full
string QM operators%
\begin{align}
\hat{X}^{M}\left(  \sigma,\varepsilon\right)  \psi\left(  x_{+},x_{-}\right)
&  =\left(  x_{+}^{M}\left(  \sigma,\varepsilon\right)  +x_{-}^{M}\left(
\sigma,\varepsilon\right)  \right)  \psi\left(  x_{+},x_{-}\right)
\label{positionInX}\\
\hat{P}_{M}\left(  \sigma\right)  \psi\left(  x_{+},x_{-}\right)   &
=-\frac{i}{2}e^{-\varepsilon\left\vert \partial_{\sigma}\right\vert }\left(
\frac{\partial\psi\left(  x_{+},x_{-}\right)  }{\partial x_{+}^{M}\left(
\sigma,\varepsilon\right)  }+\frac{\partial\psi\left(  x_{+},x_{-}\right)
}{\partial x_{-}^{M}\left(  \sigma,\varepsilon\right)  }\right)  ,
\label{momentumInX}%
\end{align}
and to compute we use the differentiation rules in Eq.(\ref{diffrulesxx}).

We now turn to the string field in the mixed phase space basis $A\left(
x_{+},p_{-}\right)  ,$ which is the Fourier transform of $\psi\left(
x_{+},x_{-}\right)  $ in the $\left(  -\right)  $ variable as in
Eq.(\ref{FourierPsiA}). The path-integral Fourier transform of $\psi\left(
x_{+}^{M}\left(  \cdot,\varepsilon\right)  ,x_{-}^{M}\left(  \cdot
,\varepsilon\right)  \right)  $ is precisely $A\left(  x\left(  \cdot
,\varepsilon\right)  ,p\left(  \cdot\right)  \right)  $%
\begin{equation}
A\left(  x\left(  \cdot,\varepsilon\right)  ,p\left(  \cdot\right)  \right)
=\int\left(  Dx_{-}\left(  \cdot,\varepsilon\right)  \right)  \exp\left[
-i\int_{0}^{\pi}d\sigma~x_{-}^{M}\left(  \sigma,\varepsilon\right)
p_{M}\left(  \sigma\right)  \right]  \psi\left(  x_{+}\left(  \cdot
,\varepsilon\right)  ,x_{-}\left(  \cdot,\varepsilon\right)  \right)
\label{Fourier}%
\end{equation}
In the Fourier exponent we divided by a factor of $2$ as compared to
Eq.(\ref{prescription1}) to take into account the double counting due to the
antisymmetry of $x_{-}\left(  \sigma,\varepsilon\right)  \cdot$ $p\left(
\sigma\right)  $ when reflected from $\pi/2.$ In the basis $A\left(
x_{+},p_{-}\right)  $ the operators $\hat{x}_{+}\left(  \sigma,\varepsilon
\right)  ,\hat{p}_{-}\left(  \sigma\right)  $ are diagonal, while the
operators $\hat{p}_{+},\hat{x}_{-}$ act like derivatives. Either by taking
Fourier transform of Eqs.(\ref{positionInX},\ref{momentumInX}) or directly
from the commutation rules (\ref{comm+-}) we derive the representations of all
operators $\hat{x}_{\pm},\hat{p}_{\pm}$ on this basis%
\begin{align}
\hat{x}_{+}\left(  \sigma,\varepsilon\right)  A\left(  x_{+},p_{-}\right)   &
=x_{+}\left(  \sigma,\varepsilon\right)  A\left(  x_{+},p_{-}\right)
,\;\;\hat{p}_{-}\left(  \sigma\right)  A\left(  x_{+},p_{-}\right)
=e^{-\varepsilon\left\vert \partial_{\sigma}\right\vert }p_{-}\left(
\sigma\right)  A\left(  x_{+},p_{-}\right) \label{x+p-A}\\
\hat{x}_{-}^{M}\left(  \sigma,\varepsilon\right)  A\left(  x_{+},p_{-}\right)
&  =\frac{i}{2}\frac{\partial A\left(  x_{+},p_{-}\right)  }{\partial
p_{-M}\left(  \sigma\right)  }\;\;,\hat{p}_{+M}\left(  \sigma\right)  A\left(
x_{+},p_{-}\right)  =-\frac{i}{2}e^{-\varepsilon\left\vert \partial_{\sigma
}\right\vert }\frac{\partial A\left(  x_{+},p_{-}\right)  }{\partial x_{+}%
^{M}\left(  \sigma,\varepsilon\right)  }. \label{x-p+A}%
\end{align}
Then the full string QM operators have the following representation%
\begin{align}
\hat{X}^{M}\left(  \sigma,\varepsilon\right)  A\left(  x_{+},p_{-}\right)   &
=\left(  x_{+}^{M}\left(  \sigma,\varepsilon\right)  +\frac{i}{2}%
\frac{\partial}{\partial p_{-M}\left(  \sigma\right)  }\right)  A\left(
x_{+},p_{-}\right)  ,\label{XA}\\
\hat{P}_{M}\left(  \sigma\right)  \psi\left(  x_{+},x_{-}\right)   &
=e^{-\varepsilon\left\vert \partial_{\sigma}\right\vert }\left(  p_{-M}\left(
\sigma\right)  -\frac{i}{2}\frac{\partial}{\partial x_{+}^{M}\left(
\sigma,\varepsilon\right)  }\right)  A\left(  x_{+},p_{-}\right)  . \label{PA}%
\end{align}

From Eq.(\ref{basic[XP]}) to Eq.(\ref{PA}) we described the QM of the string
in any background CFT and how to represent its basic quantum operators
$\hat{X},\hat{P}$ on the field $A\left(  x_{+},p_{-}\right)  $. This is
sufficient to obtain the representation of any other observable in this CFT,
such as the BRST operator, in the MSFT formalism. We will do this later, but
first we relate this differential operator representation to something more
elegant in the language of the Moyal $\star$ product.

Now we turn to the induced QM (iQM) generated by the Moyal $\star$ for string
joining for any CFT. We will show that QM is represented in terms of iQM.
First observe that using the $\star$ in Eq.(\ref{msftstar}), and the
differentiation rules (\ref{diffrulesxx}), we get a non-trivial $\star$
commutator between $x_{+}\left(  \sigma,\varepsilon\right)  $ and
$p_{-}\left(  \sigma\right)  $
\begin{align}
&  [x_{+}^{M}\left(  \sigma_{1},\varepsilon\right)  ,p_{-N}\left(  \sigma
_{2}\right)  \}_{\star}\nonumber\\
&  =x_{+}^{M}\left(  \sigma_{1},\varepsilon\right)  \star p_{-N}\left(
\sigma_{2}\right)  -\left(  -1\right)  ^{MN}p_{-N}\left(  \sigma_{2}\right)
\star x_{+}^{M}\left(  \sigma_{1},\varepsilon\right) \nonumber\\
&  =i\delta_{N}^{M}~\text{sign}\left(  \frac{\pi}{2}-\sigma_{1}\right)
\delta^{-}\left(  \sigma_{1},\sigma_{2}\right)  =i\delta_{N}^{M}~\delta
^{+}\left(  \sigma_{1},\sigma_{2}\right)  ~\text{sign}\left(  \frac{\pi}%
{2}-\sigma_{2}\right) \nonumber\\
&  \equiv i\delta_{N}^{M}\hat{\delta}_{+-}\left(  \sigma_{1},\sigma
_{2}\right)  . \label{iQM[]}%
\end{align}
This is the induced QM. It has nothing to do with ordinary quantum mechanics
since under QM the operators $\hat{x}_{+}^{M}\left(  \sigma_{1},\varepsilon
\right)  ,\hat{p}_{-N}\left(  \sigma_{2}\right)  $ commute with each other as
seen in Eq.(\ref{xpCommute}), whereas their eigenvalues do not commute under
the string-joining iQM of Eq.(\ref{iQM[]}).

The Dirac $\hat{\delta}_{+-}\left(  \sigma_{1},\sigma_{2}\right)  $ function,
that combines the sign function with the unregulated delta functions
$\delta^{\pm}$ of Eq.(\ref{delta+-}), satisfies the appropriate $\left(
nn\right)  $ or $\left(  dd\right)  $ boundary conditions consistent with the
properties of $x_{+}^{M}\left(  \sigma_{1},\varepsilon\right)  ,p_{-N}\left(
\sigma_{2}\right)  $. Furthermore, relative to the midpoint $\hat{\delta}%
_{+-}\left(  \sigma_{1},\sigma_{2}\right)  $ is symmetric in $\sigma
_{1}\rightarrow\left(  \pi-\sigma_{1}\right)  $ and antisymmetric in
$\sigma_{2}\rightarrow\left(  \pi-\sigma_{2}\right)  .$ The equivalence of the
two forms of $\hat{\delta}_{+-}\left(  \sigma_{1},\sigma_{2}\right)  $ given
above is verified by using the properties of the unregulated delta functions
as follows
\begin{align}
\hat{\delta}_{+-}\left(  \sigma_{1},\sigma_{2}\right)   &  =\text{sign}\left(
\frac{\pi}{2}-\sigma_{1}\right)  \delta^{-}\left(  \sigma_{1},\sigma
_{2}\right) \label{deltaHat1}\\
&  =\text{sign}\left(  \frac{\pi}{2}-\sigma_{1}\right)  \left(  \delta\left(
\sigma_{1},\sigma_{2}\right)  -\delta\left(  \sigma_{1},\pi-\sigma_{2}\right)
\right) \nonumber\\
&  =\text{sign}\left(  \frac{\pi}{2}-\sigma_{2}\right)  \delta\left(
\sigma_{1},\sigma_{2}\right)  -\text{sign}\left(  \frac{\pi}{2}-\left(
\pi-\sigma_{2}\right)  \right)  \delta\left(  \sigma_{1},\pi-\sigma_{2}\right)
\nonumber\\
&  =\text{sign}\left(  \frac{\pi}{2}-\sigma_{2}\right)  \delta\left(
\sigma_{1},\sigma_{2}\right)  -\text{sign}\left(  -\left(  \frac{\pi}%
{2}-\sigma_{2}\right)  \right)  \delta\left(  \sigma_{1},\pi-\sigma_{2}\right)
\nonumber\\
&  =\left(  \delta\left(  \sigma_{1},\sigma_{2}\right)  +\delta\left(
\sigma_{1},\pi-\sigma_{2}\right)  \right)  \text{sign}\left(  \frac{\pi}%
{2}-\sigma_{2}\right) \nonumber\\
&  =\delta^{+}\left(  \sigma_{1},\sigma_{2}\right)  ~\text{sign}\left(
\frac{\pi}{2}-\sigma_{2}\right)  . \label{deltaHat2}%
\end{align}
Thus, $\hat{\delta}_{+-}\left(  \sigma_{1},\sigma_{2}\right)  $ is a
distribution that has the following important properties under reflections
from the midpoint: it is even under $\sigma_{1}\rightarrow\left(  \pi
-\sigma_{1}\right)  $ and odd under $\sigma_{2}\rightarrow\left(  \pi
-\sigma_{2}\right)  $ and vanishes at both $\sigma_{1}=\pi/2$ and $\sigma
_{2}=\pi/2$. These properties can be verified from the two equivalent forms of
$\hat{\delta}_{+-}\left(  \sigma_{1},\sigma_{2}\right)  $ given above, or by
integrating with smooth functions, or by expanding in modes. This means in
Eq.(\ref{iQM[]}) that the midpoint $\bar{x}^{M}\left(  \varepsilon\right)
=x_{+}^{M}\left(  \pi/2,\varepsilon\right)  $ star-commutes with
$p_{-N}\left(  \sigma_{2}\right)  $ including $\sigma_{2}=\pi/2,$ so the
midpoint does not participate in the joining operation of strings - this is
one of the desired properties of the $\star$ product as emphasized in point 7
in section (\ref{new*}). This vital property is encoded in the $\star$ product
in Eq.(\ref{msftstar}) as well as the $\hat{\delta}_{+-}\left(  \sigma
_{1},\sigma_{2}\right)  $ that appears in the iQM.

Central features of the iQM follows from the following left and right products
of the field with the classical half phase space $x_{+}\left(  \sigma
,\varepsilon\right)  ,p_{-}\left(  \sigma\right)  $
\begin{align}
x_{+}^{M}\left(  \sigma,\varepsilon\right)  \star A\left(  x_{+},p_{-}\right)
&  =\left(  x_{+}^{M}\left(  \sigma,\varepsilon\right)  +\frac{i}%
{2}\text{sign}\left(  \frac{\pi}{2}-\sigma\right)  \partial_{p_{-M}\left(
\sigma\right)  }\right)  A\left(  x_{+},p_{-}\right) \label{starRules1}\\
A\left(  x_{+},p_{-}\right)  \star x_{+}^{M}\left(  \sigma,\varepsilon\right)
&  =A\left(  x_{+},p_{-}\right)  \left(  x_{+}^{M}\left(  \sigma
,\varepsilon\right)  -\frac{i}{2}\overleftarrow{\partial}_{p_{-M}\left(
\sigma\right)  }\text{sign}\left(  \frac{\pi}{2}-\sigma\right)  \right)
\label{starRules2}\\
p_{-M}\left(  \sigma\right)  \star A\left(  x_{+},p_{-}\right)   &  =\left(
p_{-M}\left(  \sigma\right)  -\frac{i}{2}\text{sign}\left(  \frac{\pi}%
{2}-\sigma\right)  \partial_{x_{+}^{M}\left(  \sigma,\varepsilon\right)
}\right)  A\left(  x_{+},p_{-}\right) \label{starRules3}\\
A\left(  x_{+},p_{-}\right)  \star p_{-M}\left(  \sigma\right)   &  =A\left(
x_{+},p_{-}\right)  \left(  p_{-M}\left(  \sigma\right)  +\frac{i}%
{2}\overleftarrow{\partial}_{x_{+}^{M}\left(  \sigma,\varepsilon\right)
}\text{sign}\left(  \frac{\pi}{2}-\sigma\right)  \right)  \label{starRules4}%
\end{align}
By comparing these results to Eqs.(\ref{XA},\ref{PA}) we see that the full
string QM operators $\hat{X}^{M}\left(  \sigma,\varepsilon\right)  ,\hat
{P}_{M}\left(  \sigma\right)  $ can be represented in terms of the string
joining star product in the half phase space as follows%
\begin{align}
\hat{X}^{M}\left(  \sigma,\varepsilon\right)  A\left(  x_{+},p_{-}\right)   &
=\left[
\begin{array}
[c]{l}%
\theta\left(  \frac{\pi}{2}-\sigma\right)  x_{+}^{M}\left(  \sigma
,\varepsilon\right)  \star A\left(  x_{+},p_{-}\right) \\
+\theta\left(  \sigma-\frac{\pi}{2}\right)  A\left(  x_{+},p_{-}\right)  \star
x_{+}^{M}\left(  \sigma,\varepsilon\right)  \left(  -1\right)  ^{MA}%
\end{array}
\right]  ,\label{iQMX}\\
\hat{P}_{M}\left(  \sigma\right)  A\left(  x_{+},p_{-}\right)   &
=e^{-\varepsilon\left\vert \partial_{\sigma}\right\vert }\left[
\begin{array}
[c]{l}%
\theta\left(  \frac{\pi}{2}-\sigma\right)  p_{-M}\left(  \sigma\right)  \star
A\left(  x_{+},p_{-}\right) \\
+\theta\left(  \sigma-\frac{\pi}{2}\right)  A\left(  x_{+},p_{-}\right)  \star
p_{-M}\left(  \sigma\right)  \left(  -1\right)  ^{MA}%
\end{array}
\right]  , \label{iQMP}%
\end{align}
where the sign factor $\left(  -1\right)  ^{MA}$ accounts for boson/fermion
properties of the field $A$ and the operator labeled by $M.$ Described in
words, the structures (\ref{iQMX},\ref{iQMP}) show that depending on whether
$\sigma$ is less or more than $\pi/2$ the action of the full string quantum
operators $\hat{X}^{M}\left(  \sigma,\varepsilon\right)  ,\hat{P}_{M}\left(
\sigma\right)  $ on the field is reproduced by the left or right Moyal $\star$
product with the \textit{half phase space}. In more detail, one can check that
the star product reproduces the non-derivative as well the derivative terms in
Eqs.(\ref{XA},\ref{PA}) for all $0\leq\sigma\leq\pi,$ including $\sigma
=\pi/2,$ since
\begin{align}
1  &  =\theta\left(  \frac{\pi}{2}-\sigma\right)  +\theta\left(  \sigma
-\frac{\pi}{2}\right)  ,\\
1  &  =\text{sign}\left(  \frac{\pi}{2}-\sigma\right)  \left(  \theta\left(
\frac{\pi}{2}-\sigma\right)  -\theta\left(  \sigma-\frac{\pi}{2}\right)
\right)  . \label{signPi/2}%
\end{align}
It is clear that this works correctly as long as $\sigma\neq\pi/2.$ In order
to also work correctly at $\sigma=\pi/2$ we must define carefully what values
the symbols sign$\left(  \frac{\pi}{2}-\sigma\right)  $, $\theta\left(
\frac{\pi}{2}-\sigma\right)  ,~\theta\left(  \sigma-\frac{\pi}{2}\right)  $
take at $\sigma=\pi/2.$ Thus, in our \textit{definition}, the distribution
sign$\left(  \frac{\pi}{2}-\sigma\right)  $ does not vanish at $\sigma=\pi/2,$
but rather its value at $\pi/2$ is $\pm1$ depending on the approach to the
midpoint from below or above as $\sigma\rightarrow\pi/2\mp0$. Similarly, in
our definitions, the functions $\theta\left(  \frac{\pi}{2}-\sigma\right)  $
or $\theta\left(  \sigma-\frac{\pi}{2}\right)  $ do not take the value $1/2$
at $\sigma=\pi/2$, but rather they are equal to $1$ or $0$ depending on the
approach to the midpoint from below or above as $\sigma\rightarrow\pi/2\mp0$.
Then, sign$\left(  \frac{\pi}{2}-\sigma\right)  =\theta\left(  \frac{\pi}%
{2}-\sigma\right)  -\theta\left(  \sigma-\frac{\pi}{2}\right)  ,$ takes the
values $\pm1$ as usual, except that this does not vanish at $\pi/2$ due to the
careful definition$.$ Hence Eq.(\ref{signPi/2}) is satisfied for all
$0\leq\sigma\leq\pi,$ including $\sigma=\pi/2.$

One can now check that all the rules of quantum mechanics from
Eq.(\ref{basic[XP]}) to Eq.(\ref{PA}) are correctly reproduced by the iQM
representation of the operators (\ref{iQMX},\ref{iQMP}), including at the
midpoint $\sigma=\pi/2$. From now on we do not need anymore the $\pm$ labels
on the $x_{+},p_{-}$ and we can write the representation of the quantum
operators more simply as
\begin{align}
\hat{X}^{M}\left(  \sigma,\varepsilon\right)  A\left(  x,p\right)   &
=\left\{
\begin{array}
[c]{l}%
x^{M}\left(  \sigma,\varepsilon\right)  \star A\left(  x,p\right)  ,\text{ if
}0\leq\sigma\leq\pi/2\\
A\left(  x,p\right)  \star x^{M}\left(  \sigma,\varepsilon\right)  \left(
-1\right)  ^{MA},\text{ if }\pi/2\leq\sigma\leq\pi
\end{array}
\right.  ,\label{iQMX2}\\
\hat{P}_{M}\left(  \sigma\right)  A\left(  x,p\right)   &  =\left\{
\begin{array}
[c]{l}%
\left(  e^{-\varepsilon\left\vert \partial_{\sigma}\right\vert }p_{M}\left(
\sigma\right)  \right)  \star A\left(  x,p\right)  ,\text{ if }0\leq\sigma
\leq\pi/2\\
A\left(  x,p\right)  \star\left(  e^{-\varepsilon\left\vert \partial_{\sigma
}\right\vert }p_{M}\left(  \sigma\right)  \right)  \left(  -1\right)
^{MA},\text{ if }\pi/2\leq\sigma\leq\pi
\end{array}
\right.  . \label{iQMP2}%
\end{align}
We also do not need to watch too carefully the midpoint $\bar{x}\left(
\varepsilon\right)  $ in most cases since we have seen that $\bar{x}\left(
\varepsilon\right)  $ acts trivially (like a number, or an eigenvalue) under
the $\star$ in iQM. If need be, at $\sigma=\pi/2$ we can use Eqs.(\ref{iQMX}%
,\ref{iQMP}) or equivalently Eqs.(\ref{XA},\ref{PA}) if more care is warranted
in some computations.

Indeed, note that the $\star$ in iQM in Eq.(\ref{iQMP}) does produce correctly
a derivative contribution of the momentum operator $\hat{P}_{M}\left(
\sigma\right)  $ at the midpoint with a regulator $\hat{P}_{M}\left(
\pi/2\right)  \rightarrow\left(  -i/2\right)  e^{-\varepsilon\left\vert
\partial_{\pi/2}\right\vert }\left(  \partial/\partial x\left(  \pi
/2,\varepsilon\right)  \right)  ,$ as it should be, according to QM in
Eq.(\ref{PA}). This is a bit subtle and requires more explanation. Having
pointed out earlier that the derivatives in the star product do not act on the
midpoint when considering string joining, one may wonder how the midpoint
derivative in Eq.(\ref{PA}) is reproduced from the star product representation
in Eq.(\ref{iQMP}). This subtle point is explained as follows. Consider
evaluating the star products in (\ref{iQMP}) by using (\ref{starRules3}%
,\ref{starRules4}) and concentrate on the derivative piece which takes the
form%
\begin{equation}
-\frac{i}{4}e^{-\varepsilon\left\vert \partial_{\sigma}\right\vert }\left[
\left(  \theta\left(  \frac{\pi}{2}-\sigma\right)  -\theta\left(  \sigma
-\frac{\pi}{2}\right)  \right)  \int_{0}^{\pi}d\sigma^{\prime}\delta
^{-}\left(  \sigma,\sigma^{\prime}\right)  \text{sign}\left(  \frac{\pi}%
{2}-\sigma^{\prime}\right)  \frac{\partial A}{\partial x_{+}\left(
\sigma^{\prime},\varepsilon\right)  }\right]  , \label{midpointEmmerge1}%
\end{equation}
where the delta function arises from $\partial p_{-}\left(  \sigma\right)
/\partial p_{-}\left(  \sigma^{\prime}\right)  =\delta^{-}\left(
\sigma,\sigma^{\prime}\right)  .$ Naively this expression appears to vanish at
$\sigma=\pi/2$ since $\delta^{-}\left(  \sigma,\sigma^{\prime}\right)  $
vanishes at the midpoint, and hence no midpoint contribution; but more care is
needed. The distribution in the integrand has the following property according
to Eqs.(\ref{deltaHat1}-\ref{deltaHat2}), $\delta^{-}\left(  \sigma
,\sigma^{\prime}\right)  $sign$\left(  \frac{\pi}{2}-\sigma^{\prime}\right)
=$sign$\left(  \frac{\pi}{2}-\sigma\right)  \delta^{+}\left(  \sigma
,\sigma^{\prime}\right)  .$ Both of these forms vanish at the midpoint as
argued in (\ref{deltaHat1}-\ref{deltaHat2}). However, using the second form,
the sign function sign$\left(  \frac{\pi}{2}-\sigma\right)  $ can be pulled
out of the $d\sigma^{\prime}$ integral, and after combining it with the theta
function factor in (\ref{midpointEmmerge1}) it gives an overall factor of 1
according to Eq.(\ref{signPi/2}), including at the midpoint (we emphasize the
careful definition of the sign function). In the remaining $d\sigma^{\prime}$
integrand $\delta^{+}\left(  \sigma,\sigma^{\prime}\right)  $ by itself does
not vanish at the midpoint, and Eq.(\ref{midpointEmmerge1}) yields the
following result (recall that $\delta^{+}\left(  \sigma,\sigma^{\prime
}\right)  $ has two peaks in the range $\left[  0,\pi\right]  $)%
\begin{equation}
-\frac{i}{4}e^{-\varepsilon\left\vert \partial_{\sigma}\right\vert }\left[
\int_{0}^{\pi}d\sigma^{\prime}\delta^{+}\left(  \sigma,\sigma^{\prime}\right)
\frac{\partial A}{\partial x_{+}\left(  \sigma^{\prime},\varepsilon\right)
}\right]  =-\frac{i}{2}e^{-\varepsilon\left\vert \partial_{\sigma}\right\vert
}\frac{\partial A}{\partial x_{+}\left(  \sigma,\varepsilon\right)  }.
\label{midpointEmmerge2}%
\end{equation}
This is consistent with the expected result in Eq.(\ref{PA}) which includes
the midpoint derivative. The subtle property here is that the distribution
sign$\left(  \frac{\pi}{2}-\sigma\right)  \delta^{+}\left(  \sigma
,\sigma^{\prime}\right)  =\delta^{-}\left(  \sigma,\sigma^{\prime}\right)
$sign$\left(  \frac{\pi}{2}-\sigma^{\prime}\right)  $ has no support at the
midpoint, but $\delta^{+}\left(  \sigma,\sigma^{\prime}\right)  $ by itself
does. The theta function factor was crucial to remove the sign factor
sign$\left(  \frac{\pi}{2}-\sigma\right)  $ and lead the non-trivial midpoint
contribution. This exercise makes it evident that there are circumstances in
some computations where midpoint derivatives can arise from the new star
product, and this is indeed desirable, although straightforward generic string
joining $A\star B$ is not one of those circumstances.

\section{Regulator\label{Regulator_section}}

Consider the fundamental canonical QM operators in string theory $\left(
\hat{X}^{M}\left(  \sigma\right)  ,\hat{P}_{M}\left(  \sigma\right)  \right)
$ at fixed $\tau$ for any CFT on the worldsheet. A basic tool of computation
in CFT is the operator product expansion which is a form of regularization
that controls operator products at the same point on the worldsheet. A moment
of reflection would indicate that the same regularization effect is captured
by our proposed choice of independent degrees of freedom $\left(  \hat{X}%
^{M}\left(  \sigma,\varepsilon\right)  ,\hat{P}_{M}\left(  \sigma\right)
\right)  $ where $\hat{X}^{M}\left(  \sigma,\varepsilon\right)
=e^{-\varepsilon\left\vert \partial_{\sigma}\right\vert }\hat{X}\left(
\sigma\right)  $ is regulated while $\hat{P}_{M}\left(  \sigma\right)  $ is
not. This is because $\left\vert \partial_{\sigma}\right\vert \equiv
\sqrt{-\partial_{\sigma}^{2}}$ plays the role of an approximate time
translation operator on the worldsheet for a short amount of time even when
there are background fields present in the CFT. Thus applying a Euclidean time
translation in the form $e^{-\varepsilon\left\vert \partial_{\sigma
}\right\vert }\hat{X}\left(  \sigma\right)  $ displaces the worldsheet point
from $\left(  \sigma,\tau=0\right)  $ to $\left(  \sigma,\tau=-i\varepsilon
\right)  .$ Equivalently a point on the unit circle in the complex $z$ plane
($z\equiv e^{\pm i\sigma}$) moves to the inside of the unit circle when the
Euclidean time translator $e^{-\varepsilon\left\vert \partial_{\sigma
}\right\vert }$ acts on it, namely $e^{-\varepsilon\left\vert \partial
_{\sigma}\right\vert }z=e^{-\varepsilon\left\vert \partial_{\sigma}\right\vert
}e^{\pm i\sigma}=e^{-\varepsilon\pm i\sigma}=\tilde{z},$ with $\left\vert
\tilde{z}\right\vert <1.$ In this computation we used the fact that $e^{\pm
i\sigma}$ (and similarly $e^{\pm in\sigma}$) are degenerate eigenstates of the
operator $\left\vert \partial_{\sigma}\right\vert ,$ that is
\begin{equation}
\left\vert \partial_{\sigma}\right\vert e^{\pm in\sigma}=\sqrt{-\partial
_{\sigma}^{2}}e^{\pm in\sigma}=\left\vert n\right\vert e^{\pm in\sigma}.
\end{equation}
Hence changing $\hat{X}\left(  \sigma\right)  $ to $\hat{X}^{M}\left(
\sigma,\varepsilon\right)  $ amounts precisely to what is done in operator
products when two points are slightly displaced relative to each other, one on
the unit circle and the other inside the unit circle. By defining the theory
including the regulator $\varepsilon$ as proposed, we can control all the
relevant operator products in CFT at all intermediate stages of CFT
computations. Carrying this $\varepsilon$ over to MSFT as we have done in the
previous sections insures that all MSFT computations will be finite at all
intermediate stages. Only at the end of MSFT computations we will set
$\varepsilon=0$ after a renormalization of the cubic coupling constant.

\subsection{Regulated delta functions \label{deltaSection}}

To perform computations in MSFT we must use the differentiation rules, such as
$e^{-\varepsilon\left\vert \partial_{\sigma_{2}}\right\vert }\left(  \partial
X\left(  \sigma_{1},\varepsilon\right)  /\partial X\left(  \sigma
_{2},\varepsilon\right)  \right)  $ etc., that emerged in section
(\ref{QM=iQM}) to represent the basic quantum operators $\hat{X}\left(
\sigma,\varepsilon\right)  ,\hat{P}\left(  \sigma\right)  .$ The result of
such derivatives involves various types of delta functions, with or without
regularization, that are also sensitive to D-brane boundary conditions. In
later computations there will be circumstances in which we must multiply such
delta functions with each other. Products of unregulated delta functions are
not well defined. However, in our case the parameter $\varepsilon$ provides
just the required regularization to render such products well defined. Since
the regularized deltas will be used for computation, we provide the needed
details for them below.

The basic case from which all others follow is the delta function that appears
in the QM commutation rules in Eq.(\ref{basic[XP]}) $\left[  \hat{X}%
^{M}\left(  \sigma_{1},\varepsilon\right)  ,\hat{P}_{N}\left(  \sigma
_{2}\right)  \right]  =i\delta_{N}^{M}\delta_{\varepsilon}^{M}\left(
\sigma_{1},\sigma_{2}\right)  ,$ where the $M$ on $\delta_{\varepsilon}%
^{M}\left(  \sigma_{1},\sigma_{2}\right)  $ is a reminder that in the
direction $M$ the operators satisfy either Neumann-Neumann $\left(  nn\right)
$ or Dirichlet-Dirichlet $\left(  dd\right)  $ boundary conditions.
Accordingly $\delta_{\varepsilon}^{M}\left(  \sigma_{1},\sigma_{2}\right)  $
will be either $\delta_{\varepsilon}^{nn}\left(  \sigma_{1},\sigma_{2}\right)
$ or $\delta_{\varepsilon}^{dd}\left(  \sigma_{1},\sigma_{2}\right)  .$

Thus for $\left(  nn\right)  $ we have the following unique expression,
$\delta_{\varepsilon}^{nn}\left(  \sigma_{1},\sigma_{2}\right)
=e^{-\varepsilon\left\vert \partial_{\sigma_{1}}\right\vert }$ $\delta
^{nn}\left(  \sigma_{1},\sigma_{2}\right)  ,$ that is a periodic Dirac delta
function $\delta^{nn}\left(  \sigma_{1},\sigma_{2}\right)  $ (when
$\varepsilon=0)$ which also satisfies the boundary conditions - its
\textit{derivatives} vanish at the string ends for either $\sigma_{1}=0,\pi$
or $\sigma_{2}=0,\pi.$ The regulated version is computed easily since $\cos
n\sigma_{1}$ is an eigenstate of $\left\vert \partial_{\sigma_{1}}\right\vert
,$ namely $\left\vert \partial_{\sigma_{1}}\right\vert \cos n\sigma_{1}%
=\sqrt{-\partial_{\sigma_{1}}^{2}}\cos n\sigma_{1}=\left\vert n\right\vert
\cos n\sigma_{1}.$ Hence the regulated $\delta_{\varepsilon}^{nn}\left(
\sigma_{1},\sigma_{2}\right)  $ is
\begin{equation}
\delta_{\varepsilon}^{nn}\left(  \sigma_{1},\sigma_{2}\right)  =\frac{1}{\pi
}+\frac{2}{\pi}\sum_{n\geq1}e^{-\varepsilon n}\cos n\sigma_{1}\cos n\sigma
_{2}. \label{deltann}%
\end{equation}
By writing the cosines in terms of exponentials the series turns into a
\textit{convergent} geometric series for any positive $\varepsilon,$ so that
it can be summed up and written in the following exact form, and then
approximated for small $\varepsilon$
\begin{align}
\delta_{\varepsilon}^{nn}\left(  \sigma_{1},\sigma_{2}\right)   &  =\frac
{1}{2\pi}\left(  \frac{\sinh\frac{\varepsilon}{2}\cosh\frac{\varepsilon}{2}%
}{\sinh^{2}\frac{\varepsilon}{2}+\sin^{2}\frac{\sigma_{1}-\sigma_{2}}{2}%
}+\frac{\sinh\frac{\varepsilon}{2}\cosh\frac{\varepsilon}{2}}{\sinh^{2}%
\frac{\varepsilon}{2}+\sin^{2}\frac{\sigma_{1}+\sigma_{2}}{2}}\right)
\nonumber\\
&  \simeq\frac{\varepsilon/\pi}{\varepsilon^{2}+4\sin^{2}\frac{\sigma
_{1}-\sigma_{2}}{2}}+\frac{\varepsilon/\pi}{\varepsilon^{2}+4\sin^{2}%
\frac{\sigma_{1}+\sigma_{2}}{2}}\text{ }\label{deltann2}\\
&  \simeq\delta\left(  2\sin\frac{\sigma_{1}-\sigma_{2}}{2}\right)
+\delta\left(  2\sin\left(  \frac{\sigma_{1}+\sigma_{2}}{2}\right)  \right)
\nonumber
\end{align}
Recall that only the range $0\leq\sigma_{1},\sigma_{2}\leq\pi$ matters. The
first term has a peak at $\sigma_{1}=\sigma_{2}$ when $\sigma_{1},\sigma_{2}$
are both in the range $\left[  0,\pi\right]  ,$ while the second term has no
peak in this range unless $\sigma_{1},\sigma_{2}$ are both at the end points
$0$ or $\pi$. For example if $\sigma_{2}=0$ (or $\pi$) both terms have peaks
at $\sigma_{1}=0$ (or $\pi$), but as seen with a nonzero $\varepsilon$, only
half of the area under each curve falls within the range $\left[
0,\pi\right]  ,$ so the effect is that the end points are included with the
same strength as any interior point. This is what should be expected with
Neumann-Neumann boundary conditions.

Similarly, for Dirichlet-Dirichlet boundary conditions $\delta_{\varepsilon
}^{dd}\left(  \sigma_{1},\sigma_{2}\right)  $ appears in the commutation
rules. Due to the D-brane boundaries it must vanish at both string ends
$\sigma_{1}=0,\pi$ and $\sigma_{2}=0,\pi.$ Then it is uniquely given by%
\begin{equation}
\delta_{\varepsilon}^{dd}\left(  \sigma_{1},\sigma_{2}\right)  \equiv\frac
{2}{\pi}\sum_{n=1}^{\infty}e^{-\varepsilon n}\sin n\sigma_{1}\sin n\sigma_{2}.
\label{deltadd}%
\end{equation}
Summing up the geometric series we obtain the exact form for any $\varepsilon$
and approximate forms for small $\varepsilon$
\begin{align}
\delta_{\varepsilon}^{dd}\left(  \sigma_{1},\sigma_{2}\right)   &  =\frac
{1}{2\pi}\left(  \frac{\sinh\frac{\varepsilon}{2}\cosh\frac{\varepsilon}{2}%
}{\sinh^{2}\frac{\varepsilon}{2}+\sin^{2}\frac{\sigma_{1}-\sigma_{2}}{2}%
}-\frac{\sinh\frac{\varepsilon}{2}\cosh\frac{\varepsilon}{2}}{\sinh^{2}%
\frac{\varepsilon}{2}+\sin^{2}\frac{\sigma_{1}+\sigma_{2}}{2}}\right)
\nonumber\\
&  \simeq\frac{\varepsilon/\pi}{\varepsilon^{2}+4\sin^{2}\frac{\sigma
_{1}-\sigma_{2}}{2}}-\frac{\varepsilon/\pi}{\varepsilon^{2}+4\sin^{2}%
\frac{\sigma_{1}+\sigma_{2}}{2}}\text{ }\label{deltadd2}\\
&  \simeq\delta\left(  2\sin\frac{\sigma_{1}-\sigma_{2}}{2}\right)
-\delta\left(  2\sin\left(  \frac{\sigma_{1}+\sigma_{2}}{2}\right)  \right)
\nonumber
\end{align}
The first term has a peak at $\sigma_{1}=\sigma_{2}$ while the second one has
no peak in the range $0\leq\sigma_{1},\sigma_{2}\leq\pi$ unless $\sigma
_{1},\sigma_{2}$ are both at the end points, but at either end point the peaks
of the two terms cancel each other. So there is no support at the end points.
This is what should be expected with Dirichlet-Dirichlet boundary conditions.

Now we can compute the other delta functions, $\delta_{\varepsilon}^{\pm
}\left(  \sigma_{1},\sigma_{2}\right)  =\delta_{\varepsilon}\left(  \sigma
_{1},\sigma_{2}\right)  \pm\delta_{\varepsilon}\left(  \sigma_{1},\pi
-\sigma_{2}\right)  ,$ either $\left(  nn\right)  $ or $\left(  dd\right)  ,$
that emerge in taking derivatives with respect to $x_{+}^{M}\left(
\sigma,\varepsilon\right)  $ or $p_{M}\left(  \sigma\right)  .$ These are
given by
\begin{align}
\delta_{\varepsilon}^{+nn}\left(  \sigma_{1},\sigma_{2}\right)   &  =\frac
{2}{\pi}+\frac{4}{\pi}\sum_{e\geq2}e^{-\varepsilon e}\cos e\sigma_{1}\cos
e\sigma_{2},\;e=2,4,6,\cdots\label{deltann+}\\
\delta_{\varepsilon}^{-nn}\left(  \sigma_{1},\sigma_{2}\right)   &  =\frac
{4}{\pi}\sum_{o\geq1}e^{-\varepsilon o}\cos o\sigma_{1}\cos o\sigma
_{2},\;o=1,3,5,\cdots\label{deltann-}%
\end{align}
where $\left(  e,o\right)  $ are (even,odd) positive integers. Similarly, for
$\left(  dd\right)  $ boundary conditions we have
\begin{align}
\delta_{\varepsilon}^{+dd}\left(  \sigma_{1},\sigma_{2}\right)   &  =\frac
{4}{\pi}\sum_{o\geq1}e^{-\varepsilon o}\sin o\sigma_{1}\sin o\sigma
_{2},\;o=1,3,5,\cdots\label{deltadd+}\\
\delta_{\varepsilon}^{-dd}\left(  \sigma_{1},\sigma_{2}\right)   &  =\frac
{4}{\pi}\sum_{e\geq2}e^{-\varepsilon e}\sin e\sigma_{1}\sin e\sigma
_{2},\;e=2,4,6,\cdots\label{deltadd-}%
\end{align}
$\allowbreak$ $\allowbreak$

The infinite sums can again be performed exactly. The result is obtained by
applying the instruction $\delta_{\varepsilon}^{\pm M}\left(  \sigma
_{1},\sigma_{2}\right)  =\delta_{\varepsilon}^{M}\left(  \sigma_{1},\sigma
_{2}\right)  \pm\delta_{\varepsilon}^{M}\left(  \sigma_{1},\pi-\sigma
_{2}\right)  $ to the expressions in Eqs.(\ref{deltann2},\ref{deltann2}) for
$M=\left(  nn\right)  $ or $\left(  dd\right)  $. Their fully summed exact
expressions for any $\varepsilon$ are%
\begin{align}
\delta_{\varepsilon}^{+nn}\left(  \sigma_{1},\sigma_{2}\right)   &
=\frac{2\left(  \sinh2\varepsilon\right)  \left(  \cosh2\varepsilon
-\cos2\sigma_{1}\cos2\sigma_{2}\right)  }{\pi\left(  \cosh2\varepsilon
-\cos2\left(  \sigma_{1}-\sigma_{2}\right)  \right)  \left(  \cosh
2\varepsilon-\cos2\left(  \sigma_{1}+\sigma_{2}\right)  \right)  }\\
\delta_{\varepsilon}^{-nn}\left(  \sigma_{1},\sigma_{2}\right)   &
=\frac{8\left(  \sinh\varepsilon\right)  \left(  \cos\sigma_{1}\right)
\left(  \cos\sigma_{2}\right)  \left(  1+\cosh^{2}\varepsilon-\cos^{2}%
\sigma_{1}-\cos^{2}\sigma_{2}\right)  }{\pi\left(  \cosh2\varepsilon
-\cos2\left(  \sigma_{1}-\sigma_{2}\right)  \right)  \left(  \cosh
2\varepsilon-\cos2\left(  \sigma_{1}+\sigma_{2}\right)  \right)  }\\
\delta_{\varepsilon}^{+dd}\left(  \sigma_{1},\sigma_{2}\right)   &
=\frac{8\left(  \sinh\varepsilon\right)  \left(  \sin\sigma_{1}\right)
\left(  \sin\sigma_{2}\right)  \left(  \cos^{2}\sigma_{1}+\cos^{2}\sigma
_{2}+\sinh^{2}\varepsilon\right)  }{\pi\left(  \cosh2\varepsilon-\cos2\left(
\sigma_{1}-\sigma_{2}\right)  \right)  \left(  \cosh2\varepsilon-\cos2\left(
\sigma_{1}+\sigma_{2}\right)  \right)  }\\
\delta_{\varepsilon}^{-dd}\left(  \sigma_{1},\sigma_{2}\right)   &
=\frac{2\left(  \sinh2\varepsilon\right)  \left(  \sin2\sigma_{1}\right)
\left(  \sin2\sigma_{2}\right)  }{\pi\left(  \cosh2\varepsilon-\cos2\left(
\sigma_{1}-\sigma_{2}\right)  \right)  \left(  \cosh2\varepsilon-\cos2\left(
\sigma_{1}+\sigma_{2}\right)  \right)  }%
\end{align}
From this we see that $\delta_{\varepsilon}^{\pm M}\left(  \sigma_{1}%
,\sigma_{2}\right)  $ have two peaks in the range $0\leq\sigma_{1},\sigma
_{2}\leq\pi,$ one at $\sigma_{1}=\sigma_{2}$ and the other at $\sigma_{1}%
=\pi-\sigma_{2}.$ For small $\varepsilon,$ the $\left(  nn\right)  ,\left(
dd\right)  $ distributions are almost the same for most of the range, but they
differ close to the end points as seen by comparing the plots in Figs.(1,2),
namely $\delta_{\varepsilon}^{\pm dd}\left(  \sigma_{1},\sigma_{2}\right)  $
vanishes at the end points. In the case of $\delta_{\varepsilon}^{+M}\left(
\sigma_{1},\sigma_{2}\right)  $ both peaks are positive, but in the case of
$\delta_{\varepsilon}^{-M}\left(  \sigma_{1},\sigma_{2}\right)  $ the second
peak is negative; so when the peaks are at the midpoint $\sigma_{1}%
=\pi/2=\sigma_{2},$ the peaks in $\delta_{\varepsilon}^{+M}\left(  \sigma
_{1},\sigma_{2}\right)  $ add, while the peaks in $\delta_{\varepsilon}%
^{-M}\left(  \sigma_{1},\sigma_{2}\right)  $ cancel each other. Finally, when
the peaks are all the way at the end points, $\delta_{\varepsilon}^{\pm
nn}\left(  \sigma_{1},\sigma_{2}\right)  $ have support with half of the area
under each peak at each end point, but $\delta_{\varepsilon}^{\pm dd}\left(
\sigma_{1},\sigma_{2}\right)  $ vanishes at each end point$.$ These properties
are illustrated in Figs.(1,2) for a finite but small value of $\varepsilon.$
As $\varepsilon$ approaches zero the peaks become very tall and very narrow
while the plots for $\left(  nn\right)  $ and $\left(  dd\right)  $ appear to
converge to each other and become the same. But they are actually different
from each other exactly at the end points even when $\varepsilon=0.$

\begin{center}
\includegraphics[
natheight=2.063400in,
natwidth=3.409100in,
height=2.0634in,
width=3.4091in
]%
{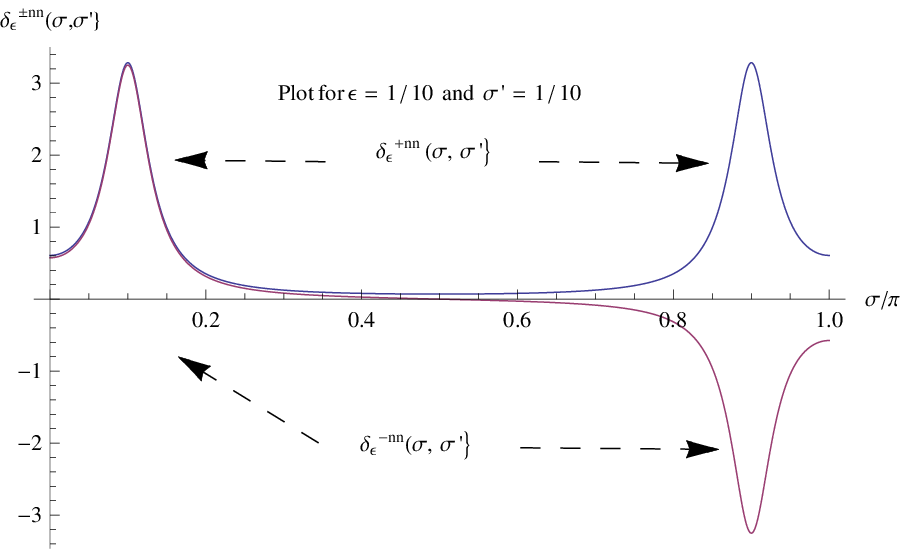} %
\\
Fig.(1) - Plot of $\delta_{\varepsilon}^{\pm nn}\left(  \sigma,\sigma^{\prime
}\right)  $ for $\varepsilon=1/10$ and $\sigma^{\prime}=1/10$%
\label{fig1}%
\end{center}

\medskip

\begin{center}
\includegraphics[
natheight=2.070400in,
natwidth=3.419500in,
height=2.0704in,
width=3.4195in
]%
{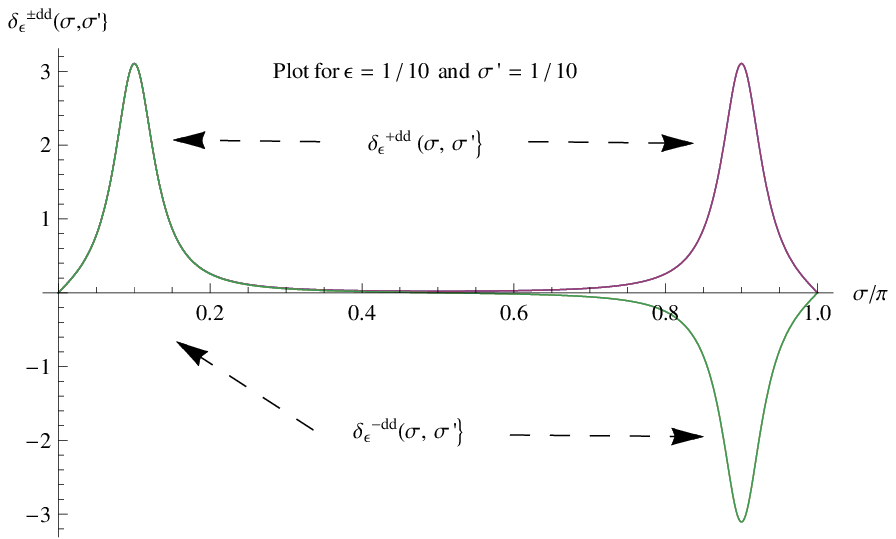} %
\\
Fig.(2) - Plot of $\delta_{\varepsilon}^{\pm dd}\left(  \sigma,\sigma^{\prime
}\right)  $ for $\varepsilon=1/10$ and $\sigma^{\prime}=1/10$%
\label{fig2}%
\end{center}

\subsection{Midpoint not treated separately}

We have seen that the new $\star$ product (\ref{msftstar}) does not treat the
midpoint in a special way, nevertheless is able to subtly exclude the midpoint
from the star product in the process of string joining. This desirable outcome
seems natural but it took a lot of effort to reach this stage: after going
through many several alternative formalisms in which the midpoint was
explicitly separated as suggested in the discussion after Eq.(\ref{Wproduct}),
we eventually discovered that there is a better way of choosing the
independent variables to label the string field, as presented above, and
\textit{then} the midpoint need not be treated separately from the rest.

To clarify this point let us show how one could setup a formalism in which the
midpoint is treated differently from the rest. The regulated delta functions
are very useful to provide the following well defined separation
\begin{equation}
x_{+}\left(  \sigma,\varepsilon\right)  =\tilde{x}\left(  \sigma
,\varepsilon\right)  +\frac{\delta_{\varepsilon}\left(  \sigma,\pi/2\right)
}{\delta_{\varepsilon}\left(  0\right)  }\bar{x}\left(  \varepsilon\right)
;\;\delta_{\varepsilon}\left(  0\right)  \equiv\delta_{\varepsilon}\left(
\pi/2,\pi/2\right)  \simeq\frac{1}{\pi\varepsilon}%
\end{equation}
where the midpoint is $\bar{x}\left(  \varepsilon\right)  =x\left(
\pi/2,\varepsilon\right)  .$ The $\tilde{x}\left(  \sigma,\varepsilon\right)
$ which vanishes at the midpoint, $\tilde{x}\left(  \pi/2,\varepsilon\right)
=0,$ is the rest of the symmetric $x_{+}\left(  \sigma,\varepsilon\right)  .$
Then treating $\tilde{x}\left(  \sigma,\varepsilon\right)  ,\bar{x}\left(
\varepsilon\right)  $ as independent variables, and using the chain rule, we
construct the derivative representation of the canonical variable $\hat{P}%
_{+}\left(  \sigma\right)  $ in Eq.(\ref{p+-Psi})%
\begin{equation}
\hat{P}_{+}\left(  \sigma\right)  \rightarrow\frac{-i}{2}\left(
e^{-\varepsilon\left\vert \partial_{\sigma}\right\vert }\frac{\partial
}{\partial x_{+}\left(  \sigma,\varepsilon\right)  }\right)  =\frac{-i}%
{2}\left(  e^{-\varepsilon\left\vert \partial_{\sigma}\right\vert }%
\frac{\partial}{\partial\tilde{x}\left(  \sigma,\varepsilon\right)  }\right)
-i\frac{\delta_{\varepsilon}\left(  \sigma,\pi/2\right)  }{\delta
_{\varepsilon}\left(  0\right)  }\frac{\partial}{\partial\bar{x}\left(
\varepsilon\right)  }. \label{p+xxtilde}%
\end{equation}
Then the differentiation rules become more complicated, such as
\begin{equation}
e^{-\varepsilon\left\vert \partial_{\sigma}\right\vert }\frac{\partial
\tilde{x}\left(  \sigma^{\prime},\varepsilon\right)  }{\partial\tilde
{x}\left(  \sigma,\varepsilon\right)  }=\delta_{\varepsilon}^{+}\left(
\sigma,\sigma^{\prime}\right)  -2\frac{\delta_{\varepsilon}\left(  \sigma
,\pi/2\right)  \delta_{\varepsilon}\left(  \sigma^{\prime},\pi/2\right)
}{\delta_{\varepsilon}\left(  0\right)  },
\end{equation}
which is consistent with vanishing at either $\sigma=\pi/2$ or $\sigma
^{\prime}=\pi/2$.

Continuing in this way the star product is constructed just like
Eq.(\ref{msftstar}) but with $\partial/\partial\tilde{x}\left(  \sigma
,\varepsilon\right)  $ appearing instead of $\partial/\partial x\left(
\sigma,\varepsilon\right)  ,$ so that it conforms to the separation of the
midpoint implied by string joining in Eq.(\ref{Wproduct}). Indeed, such a star
product is guaranteed not to touch the midpoint. This is because $\bar
{x}\left(  \varepsilon\right)  $ is independent of $\tilde{x}\left(
\sigma,\varepsilon\right)  $ and therefore $\partial\bar{x}\left(
\varepsilon\right)  /\partial\tilde{x}\left(  \sigma,\varepsilon\right)  =0$
insures that the midpoint is unaffected by the $\star.$

This reformulation can certainly be carried on, as we did for quite a while
during our investigation, and wasted quite a bit of time and effort. The
formalism became messy, cumbersome and obscure on some issues. However, we
finally noticed that the star product (\ref{msftstar}) does the same job for
string joining whether written in terms of $\partial/\partial x\left(
\sigma,\varepsilon\right)  $ or $\partial/\partial\tilde{x}\left(
\sigma,\varepsilon\right)  .$ This is because, as seen from (\ref{p+xxtilde}),
the difference in the star product (i.e. constructed with $\partial/\partial
x\left(  \sigma,\varepsilon\right)  $ as compared to $\partial/\partial
\tilde{x}\left(  \sigma,\varepsilon\right)  )$ comes from the second term on
the right hand side in the following equation
\begin{equation}
\frac{\partial}{\partial x\left(  \sigma,\varepsilon\right)  }\cdot
\frac{\partial}{\partial p\left(  \sigma\right)  }=\frac{\partial}%
{\partial\tilde{x}\left(  \sigma,\varepsilon\right)  }\cdot\frac{\partial
}{\partial p\left(  \sigma\right)  }+2\frac{\delta\left(  \sigma,\pi/2\right)
}{\delta_{\varepsilon}\left(  0\right)  }\frac{\partial}{\partial\bar
{x}\left(  \varepsilon\right)  }\cdot\frac{\partial}{\partial p\left(
\sigma\right)  } \label{dropMidpoint}%
\end{equation}
where the nonregulated delta appears in the numerator of the second term on
the right hand side because $e^{-\varepsilon\left\vert \partial_{\sigma
}\right\vert }$ has been removed from (\ref{p+xxtilde}). However the extra
piece drops out under the integral $\int d\sigma$ in the star product,
$\delta\left(  \sigma,\pi/2\right)  \left(  \partial/\partial p\left(
\sigma\right)  \right)  \rightarrow0,$ since formally $\partial/\partial
p\left(  \pi/2\right)  =0.$ This shows that the string-joining $\star$ in
(\ref{msftstar}) could avoid the midpoint even though $\partial/\partial
x\left(  \sigma,\varepsilon\right)  $ in Eq.(\ref{dropMidpoint}) appears to
include it. We are careful to say that this argument is formal because there
are delicate circumstances in which there is a midpoint contribution from the
star product as noted in Eqs.(\ref{midpointEmmerge1}-\ref{midpointEmmerge2}).
However, this is a desirable behavior of the star product in such
circumstances, hence we concluded that there is no need to separate the
midpoint from the rest of $x\left(  \sigma,\varepsilon\right)  .$ The $\star$
formalism in terms of the full $x\left(  \sigma,\varepsilon\right)  $ greatly
simplifies and becomes much easier for computations while providing new
insights as will be seen in what follows.

\section{Representations of CFT operators in MSFT\label{repr_msft}}

A key ingredient in the construction of SFT is the BRST operator $\hat{Q}_{B}$
that appears in the quadratic kinetic term. Constructing the representation of
the BRST operator in the convenient space $\left(  x\left(  \sigma
,\varepsilon\right)  ,p\left(  \sigma\right)  \right)  $ is the remaining task
to construct the MSFT action.

The BRST operator $\hat{Q}_{B}$ can be associated to any exact conformal field
theory (CFT) with any set of background fields that satisfy the exact CFT
conditions. To proceed with our formulation we first define the basic
(unregulated) canonical conjugates $\hat{X}^{M}\left(  \sigma\right)  $%
,$\hat{P}_{M}\left(  \sigma\right)  $ both for the string and ghost
degrees of freedom from the Lagrangian for the CFT. Recall that
$X^{M}$ has contravariant indices and $P_{M}\left(
\sigma,\tau\right)  =\partial S_{CFT}/\partial \left(
\partial_{\tau}X^{M}\left(  \sigma,\tau\right)  \right)  $ has
covariant indices; there is no metric involved in lowering the index
for $P_{M},$ it has a contravariant $M$ index for any set of
background fields in the CFT. Next consider for this CFT the
corresponding stress tensor $T_{\pm \pm}\left(  \sigma\right)  ,$
BRST current $J_{\pm B}\left(  \sigma\right)  $ and BRST operator
$Q_{B}$ , and if desired any vertex operator, but written in terms
of these canonical operators. Furthermore, perform normal ordering
and insert the regulator $\varepsilon$ so that
$\hat{T}_{\pm\pm}\left( \sigma,\varepsilon\right)  ,\hat{J}_{\pm
B}\left(  \sigma,\varepsilon\right) ,\hat{Q}_{B}\left(
\varepsilon\right)  $ are well defined as quantum operators. In
particular, insure that $\left(  \hat{Q}_{B}\left(
\varepsilon\right)  \right)  ^{2}=0$ as an operator in CFT when
$\varepsilon \rightarrow0$. As outlined in the previous section, the
regulator $\varepsilon$ is basically equivalent to the regulator
implied in operator products in a CFT; with our prescription, the
regulator is built in so one can proceed to computations
algebraically, using only the properties of the operators, without
any further reference to the CFT. We will illustrate this with an
example below.

\subsection{Map from CFT to MSFT and operator products \label{map}}

Once the steps above are performed for the CFT by using standard CFT methods,
the next step is to compute the representation of these operators, in
particular $\hat{Q}_{B}\left(  \varepsilon\right)  ,$ on the string field in
our basis $A\left(  x^{M}\left(  \cdot,\varepsilon\right)  ,p_{M}\left(
\cdot\right)  \right)  .$ With our setup this step is straightforward because
all we need to do is replace every operator $\hat{X}^{M}\left(  \sigma\right)
,\hat{P}_{M}\left(  \sigma\right)  $ that appears in the CFT operators by
their representations given in Eqs.(\ref{XA},\ref{PA}) as differential
operators. But an equivalent and a much more elegant representation is the
corresponding Moyal $\star$ representation in Eqs.(\ref{iQMX2},\ref{iQMP2}).
Hence a CFT operator of the form $\hat{O}\left(  \hat{X}^{M}\left(
\sigma\right)  ,\hat{P}_{M}\left(  \sigma\right)  \right)  ,$ where $\hat{O}$
is some function of the canonical variables, will act on the string field $A$
as follows%
\begin{equation}
\hat{O}\left(  \hat{X}\left(  \sigma,\varepsilon\right)  ,\hat{P}\left(
\sigma\right)  \right)  A\left(  x,p\right)  =\left\{
\begin{array}
[c]{l}%
O_{\star}\left(  x\left(  \sigma,\varepsilon\right)  ,e^{-\varepsilon
\left\vert \partial_{\sigma}\right\vert }p\left(  \sigma\right)  \right)
\star A\text{ for }0\leq\sigma\leq\frac{\pi}{2},\\
A\star O_{\star}\left(  x\left(  \sigma,\varepsilon\right)  ,e^{-\varepsilon
\left\vert \partial_{\sigma}\right\vert }p\left(  \sigma\right)  \right)
\left(  -1\right)  ^{\left\vert A\right\vert \left\vert O\right\vert }\text{
for }\frac{\pi}{2}\leq\sigma\leq\pi.
\end{array}
\right.  \label{O}%
\end{equation}
Here $\left(  -1\right)  ^{\left\vert A\right\vert \left\vert O\right\vert }$
is the sign for bose/fermi generalization. The function $O_{\star}\left(
x\left(  \sigma,\varepsilon\right)  ,e^{-\varepsilon\left\vert \partial
_{\sigma}\right\vert }p\left(  \sigma\right)  \right)  $ is star multiplied on
the left or right of $A$ depending on the value of the local worldsheet
parameter $\sigma.$ The functional form of $O_{\star}\left(  x,e^{-\varepsilon
\left\vert \partial_{\sigma}\right\vert }p\right)  $ is identical to the
functional form of $\hat{O}\left(  \hat{X},\hat{P}\right)  .$ Within the
function $O_{\star}\left(  x,e^{-\varepsilon\left\vert \partial_{\sigma
}\right\vert }p\right)  $ there are star products among the factors of
$x\left(  \sigma,\varepsilon\right)  ~$or $e^{-\varepsilon\left\vert
\partial_{\sigma}\right\vert }p\left(  \sigma\right)  $ which must appear in
the same order as the original quantum ordered CFT operators in $\hat
{O}\left(  \hat{X},\hat{P}\right)  $. This representation is possible because
the Moyal $\star$ product is an associative product just like products of
quantum operators are also associative. This map from CFT operators $\hat
{O}\left(  \hat{X},\hat{P}\right)  $ to their MSFT representations $O_{\star
}\left(  x,e^{-\varepsilon\left\vert \partial_{\sigma}\right\vert }p\right)  $
follows directly from the map between QM to iQM and vice-versa.

If all the star products within $O_{\star}\left(  x,e^{-\varepsilon\left\vert
\partial_{\sigma}\right\vert }p\right)  $ are evaluated, it reduces to a
classical function of $x\left(  \sigma,\varepsilon\right)  ,e^{-\varepsilon
\left\vert \partial_{\sigma}\right\vert }p\left(  \sigma\right)  .$ The
classical $O\left(  x,e^{-\varepsilon\left\vert \partial_{\sigma}\right\vert
}p\right)  $ obtained in this way is a field just like $A\left(  x,p\right)
$. Hence the representation of the CFT operator $\hat{O}$ reduces to the Moyal
$\star$ product between two fields as follows
\begin{equation}
\hat{O}\left(  \hat{X}\left(  \sigma,\varepsilon\right)  ,\hat{P}\left(
\sigma\right)  \right)  A\left(  x,p\right)  =\left\{
\begin{array}
[c]{l}%
O\left(  x,e^{-\varepsilon\left\vert \partial_{\sigma}\right\vert }p\right)
\star A\left(  x,p\right)  \text{ \ for }0\leq\sigma\leq\frac{\pi}{2},\\
A\left(  x,p\right)  \star O\left(  x,e^{-\varepsilon\left\vert \partial
_{\sigma}\right\vert }p\right)  \left(  -1\right)  ^{\left\vert A\right\vert
\left\vert O\right\vert }\text{ for }\frac{\pi}{2}\leq\sigma\leq\pi,
\end{array}
\right.  \label{Ofield}%
\end{equation}
where the functional form of $\hat{O}\left(  x,p\right)  $ is closely related
to $O_{\star}\left(  x,e^{-\varepsilon\left\vert \partial_{\sigma}\right\vert
}p\right)  $ as just described, while $O_{\star}\left(  x,e^{-\varepsilon
\left\vert \partial_{\sigma}\right\vert }p\right)  $ has an identical form to
the CFT operator $\hat{O}\left(  \hat{X},\hat{P}\right)  $. This transparent
relationship between any CFT operator and its representation in MSFT is not
only elegant, but is also useful for practical computations in string field
theory in both flat and curved spaces. The reason is that now the mathematics
is algebraically the same as usual quantum mechanics and we can use all we
know in QM both mathematically and intuitively to perform computations in MSFT.

For clarity we provide an example of the correspondence between CFT operators
and their representation as functions with star products. Consider the normal
ordered $T_{01}$ component of the matter energy-momentum tensor for the string
in flat space $\hat{T}_{01}\left(  \hat{X},\hat{P}\right)  =\frac{1}{4}\left(
:\pi\hat{P}\left(  \sigma\right)  \partial_{\sigma}\hat{X}\left(
\sigma,\varepsilon\right)  :\right)  $, where in addition to normal ordering
we also introduced the regulator $\varepsilon.$ First define the normal
ordering and then apply this operator on a state in Moyal space in the case
$\sigma\leq\pi/2$%
\begin{align}
&  \hat{T}_{01}\left(  \hat{X}\left(  \sigma,\varepsilon\right)  ,\hat
{P}\left(  \sigma\right)  \right)  A\left(  x,p\right) \nonumber\\
&  =\frac{\pi}{4}\left(  :\hat{P}\left(  \sigma\right)  \partial_{\sigma}%
\hat{X}_{\varepsilon}\left(  \sigma,\varepsilon\right)  :\right)  A\left(
x,p\right)  =\frac{\pi}{4}\left(  \hat{P}\left(  \sigma\right)  \partial
_{\sigma}\hat{X}_{\varepsilon}\left(  \sigma,\varepsilon\right)
-\Delta^{\prime}\left(  \varepsilon\right)  \right)  A\left(  x,p\right)
\nonumber\\
&  =\frac{\pi}{4}\left(  e^{-\varepsilon\left\vert \partial_{\sigma
}\right\vert }p\left(  \sigma\right)  \star\partial_{\sigma}x\left(
\sigma,\varepsilon\right)  -\Delta^{\prime}\left(  \varepsilon\right)
\right)  \star A\left(  x,p\right) \\
&  =\frac{\pi}{4}\left(  :e^{-\varepsilon\left\vert \partial_{\sigma
}\right\vert }p\left(  \sigma\right)  \star\partial_{\sigma}x\left(
\sigma,\varepsilon\right)  :\right)  \star A\left(  x,p\right) \nonumber\\
&  =T_{01}^{\star}\left(  x,e^{-\varepsilon\left\vert \partial_{\sigma
}\right\vert }p\right)  \star A\left(  x,p\right) \nonumber
\end{align}
where we denoted the normal ordering constant, $\Delta^{\prime}\left(
\varepsilon\right)  =\lim_{\sigma^{\prime}\rightarrow\sigma}\partial
_{\sigma^{\prime}}\langle\hat{P}\left(  \sigma\right)  \hat{X}\left(
\sigma^{\prime},\varepsilon\right)  \rangle,$ which is zero in this flat
spacetime example. The $\star$ product within the operator $T_{01}^{\star
}\left(  x,e^{-\varepsilon\left\vert \partial_{\sigma}\right\vert }p\right)
=\frac{\pi}{4}:e^{-\varepsilon\left\vert \partial_{\sigma}\right\vert
}p\left(  \sigma\right)  \star\partial_{\sigma}x\left(  \sigma,\varepsilon
\right)  :$ can be evaluated to finally construct the corresponding classical
field $T_{01}\left(  x,p\right)  $, although this step may not be convenient
to perform in some cases. For example for $\hat{Q}_{B}$ it is more transparent
and easier to perform computations with $Q_{B}^{\star}$ than with the
corresponding classical function $Q_{B}.$

It is worth to note that\emph{ }the procedure in Eq. (\ref{O}) depends only on
the canonical structure, therefore the expressions are valid for any general
CFT, including any set of background fields.

The transparent relationship of the present MSFT formalism to conformal field
theory is now apparent. For example, operator products in MSFT are computed
with the same procedure in CFT by simply replacing products of first quantized
operators in CFT by their star product counterparts in MSFT applied on
$A\left(  x,p\right)  $, such as
\begin{align}
&  \hat{O}^{1}\left(  \hat{X}\left(  \sigma_{1},\varepsilon\right)  ,\hat
{P}\left(  \sigma_{1}\right)  \right)  \hat{O}^{2}\left(  \hat{X}\left(
\sigma_{2},\varepsilon\right)  ,\hat{P}\left(  \sigma_{2}\right)  \right)
A\left(  x,p\right) \nonumber\\
&  =O_{\star}^{1}\left(  x\left(  \sigma_{1},\varepsilon\right)
,e^{-\varepsilon\left\vert \partial_{\sigma_{1}}\right\vert }p\left(
\sigma_{1}\right)  \right)  \star O_{\star}^{2}\left(  x\left(  \sigma
_{2},\varepsilon\right)  ,e^{-\varepsilon\left\vert \partial_{\sigma_{2}%
}\right\vert }p\left(  \sigma_{2}\right)  \right)  \star A\left(  x,p\right)
,
\end{align}
where we assumed both $\sigma_{1}$ and $\sigma_{2}$ are smaller than $\pi/2.$
If one of them is larger than $\pi/2$ it would appear on the right side of
$A.$ The QM to iQM map we have constructed above shows that computations of
operator products in CFT, $\hat{O}^{i}\hat{O}^{j}=c^{ijk}\hat{O}^{k}$, would
be reproduced one to one in MSFT by using the Moyal $\star$ star product and
yield the same operator product coefficients $c^{ijk}$ in
\begin{equation}
\hat{O}_{\star}^{i}\star\hat{O}_{\star}^{j}=c^{ijk}\hat{O}_{\star}^{k}.
\label{opProducts}%
\end{equation}
Hence we can take over all such results that are already computed in CFT and
directly use them in MSFT without any further effort.

In particular, consider the stress tensor $T_{\pm\pm}\left(  \sigma\right)  $
for any CFT. The CFT operator products $\hat{T}_{\pm\pm}\left(  \sigma
_{1}\right)  \hat{T}_{\pm\pm}\left(  \sigma_{2}\right)  $ are directly
reproduced one to one by using the star products in MSFT $T_{\pm\pm}^{\star
}\left(  x\left(  \sigma_{1},\varepsilon\right)  ,e^{-\varepsilon\left\vert
\partial_{\sigma_{1}}\right\vert }p\left(  \sigma_{1}\right)  \right)  \star
T_{\pm\pm}^{\star}\left(  x\left(  \sigma_{2},\varepsilon\right)
,e^{-\varepsilon\left\vert \partial_{\sigma_{2}}\right\vert }p\left(
\sigma_{2}\right)  \right)  $ when $\sigma_{1},\sigma_{2}$ are both
on either side of the midpoint, and including the midpoint.

\subsection{Ghosts\label{gho}}

Up to now we have treated the ghosts in a unified notation as the fermionic
part of the OSp$\left(  d|2\right)  $ vectors. In this section we are going to
give the explicit connection to the $B_{\pm\pm}\left(  \sigma,\tau\right)
,C^{\pm}\left(  \sigma,\tau\right)  $ ghosts of conformal field theory. As
usual we use capital letters $B,C$ to denote the CFT quantities and reserve
low case letters $b,c$ for MSFT labels. Since we have the same set of ghosts
for any set of conformal background fields in the matter sector of any CFT, we
can consider the ghost space as being always in a flat background. Recall the
mode expansion%
\begin{equation}
\hat{B}_{\pm\pm}\left(  \sigma,\tau\right)  =\sum_{n=-\infty}^{\infty}\hat
{b}_{n}e^{-in\left(  \tau\pm\sigma\right)  },\;\hat{C}^{\pm}\left(
\sigma,\tau\right)  =\sum_{n=-\infty}^{\infty}\hat{c}_{n}e^{-in\left(  \tau
\pm\sigma\right)  }. \label{BCseries}%
\end{equation}
Let $\hat{B}_{\pm\pm}\left(  \sigma\right)  ,\hat{C}^{\pm}\left(
\sigma\right)  $ denote the full string first quantized operators at $\tau=0$.
Hence $\hat{B}_{\pm\pm}\left(  \sigma\right)  =\hat{B}\left(  \pm
\sigma\right)  $ and $\hat{C}^{\pm}\left(  \sigma\right)  =\hat{C}\left(
\pm\sigma\right)  ,$ so we have just two operators $\hat{B}\left(
\sigma\right)  ,\hat{C}\left(  \sigma\right)  $ that are each other's
canonical conjugates according to the first quantization of the string. After
expanding $e^{\mp in\sigma}=\cos n\sigma\mp i\sin n\sigma,$ we define
position-momentum operators $X^{b,c}\left(  \sigma\right)  $, $P_{b,c}\left(
\sigma\right)  $ as the parts of $\hat{B},\hat{C}$ associated with the
$cosine$ or $sine$ series as follows%
\begin{equation}
\hat{B}\left(  \pm\sigma\right)  =\left(  -i\hat{X}^{b}\left(  \sigma\right)
\pm\pi\partial_{\sigma}\hat{P}_{c}\left(  \sigma\right)  \right)  ,\;\;\hat
{C}\left(  \pm\sigma\right)  =\left(  \pi\hat{P}_{b}\left(  \sigma\right)  \mp
i\left\vert \partial_{\sigma}\right\vert ^{-2}\partial_{\sigma}\hat{X}%
^{c}\left(  \sigma\right)  \right)  . \label{BCXP}%
\end{equation}
The $\hat{X}^{c,b}\left(  \sigma\right)  ,\hat{P}_{c,b}\left(  \sigma\right)
$ are $cosine$ series, just like the matter sector $\left(  X^{\mu}\left(
\sigma\right)  ,P_{\mu}\left(  \sigma\right)  \right)  $ when the latter are
in flat space,
\begin{equation}
\hat{X}^{b,c}\left(  \sigma\right)  =\hat{X}_{0}^{b,c}+\sqrt{2}\sum_{n\geq
1}\hat{X}_{n}^{b,c}\cos n\sigma,\;\pi\hat{P}_{b,c}\left(  \sigma\right)
=\hat{P}_{\left(  b,c\right)  0}+\sqrt{2}\sum_{n\geq1}\hat{P}_{\left(
b,c\right)  n}\cos n\sigma,
\end{equation}
but the combinations that appear in $\hat{B}\left(  \sigma\right)  ,\hat
{C}\left(  \sigma\right)  $, namely $\left\vert \partial_{\sigma}\right\vert
^{-2}\partial_{\sigma}\hat{X}^{c}\left(  \sigma\right)  $ or $\pi
\partial_{\sigma}\hat{P}_{c}\left(  \sigma\right)  ,$ are $sine$ series since
$\left\vert \partial_{\sigma}\right\vert ^{k}\partial_{\sigma}\cos
n\sigma=-n^{k+1}\sin n\sigma$ for any $k.$ The correspondence for the modes is
(recall $X^{b,c}$ are antihermitian while $P_{b,c}$ are hermitian as explained
following Eq.(\ref{superxp}) )
\begin{align}
\hat{X}_{0}^{b}  &  =i\hat{b}_{0},\;\hat{X}_{n\geq1}^{b}=\frac{i}{\sqrt{2}%
}\left(  \hat{b}_{n}+\hat{b}_{-n}\right)  ,\;\hat{X}_{n\geq1}^{c}=-\frac
{n}{\sqrt{2}}\left(  \hat{c}_{n}-\hat{c}_{-n}\right)  ,\label{XcbModes}\\
\hat{P}_{b0}  &  =\hat{c}_{0},\;\;\hat{P}_{b,n\geq1}=\frac{1}{\sqrt{2}}\left(
\hat{c}_{n}+\hat{c}_{-n}\right)  ,\;\hat{P}_{c,n\geq1}=\frac{i}{\sqrt{2}%
n}\left(  \hat{b}_{n}-\hat{b}_{-n}\right)  , \label{PcbModes}%
\end{align}
and their inverse is%
\begin{align}
\hat{b}_{0}  &  =-i\hat{X}_{0}^{b},\;\hat{b}_{n\geq1}=\frac{i}{\sqrt{2}%
}\left(  -\hat{X}_{n\geq1}^{b}-n\hat{P}_{c,n\geq1}\right)  ,\;\hat{b}_{\left(
-n\leq1\right)  }=\frac{i}{\sqrt{2}}\left(  -\hat{X}_{n\geq1}^{b}+n\hat
{P}_{c,n\geq1}\right)  ,\\
\hat{c}_{0}  &  =\hat{P}_{b0},\;\hat{c}_{n\geq1}=\frac{1}{\sqrt{2}}\left(
-\frac{1}{n}\hat{X}_{n\geq1}^{c}+\hat{P}_{b,n\geq1}\right)  ,\;\hat
{c}_{\left(  -n\leq1\right)  }=\frac{1}{\sqrt{2}}\left(  \frac{1}{n}\hat
{X}_{n\geq1}^{c}+\hat{P}_{b,n\geq1}\right)  .
\end{align}
The remaining zero modes ($\hat{X}_{0}^{c},\hat{P}_{0c})$ drop out in these
expressions for $\hat{B}\left(  \sigma\right)  ,\hat{C}\left(  \sigma\right)
$ because $\left\vert \partial_{\sigma}\right\vert ^{k}\partial_{\sigma}$
applied on a constant is zero. So both ($\hat{X}_{0}^{c},\hat{P}_{0c})$ may be
taken as zero or they may be treated as additional non-vanishing zero modes in
intermediate stages of the formalism. In any case they drop out in the
relevant structures of the MSFT dynamics$.$

We now introduce the regulated operators for ghosts, $\hat{X}^{c,b}\left(
\sigma,\varepsilon\right)  ,\hat{P}_{c,b}\left(  \sigma\right)  ,$ namely
$\hat{X}^{c,b}\left(  \sigma,\varepsilon\right)  \equiv e^{-\varepsilon
\left\vert \partial_{\sigma}\right\vert }\hat{X}^{c,b}\left(  \sigma\right)
,$ while $\hat{P}_{c,b}\left(  \sigma\right)  $ remain unregulated, in
parallel to the matter sector. Since the $\hat{X}^{c,b}\left(  \sigma
,\varepsilon\right)  ,\hat{P}_{c,b}\left(  \sigma\right)  $ satisfy $\left(
nn\right)  $ boundary conditions, their QM anticommutation rules are
($m=c,b$)
\begin{equation}
\left\{  \hat{X}^{m}\left(  \sigma,\varepsilon\right)  ,\hat{P}_{m^{\prime}%
}\left(  \sigma^{\prime}\right)  \right\}  =i\delta_{m^{\prime}}^{m}%
~\delta_{\varepsilon}^{nn}\left(  \sigma,\sigma^{\prime}\right)  .
\label{XbcPbcQM}%
\end{equation}
Note that $i$ appears on the right hand side of Eq.(\ref{XbcPbcQM}) in accord
with the comments following Eq.(\ref{superxp}). The corresponding
anticommutation rules for the modes $\hat{X}_{n}^{\left(  b,c\right)  }%
,\hat{P}_{\left(  b,c\right)  n}$ are consistent with the anticommutation
rules for the ghost operators $\hat{B}\left(  \sigma\right)  ,\hat{C}\left(
\sigma\right)  $ or the ghost modes $\left\{  \hat{b}_{\pm n},\hat{c}_{\mp
n^{\prime}}\right\}  =\delta_{nn^{\prime}}.$

Just like the matter sector, the ghosts $X^{m}\left(  \sigma,\varepsilon
\right)  ,P_{m}\left(  \sigma\right)  $ are split into their even and odd
parts $x_{\pm}^{m}\left(  \sigma,\varepsilon\right)  ,p_{\pm m}\left(
\sigma\right)  $ as in Eqs.(\ref{x+-},\ref{p+-}) and then treated in a unified
way with the bosons as part of the OSp$\left(  d|2\right)  $ supervectors as
we did in all previous sections. Then the string field is labeled as $A\left(
x_{+}^{M}\left(  \sigma,\varepsilon\right)  ,p_{-M}\left(  \sigma\right)
\right)  $ including the eigenvalues of the simultaneous ghost observables
$\left(  \hat{x}_{+}^{b},\hat{p}_{-b},\hat{x}_{+}^{c},\hat{p}_{-c}\right)  .$
Next we compute the representation of the full ghost operators on the string
field which may be labeled by the eigenvalues as $A\left(  x_{+}%
^{b,c},p_{-b,c}\right)  $ or equivalently as $A\left(  b,c\right)  $ where
$\left(  b,c\right)  $ refer to Eqs.(\ref{bcxp}-\ref{cseries}) below. This
corresponds to specializing Eqs.(\ref{XA},\ref{PA}), Eqs.(\ref{iQMX}%
,\ref{iQMP}) and Eqs.(\ref{iQMX2},\ref{iQMP2}) to the case $M=m=\left(
b,c\right)  .$ From these it is useful to extract the $\star$ representation
of the regulated full string ghost operators%
\begin{align}
\hat{B}\left(  \pm\sigma,\varepsilon\right)   &  =\left(  -ie^{-\varepsilon
\left\vert \partial_{\sigma}\right\vert }\hat{X}^{b}\left(  \sigma
,\varepsilon\right)  \pm\pi\partial_{\sigma}\hat{P}_{c}\left(  \sigma\right)
\right)  ,\label{BCXP-regulated}\\
\hat{C}\left(  \pm\sigma,\varepsilon\right)   &  =\left(  \pi\hat{P}%
_{b}\left(  \sigma\right)  \mp ie^{-\varepsilon\left\vert \partial_{\sigma
}\right\vert }\left\vert \partial_{\sigma}\right\vert ^{-2}\partial_{\sigma
}\hat{X}^{c}\left(  \sigma,\varepsilon\right)  \right)  ,
\end{align}
as follows%
\begin{equation}
\hat{B}\left(  +\sigma,\varepsilon\right)  A\left(  b,c\right)  =\left\{
\begin{array}
[c]{l}%
b\left(  \sigma,\varepsilon\right)  \star A\left(  b,c\right)  ,\text{ if
}0\leq\sigma\leq\pi/2\\
A\left(  b,c\right)  \star b\left(  \sigma,\varepsilon\right)  \left(
-1\right)  ^{A},\text{ if }\pi/2\leq\sigma\leq\pi
\end{array}
\right.  , \label{BonA}%
\end{equation}
and%
\begin{equation}
\hat{C}\left(  +\sigma,\varepsilon\right)  A\left(  b,c\right)  =\left\{
\begin{array}
[c]{l}%
c\left(  \sigma,\varepsilon\right)  \star A\left(  b,c\right)  ,\text{ if
}0\leq\sigma\leq\pi/2\\
A\left(  b,c\right)  \star c\left(  \sigma,\varepsilon\right)  \left(
-1\right)  ^{A},\text{ if }\pi/2\leq\sigma\leq\pi
\end{array}
\right.  . \label{ConA}%
\end{equation}
The low case $b\left(  \sigma,\varepsilon\right)  ,c\left(  \sigma
,\varepsilon\right)  $ correspond to the following combinations of $\left(
x_{+}^{b,c},p_{-b,c}\right)  $%
\begin{equation}%
\begin{array}
[c]{l}%
b\left(  \sigma,\varepsilon\right)  =e^{-\varepsilon\left\vert \partial
_{\sigma}\right\vert }\left(  -ix_{+}^{b}\left(  \sigma,\varepsilon\right)
+\pi\partial_{\sigma}p_{-c}\left(  \sigma\right)  \right)  ,\\
c\left(  \sigma,\varepsilon\right)  =e^{-\varepsilon\left\vert \partial
_{\sigma}\right\vert }\left(  \pi p_{-b}\left(  \sigma\right)  -i\left\vert
\partial_{\sigma}\right\vert ^{-2}\partial_{\sigma}x_{+}^{c}\left(
\sigma,\varepsilon\right)  \right)  .
\end{array}
\; \label{bcxp}%
\end{equation}
When written in terms of the modes, this gives the regulated $b\left(
\sigma,\varepsilon\right)  ,c\left(  \sigma,\varepsilon\right)  $ in terms of
the unregulated odd/even modes of the ghosts in Eq.(\ref{BCseries})%
\begin{align}
b\left(  \sigma,\varepsilon\right)   &  =b_{0}+\sum_{o\geq1}\left(
b_{e}+b_{-e}\right)  e^{-2\varepsilon e}\cos e\sigma-i\sum_{e\geq2}\left(
b_{o}-b_{-o}\right)  e^{-\varepsilon o}\sin o\sigma,\label{bseries}\\
c\left(  \sigma,\varepsilon\right)   &  =\sum_{e\geq2}\left(  c_{o}%
+c_{-o}\right)  e^{-\varepsilon o}\cos o\sigma-i\sum_{o\geq1}\left(
c_{e}-c_{-e}\right)  e^{-2\varepsilon e}\sin e\sigma, \label{cseries}%
\end{align}
where $e=2,4,6,\cdots$ are positive even integers and $o=1,3,5,\cdots$ are
positive odd integers. Clearly $\left(  b\left(  \sigma,\varepsilon\right)
,c\left(  \sigma,\varepsilon\right)  \right)  $ is only half of the ghost
phase space in (\ref{BCseries}). Note that $c\left(  \sigma,\varepsilon
\right)  $ has no zero mode as mentioned after Eq.(\ref{PcbModes}).
Furthermore, $b$ is even while $c$ is odd under reflections from the midpoint
\begin{equation}
b\left(  \sigma,\varepsilon\right)  =b\left(  \pi-\sigma,\varepsilon\right)
,\;c\left(  \sigma,\varepsilon\right)  =-c\left(  \pi-\sigma,\varepsilon
\right)  . \label{bc-oddeven}%
\end{equation}
The $\left(  b\left(  \sigma,\varepsilon\right)  ,c\left(  \sigma
,\varepsilon\right)  \right)  $ given above are constructed from the
eigenvalues of the ghost operators, $\hat{b}_{n}$ and $\hat{c}_{n},$ in such
combinations that clearly anticommute with each other under the standard QM
rules. By contrast, under the string-joining $\star$ product in
(\ref{msftstar}) $\left(  b\left(  \sigma,\varepsilon\right)  ,c\left(
\sigma,\varepsilon\right)  \right)  $ do not commute with each other in the
induced iQM. Using Eq.(\ref{iQM[]}), and (\ref{bcxp}), we compute that they
satisfy the following iQM anticommutation rule%
\begin{equation}
\left\{  b\left(  \sigma_{1},\varepsilon\right)  ,c\left(  \sigma
_{2},\varepsilon\right)  \right\}  _{\star}=2\pi e^{-\varepsilon\left\vert
\partial_{\sigma_{1}}\right\vert }e^{-\varepsilon\left\vert \partial
_{\sigma_{2}}\right\vert }\hat{\delta}^{nn+dd}\left(  \sigma_{1},\sigma
_{2}\right)  , \label{bcstar}%
\end{equation}
where
\begin{equation}
\hat{\delta}^{nn+dd}\left(  \sigma_{1},\sigma_{2}\right)  =\frac{1}{2}\left(
\delta^{+nn}\left(  \sigma_{1},\sigma_{2}\right)  \text{sign}\left(  \frac
{\pi}{2}-\sigma_{2}\right)  +\text{sign}\left(  \frac{\pi}{2}-\sigma
_{1}\right)  \delta^{-dd}\left(  \sigma_{1},\sigma_{2}\right)  \right)  .
\end{equation}
where $\delta^{+nn}\left(  \sigma_{1},\sigma_{2}\right)  ,$ $\delta
^{-dd}\left(  \sigma_{1},\sigma_{2}\right)  $ are given in (\ref{deltann+}%
-\ref{deltadd-}) and figures Fig. 1,2. The $sign$ functions that appear in
$\hat{\delta}^{nn+dd}\left(  \sigma_{1},\sigma_{2}\right)  $ are essential.
This expression obeys similar relations to Eqs.(\ref{deltaHat1}%
-\ref{deltaHat2}). From the midpoint properties of $\hat{\delta},$ namely
$\hat{\delta}\left(  \pi/2,\sigma_{2}\right)  =0=\hat{\delta}\left(
\sigma_{1},\pi/2\right)  ,$ we see that the midpoint ghost degrees of freedom
act trivially under the string joining $\star$ product, just as desired.

In practical computations sometimes it is useful to use the star
product directly in terms of $b,c$ as in Eq.(\ref{bcstar}) or
sometimes revert back to $x_{+}^{b,c},p_{-b,c}$ through
Eq.(\ref{bcxp}) and write everything in terms of $x^{M}\left(
\sigma,\varepsilon\right)  ,p_{M}\left(  \sigma\right)  $ to take
advantage of the OSp$\left(  d|2\right)  $ symmetry of the $\star$
product. The $\left(  b,c\right)  $ basis is useful for constructing
the representation of the BRST operator as in the next section.

\subsection{Stress tensor, BRST current and BRST operator\label{BRSTfield}}

In this section we discuss the BRST charge and the associated
building blocks, stress tensor and BRST current, in the MSFT
formalism. The plan is to first define the regulated first quantized
operators in QM and then construct their iQM representations in
terms of only the string-joining Moyal star product

\subsubsection{Regulated QM operators}

The BRST operator in QM is defined as an integral over the left and right
moving BRST currents for the full string \cite{GSW}%
\begin{equation}
\hat{Q}_{B}\left(  \varepsilon\right)  =\frac{1}{\pi}\int_{0}^{\pi}\hat{J}%
_{B}\left(  \sigma,\varepsilon\right)  d\sigma,\;\text{with }\hat{J}%
_{B}\left(  \sigma,\varepsilon\right)  \equiv\frac{1}{2}\left(  \hat{J}%
_{+}^{B}\left(  \sigma,\varepsilon\right)  +\hat{J}_{-}^{B}\left(
\sigma,\varepsilon\right)  \right)  , \label{Q_int}%
\end{equation}
where we also introduce the regulator $\varepsilon$ as shown below. In the QM
operator formalism the regulated and normal ordered current is defined as%
\begin{equation}
\hat{J}_{\pm}^{B}\left(  \sigma,\varepsilon\right)  =:\left(
\begin{array}
[c]{c}%
2\hat{C}\left(  \pm\sigma,\varepsilon\right)  \left(  \hat{T}_{\pm\pm}%
^{m}\left(  \sigma,\varepsilon\right)  +\frac{1}{2}\hat{T}_{\pm\pm}%
^{gh}\left(  \sigma,\varepsilon\right)  \right)  +a\hat{C}\left(  \pm
\sigma,\varepsilon\right) \\
+\frac{3}{2}\left(  -i\partial_{\pm\sigma}\right)  \left(  -i\partial
_{\pm\sigma}+1\right)  \hat{C}\left(  \pm\sigma,\varepsilon\right)
\end{array}
\right)  :,~~~~a=-1
 \label{JBqm}
\end{equation}
When $\varepsilon\rightarrow0$ this agrees with standard definitions, e.g. see
\cite{GSW}, \cite{Pochinski}$.$ The total derivative term in the second line
drops out in the computation of the integral BRST operator, but is needed to
insure that $\hat{J}_{\pm}^{B}$ is a conformal tensor. The constant
coefficient $a$ arises from normal ordering, and is fixed to $a=-1$ by
requiring the BRST operator to satisfy $\hat{Q}_{B}^{2}=0.$

To provide an example of the consistent regularization, we take the case of
the flat space CFT with trivial background fields, where the regularized
operator for the matter energy-momentum tensor\emph{ }$\hat{T}_{\pm\pm}%
^{m}\left(  \sigma,\varepsilon\right)  $ takes the form\emph{ }%
\begin{equation}
\hat{T}_{\pm\pm}^{m}\left(  \sigma,\varepsilon\right)  =:\frac{1}{4}\left[
\pi\hat{P}^{\mu}\left(  \sigma\right)  \pm e^{-\varepsilon\left\vert
\partial_{\sigma}\right\vert }\partial_{\sigma}\hat{X}^{\mu}\left(
\sigma,\varepsilon\right)  \right]  ^{2}: \label{flatTm}%
\end{equation}
with the regulator $\varepsilon$ included. There is also\emph{ }an identical
form for\emph{ } the regulated ghost stress tensor, which is the same for all
CFTs, as given below in Eq.(\ref{TghostSp2}). In these expressions we could
have written $e^{-\varepsilon\left\vert \partial_{\sigma}\right\vert }\hat
{X}\left(  \sigma,\varepsilon\right)  =\hat{X}\left(  \sigma,2\varepsilon
\right)  $, but we should keep it as given in (\ref{flatTm}) because we are
committed to the notation that the independent degrees of freedom are
designated as $\hat{X}\left(  \sigma,\varepsilon\right)  .$ The reader
familiar with string theory can verify that the expressions above revert to
the familiar unregulated expressions in the $\varepsilon\rightarrow0$ limit
\cite{GSW}.

Next we introduce the regulated ghost stress tensor $\hat{T}_{\pm\pm}%
^{gh}\left(  \sigma,\varepsilon\right)  $ in terms of the regulated $\hat
{C}\left(  \pm\sigma,\varepsilon\right)  ,\hat{B}\left(  \pm\sigma
,\varepsilon\right)  $ operators of Eq.(\ref{BCXP-regulated}), and compute it
as follows
\begin{align}
\hat{T}_{\pm\pm}^{gh}  &  =:i\left(  \partial_{\pm\sigma}\hat{C}\left(
\pm\sigma,\varepsilon\right)  \hat{B}\left(  \pm\sigma,\varepsilon\right)
+\frac{1}{2}\hat{C}\left(  \pm\sigma,\varepsilon\right)  \partial_{\pm\sigma
}\hat{B}\left(  \pm\sigma,\varepsilon\right)  \right)  :\\
&  =:\frac{i}{2}\left(  \partial_{\pm\sigma}\hat{C}\left(  \pm\sigma
,\varepsilon\right)  \hat{B}\left(  \pm\sigma,\varepsilon\right)
+\partial_{\pm\sigma}\left(  \hat{C}\left(  \pm\sigma,\varepsilon\right)
\hat{B}\left(  \pm\sigma,\varepsilon\right)  \right)  \right)  :\nonumber\\
&  =\frac{i}{2}:\left(
\begin{array}
[c]{c}%
e^{-\varepsilon\left\vert \partial_{\sigma}\right\vert }\hat{X}^{c}\left(
\sigma,\varepsilon\right) \\
\mp\pi i\partial_{\sigma}\hat{P}_{b}\left(  \sigma\right)
\end{array}
\right)  \left(
\begin{array}
[c]{c}%
e^{-\varepsilon\left\vert \partial_{\sigma}\right\vert }\hat{X}^{b}\left(
\sigma,\varepsilon\right) \\
\pm i\pi\partial_{\sigma}\hat{P}_{c}\left(  \sigma\right)
\end{array}
\right)  :-\frac{1}{2}i\partial_{\pm\sigma}J_{\pm}^{gh}\left(  \sigma
,\varepsilon\right)  .
\end{align}
where, $J_{\pm}^{gh}\left(  \sigma,\varepsilon\right)  =\hat{C}\left(
\pm\sigma,\varepsilon\right)  \hat{B}\left(  \pm\sigma,\varepsilon\right)  ,$
is the ghost number current density\footnote{The regulated ghost number
operator is (after dropping total derivatives in the integrations below)%
\begin{align}
\hat{N}_{gh}  &  =\frac{1}{2\pi}\int_{0}^{\pi}d\sigma\sum_{\pm}:\hat{C}\left(
\pm\sigma,\varepsilon\right)  \hat{B}\left(  \pm\sigma,\varepsilon\right)
:\nonumber\\
&  =\frac{1}{\pi}\int_{0}^{\pi}d\sigma:\left(  ie^{-\varepsilon\left\vert
\partial_{\sigma}\right\vert }\hat{X}^{b}\left(  \sigma,\varepsilon\right)
\pi\hat{P}_{b}\left(  \sigma\right)  -ie^{-\varepsilon\left\vert
\partial_{\sigma}\right\vert }\hat{X}^{c}\left(  \sigma,\varepsilon\right)
\pi\hat{P}_{c}\left(  \sigma\right)  \right)  :\\
&  =g_{0}+i\left(  \hat{X}_{0}^{b}\hat{P}_{0b}-\hat{X}_{0}^{c}\hat{P}%
_{0c}+\cdots\right) \\
&  =g_{0}-x_{0}^{b}\frac{\partial}{\partial x_{0}^{b}}+x_{0}^{c}\frac
{\partial}{\partial x_{0}^{c}}+\cdots,
\end{align}
where $g_{0}$ is a charge associated with the string state. For the string
field $A$ that appears in the action we have $g_{0}=2$. See also below.\emph{
}}. $\hat{T}_{\pm\pm}^{gh}\left(  \sigma,\varepsilon\right)  $ can be written
in a more appealing Sp$\left(  2\right)  $-invariant form by raising and
lowering indices in the ghost sector, $\hat{P}^{m}=-i\epsilon^{mm^{\prime}%
}\hat{P}_{m^{\prime}}$ etc. (namely $P^{b}=-iP_{c}$ and $P^{c}=iP_{b}$), using
the antisymmetric Sp$\left(  2\right)  $ metric $\eta_{mm^{\prime}}\equiv
i\epsilon_{mm^{\prime}},$ which satisfies
\begin{equation}
\left(  -i\epsilon^{mm^{\prime}}\right)  \left(  i\epsilon_{m^{\prime}%
n}\right)  =\delta_{n}^{m}:\epsilon^{bc}=-\epsilon^{cb}=-\epsilon
_{bc}=\epsilon_{cb}=1. \label{sp2metric}%
\end{equation}
Then we can write the manifestly Sp$\left(  2\right)  $ invariant expression
for $\hat{T}_{\pm\pm}^{gh}$
\begin{equation}
\hat{T}_{\pm\pm}^{gh}=\frac{1}{4}\left(  i\epsilon_{mm^{\prime}}\right)
:\left(
\begin{array}
[c]{c}%
\pi\partial_{\sigma}\hat{P}^{m}\left(  \sigma\right) \\
\mp e^{-\varepsilon\left\vert \partial_{\sigma}\right\vert }\hat{X}^{m}\left(
\sigma,\varepsilon\right)
\end{array}
\right)  \left(
\begin{array}
[c]{c}%
\pi\partial_{\sigma}\hat{P}^{m^{\prime}}\left(  \sigma\right) \\
\mp e^{-\varepsilon\left\vert \partial_{\sigma}\right\vert }\hat{X}%
^{m^{\prime}}\left(  \sigma,\varepsilon\right)
\end{array}
\right)  :-\frac{1}{2}i\partial_{\pm\sigma}J_{\pm}^{gh}, \label{TghostSp2}%
\end{equation}

It is now interesting to note the similarities and differences between the
ghost stress tensor and the matter stress tensor in flat space given in
Eq.(\ref{flatTm}). Setting aside the extra total derivative term $\frac{1}%
{2}\partial_{\pm\sigma}J_{\pm}^{gh},$ the structure of $\hat{T}_{\pm\pm}^{gh}$
is similar to $\hat{T}_{\pm\pm}^{matter}$ in flat space except for the
following differences: the metric in flat space is the Minkowski metric
$\eta_{\mu\nu}$ whereas the metric in ghost space is the Sp$\left(  2\right)
$ metric $\eta_{mn}=i\epsilon_{mn}$; furthermore the $\partial_{\sigma}$
derivative structure is different; however had the derivative structure been
the same as the bosonic sector then there would have been an OSp$\left(
d|2\right)  $ symmetry in the kinetic energy operator (see however section
(\ref{alpha-gen}) below for an improved supersymmetric basis). Consider the
zeroth Virasoro operator $L_{0}=\frac{1}{\pi}\int_{0}^{\pi}d\sigma\sum_{\pm
}\left(  T_{\pm\pm}^{matter}+\hat{T}_{\pm\pm}^{gh}\right)  $ which is the
kinetic energy operator in the Siegel gauge as seen below. Doing integration
by parts in the $\sigma$ integral, and recalling $\left\vert \partial_{\sigma
}\right\vert =\sqrt{-\partial_{\sigma}^{2}}$, then $\hat{L}_{0}$ for the flat
CFT case takes the form%
\begin{equation}
\hat{L}_{0}^{\left(  flat\right)  }=\frac{1}{\pi}\int_{0}^{\pi}d\sigma:\left(
\begin{array}
[c]{c}%
\frac{1}{2}\eta_{\mu\nu}\left(  \pi^{2}\hat{P}^{\mu}\hat{P}^{\nu}+\hat{X}%
^{\mu}\left\vert \partial_{\sigma}\right\vert ^{2}\hat{X}^{\nu}\right) \\
+\frac{1}{2}i\epsilon_{mn}\left(  \pi^{2}\hat{P}^{m}\left\vert \partial
_{\sigma}\right\vert ^{2}\hat{P}^{n}+\hat{X}^{m}\hat{X}^{n}\right)
\end{array}
\right)  :, \label{LoOper}%
\end{equation}
This shows more clearly how the OSp$\left(  d|2\right)  $ symmetry of the star
product in the cubic term of the action is broken in the quadratic term down
to OSp$\left(  d\right)  \times$Sp$\left(  2\right)  $.

For the example of flat CFT the regulated BRST current takes the form%
\begin{equation}
\hat{J}_{B}^{(flat)}\left(  \sigma,\varepsilon\right)  =:\left(
\begin{array}
[c]{c}%
\pi\hat{P}_{b}\left[  \left(  \pi^{2}\hat{P}^{\mu}\hat{P}_{\mu}+\hat
{X}^{\prime\mu}\hat{X}_{\mu}^{\prime}\right)  +\left(  i\pi^{2}\hat{P}%
_{b}^{\prime}\hat{P}_{c}^{\prime}+i\hat{X}^{c}\hat{X}^{b}\right)  +1\right] \\
+i\pi\partial_{\sigma}^{-1}\hat{X}^{c}\left[  \hat{X}^{\prime\mu}\hat{P}_{\mu
}+\hat{P}_{c}^{\prime}\hat{X}^{c}+\hat{P}_{b}^{\prime}\hat{X}^{b}\right]
+\partial_{\sigma}\left(  ...\right)
\end{array}
\right)  :, \label{flatJB}%
\end{equation}
where all $\hat{X}^{M}$ should be replaced by their regulated form
$e^{-\varepsilon\left\vert \partial_{\sigma}\right\vert }\hat{X}^{M}\left(
\sigma,\varepsilon\right)  $ while all $\hat{P}_{M}$ remain unregulated
$\hat{P}_{M}\left(  \sigma\right)  .$ The BRST operator is the integral of
this current, $\hat{Q}_{B}^{\left(  flat\right)  }=\int_{0}^{\sigma}%
d\sigma\hat{J}_{B}^{(flat)}\left(  \sigma,\varepsilon\right)  .$ As
is well known, the vanishing $\hat{Q}_{B}^{2}=0$ requires $d=26.$

\subsubsection{iQM Representation of Regulated QM Operators}

Having defined the regulated version of the first quantized operators, next we
write the iQM representations of all these QM operators when they act on the
string field $A\left(  x,p\right)  $ by following the prescription given in
the previous section in Eq.(\ref{O}). In particular, we are interested in all
values of $\sigma$ below or above the midpoint. Hence, for the BRST current we
have (with step functions as in Eqs.(\ref{iQMX},\ref{iQMP}))%

\begin{equation}
\hat{J}_{B}\left(  \sigma,\varepsilon\right)  A\left(  x,p\right)  =\left[
\begin{array}
[c]{c}%
\theta\left(  \frac{\pi}{2}-\sigma\right)  j_{\star}\left(  \sigma
,\varepsilon\right)  \star A\left(  x,p\right) \\
+\theta\left(  \sigma-\frac{\pi}{2}\right)  A\left(  x,p\right)  \star
j_{\star}\left(  \sigma,\varepsilon\right)  \left(  -1\right)  ^{A}%
\end{array}
\right]  \label{J_star}%
\end{equation}
The iQM image of the BRST current $j_{\pm\star}\left(
\sigma,\varepsilon \right)  $ is expressed in terms of the half
phase space in both the matter and ghost sectors. In the previous
sections these were lumped together as supervectors $x^{M}\left(
\sigma,\varepsilon\right)  ,p_{M}\left( \sigma\right)  $ where the
$x^{b,c}\left(  \sigma,\varepsilon\right)  $ and $p_{b,c}\left(
\sigma\right)  $ components referred to the ghosts. As explained in
the next section, a combination of $x^{b,c},p_{b,c}$ is written in
terms of $b\left(  \sigma,\varepsilon\right)  ,c\left(  \sigma
,\varepsilon\right)  $ that form the half phase space more directly
related to the operators $\hat{B},\hat{C}$ as given in
Eqs.(\ref{BonA},\ref{ConA}) which are the equivalent of
Eq.(\ref{iQMX},\ref{iQMP}) in just the ghost sector. It is
suggestive to write the Moyal basis expression for
$J_{\pm\star}\left( \sigma,\varepsilon\right)  $ in terms of
$b\left(  \sigma,\varepsilon\right) ,c\left(
\sigma,\varepsilon\right)  $ rather than $x^{b,c}\left(
\sigma,\varepsilon\right)  $ and $p_{b,c}\left(  \sigma\right)  $ as
follows so that the iQM parallel to the operator QM version in
(\ref{JBqm}) is evident
\begin{equation}
j_{\star}\left(  \sigma,\varepsilon\right)  =\sum_{\pm}:\left(
\begin{array}
[c]{c}%
2c\left(  \pm\sigma,\varepsilon\right)  \left(  T_{\pm\pm\star}^{m}\left(
\sigma,\varepsilon\right)  +\frac{1}{2}T_{\pm\pm\star}^{gh}\left(
\sigma,\varepsilon\right)  \right)  +ac\left(  \pm\sigma,\varepsilon\right) \\
+\frac{3}{2}\left(  -i\partial_{\pm\sigma}\right)  \left(  \left(
-i\partial_{\pm\sigma}\right)  +1\right)  c\left(  \pm\sigma,\varepsilon
\right)
\end{array}
\right)  :~. \label{jstar}%
\end{equation}
Taking the integral of both sides of Eq.(\ref{J_star}), we obtain the
representation of the QM BRST charge in the MSFT formalism%

\begin{align}
\hat{Q}_{B}A\left(  x,p\right)   &  =\left(  \frac{1}{2\pi}\int_{0}^{\pi
/2}d\sigma j_{\star}\left(  \sigma,\varepsilon\right)  \right)  \star A\left(
x,p\right)  +\left(  -1\right)  ^{A}A\left(  x,p\right)  \star\left(  \frac
{1}{2\pi}\int_{\pi/2}^{\pi}d\sigma j_{\star}\left(  \sigma,\varepsilon\right)
\right) \nonumber\\
&  =Q\left(  x,p\right)  \star A\left(  x,p\right)  -\left(  -1\right)
^{A}A\left(  x,p\right)  \star Q\left(  x,p\right)  . \label{Q_star}%
\end{align}
The relative minus sign in the second term on the second line arises because
$j_{\star}\left(  \sigma,\varepsilon\right)  $ is antisymmetric relative to
the midpoint $j_{\star}\left(  \pi-\sigma,\varepsilon\right)  =-j_{\star
}\left(  \sigma,\varepsilon\right)  ,$ leading to
\begin{equation}
Q\left(  x,p\right)  \equiv\frac{1}{2\pi}\int_{0}^{\pi/2}d\sigma j_{\star
}\left(  \sigma,\varepsilon\right)  =-\frac{1}{2\pi}\int_{\pi/2}^{\pi}d\sigma
j_{\star}\left(  \sigma,\varepsilon\right)  =\frac{1}{2\pi}\int_{0}^{\pi
}d\sigma\text{sign}\left(  \frac{\pi}{2}-\sigma\right)  j_{\star}\left(
\sigma,\varepsilon\right)  . \label{QisJint}%
\end{equation}
Hence, the representation of the QM BRST operator reduces to the
super-commutator (\ref{Q_star}) in iQM in the MSFT formalism. It must be
emphasized that the integral that defines the string field $Q\left(
x,p\right)  $ in Eq.(\ref{QisJint}) is over the half string. For the CFT with
flat background $Q\left(  x,p\right)  $ is given by (normal ordering is
implied, and $^{\prime}$ means $\partial_{\sigma}$)%
\begin{equation}
Q\left(  x,p\right)  =\frac{1}{2\pi}\int_{0}^{\pi/2}d\sigma\left[
\begin{array}
[c]{c}%
\pi p_{b}\left(  \pi^{2}p^{\mu}p_{\mu}+x^{\prime\mu}x_{\mu}^{\prime}+i\pi
^{2}p_{b}^{\prime}p_{c}^{\prime}+ix^{c}x^{b}\right) \\
+i\partial_{\sigma}^{-1}x^{c}\left(  \pi p_{\mu}x^{\prime\mu}+\pi
p_{c}^{\prime}x^{c}+\pi p_{b}^{\prime}x^{b}\right)
\end{array}
\right]  \label{Qfield-flat}%
\end{equation}
where all $x^{M}$ ($p_{M}$) are symmetric (antisymmetric) under reflections
from the midpoint and appear everywhere in $Q\left(  x,p\right)  $ in the
regulated forms
\begin{equation}
x^{M}\rightarrow e^{-\varepsilon\left\vert \partial_{\sigma}\right\vert }%
x_{+}^{M}\left(  \sigma,\varepsilon\right)  ,\;p_{M}\rightarrow
e^{-\varepsilon\left\vert \partial_{\sigma}\right\vert }p_{-M}\left(
\sigma\right)  . \label{regulatedxp}%
\end{equation}

For the general CFT the matter part in (\ref{Qfield-flat}) is modified as
follows, while the ghost sector in (\ref{Qfield-flat}) is common to all CFTs
so it remains unchanged. We begin with the matter stress tensor $T^{matter}%
\left(  \sigma,\varepsilon\right)  $ for the desired CFT written in terms of
phase space $\left(  X^{\mu},P_{\mu}\right)  $ by using the standard canonical
procedure (replacing all velocities by momenta). Recall that in the general
$CFT$ all positions are defined with an upper index $X^{\mu}$ while the lower
index in $P_{\mu}$ naturally follows from the CFT action $S_{CFT}\left(
X,\partial X\right)  $ by using the standard canonical procedure $P_{\mu
}=\partial S_{CFT}/\partial\left(  \partial_{\tau}X^{\mu}\right)  $. All
operators $\left(  \hat{X}^{\mu},\hat{P}_{\mu}\right)  $ in $\hat{T}%
^{matter}\left(  \sigma,\varepsilon\right)  $ are quantum ordered to insure
$\hat{T}^{matter}$ is normal ordered and has the correct quantum properties as
the generator of conformal transformations on the worldsheet. After this step
the operators $\left(  \hat{X}^{\mu},\hat{P}_{\mu}\right)  $ are replaced by
half of the phase space $\left(  x^{\mu},p_{\mu}\right)  $ in the regulated
form shown in (\ref{regulatedxp}), in the same order as the operators but with
star products $\star$ inserted in between non-commuting factors. Then this
$\hat{T}^{matter}\left(  \sigma,\varepsilon\right)  $ is inserted in the BRST
operator to obtain the BRST field $Q\left(  x,p\right)  $ that modifies
(\ref{Qfield-flat}) to a general CFT.

As explained just before Eq.(\ref{Ofield}) by evaluating all the star products
within $Q\left(  x,p\right)  $ it can be expressed as a classical function
$Q\left(  x,p\right)  $ which is a string field just like $A\left(
x,p\right)  .$ For example, for the flat CFT, evaluating all the star products
in (\ref{jstar}) is equivalent to forgetting all the $\star$'s and replacing
the constant $a$ by a shifted value that depends on the regulator
$\varepsilon.$

In the operator formalism the normal ordered BRST charge in any CFT is quantum
ordered to be nilpotent
\begin{equation}
\hat{Q}_{B}^{2}=0.
\end{equation}
For the flat CFT background this condition is satisfied at the critical
dimension $d=26$ and intercept $a=-1$. In the MSFT approach the nilpotency
property for any CFT has the following interesting consequence by using the
associativity property of the $\star$
\begin{align}
0  &  =\hat{Q}_{B}^{2}A=\hat{Q}_{B}\left(  Q\star A-\left(  -1\right)
^{A}A\star Q\right) \nonumber\\
&  =Q\star\left(  Q\star A-\left(  -1\right)  ^{A}A\star Q\right)  -\left(
-1\right)  ^{A+1}\left(  Q\star A-\left(  -1\right)  ^{A}A\star Q\right)
\star Q\nonumber\\
&  =\left(  Q\star Q\right)  \star A-A\star\left(  Q\star Q\right) \nonumber\\
&  =\left[  \left(  Q\star Q\right)  ,A\right]  _{\star}%
\end{align}
Since this commutator must vanish for any $A,$ we conclude that the star
product of the fields $Q\star Q$ must be a constant, and possibly zero. To
figure out what the constant is we need to compute the following
star-anticommutator%
\begin{equation}
Q\star Q=\frac{1}{2}\left\{  Q,Q\right\}  _{\star}=\frac{1}{4\pi}\int_{0}%
^{\pi/2}d\sigma_{1}\int_{0}^{\pi/2}d\sigma_{2}\left\{  j_{\star}\left(
\sigma_{1},\varepsilon\right)  ,j_{\star}\left(  \sigma_{2},\varepsilon
\right)  \right\}  _{\star}. \label{intRegions}%
\end{equation}
This anticommutator naively would be zero for two fermions; however the star
product turns them essentially into non-anticommuting operators in iQM such
that the anticommutator has support only in sharply local regions where
$\sigma_{1}$ approaches $\sigma_{2}.$ Given the map we have established
between iQM and QM, the exact parallel of this iQM computation can be done in
the operator QM version in any CFT. Therefore, we can simply take over the
known universal results for the \textit{local} operator products $\left\{
\hat{J}_{B}\left(  \sigma_{1},\varepsilon\right)  ,\hat{J}_{B}\left(
\sigma_{2},\varepsilon\right)  \right\}  $ for the BRST current in any CFT
[e.g. \cite{Pochinski} Eq.(4.3.10)], restrict $\sigma_{1},\sigma_{2}$ to only
the left (or right) half of the string, and integrate $\sigma_{1},\sigma_{2}$
in the range for half of the string $\left[  0,\pi/2\right]  $. The same local
computation was used to prove $\hat{Q}_{B}^{2}=0$ but in that case the range
of the integrals is $\left[  0,\pi\right]  .$ The result of the half-string
integrals is the same as the full string integrals because the support comes
only from sharply local regions$,$ in our case within the region $0\leq
\sigma_{1},\sigma_{2}\leq\pi/2.$ The non-trivial anticommutator has three
terms \cite{Pochinski}: $\hat{Q}_{B}^{2}=\int\int\sum_{i=0}^{2}\hat{\alpha
}_{i}\left(  \sigma_{1}\right)  \delta^{\left(  i\right)  }\left(  \sigma
_{1},\sigma_{2}\right)  ,$ where $\delta^{\left(  i\right)  }$ are the zeroth,
first and second derivatives of the delta function $i=0,1,2.$ The integrals
vanish for $i=1,2,$ while the coefficient $\alpha_{0}$ is proportional to
$\left(  c-26\right)  \left(  \hat{C}\partial_{\sigma}^{2}\hat{C}\right)  $ so
it vanishes for any critical CFT $c=26$. Therefore the field $Q\left(
x,p\right)  $ is actually nilpotent provided the corresponding CFT satisfies
the criticality conditions, namely the same conditions for which $\hat{Q}%
_{B}^{2}=0.$ Hence, for the flat CFT
\begin{equation}
Q\left(  x,p\right)  \star Q\left(  x,p\right)  =0,\text{ iff }c=26,\text{ and
}a=1,
\end{equation}
where $c$ is the central charge of the CFT and $a$ is the \textquotedblleft
intercept\textquotedblright. For the general CFT, provided the background
fields have the correct properties at the quantum level for the theory to be a
CFT, namely $\hat{Q}_{B}^{2}=0$ for the operator, then the corresponding
$Q\left(  x,p\right)  $ constructed with the prescription above will also
satisfy $Q\left(  x,p\right)  \star Q\left(  x,p\right)  =0$ under the
string-joining Moyal $\star$ product.

We have established that the BRST QM operator $\hat{Q}_{B}$ is
represented in iQM as a super-inner product with the nilpotent field
$Q\left(  x,p\right)  $ as given in Eq.(\ref{QisJint}).

\subsection{The MSFT action}

To construct the action for MSFT we need three ingredients: the star product,
the BRST operator, and a rule for integration that has the property $\int
\hat{Q}_{B}A=0.$ We already have the first two, now we define the rule for
integration which is equivalent to a supertrace or a super-integral over all
the phase space degrees of freedom in the induced quantum mechanics%
\begin{equation}
\text{Str}\left(  A\right)  \equiv\int\left(  Dx_{+}^{M}\left(  \cdot
,\varepsilon\right)  \right)  \left(  Dp_{-M}\left(  \cdot\right)  \right)
~A\left(  x,p\right)  . \label{supertrace}%
\end{equation}
This super-integral in phase space is indeed a supertrace of an operator $A$
in iQM, as is well known in the Moyal product literature. So using the notions
of supertrace and its cyclic properties, we immediately know that the
supertrace of a supercommutator is always zero, hence the desired property
$\int\hat{Q}_{B}A=0$ is satisfied
\begin{equation}
\int\hat{Q}_{B}A=\text{Str}\left(  [Q,A\}_{\star}\right)  =0.
\end{equation}
To see this result directly as the property of the phase space integral, we
compute $\int\hat{Q}_{B}A$ by using the iQM representation (\ref{Q_star}) of
the BRST operator in terms of the field $Q\left(  x,p\right)  $ as follows%
\begin{equation}
\int\hat{Q}_{B}A=\int\left(  DxDp\right)  \left(  Q\left(  x,p\right)  \star
A\left(  x,p\right)  -\left(  -1\right)  ^{A}A\left(  x,p\right)  \star
Q\left(  x,p\right)  \right)  .
\end{equation}
Under the integral the $\star$ can be removed when there are only two fields
because each non-trivial piece of the Moyal $\star$ (\ref{msftstar}) can be
rewritten as a total derivative and would lead to vanishing terms at the
boundaries of phase space where $A\left(  x,p\right)  $ is assumed to vanish.
The remaining ordinary product between the two classical string fields
$\left(  QA-\left(  -1\right)  ^{A}AQ\right)  $ is trivially zero since the
second term cancels the first term after interchanging the orders of $A$ and
$Q$. Hence, once again we have proven $\int\hat{Q}_{B}A=0,$ so we have chosen
a good integration rule.

Now we can convert the cubic action for open string field theory proposed by
Witten \cite{witten}
\begin{equation}
S\left(  \Psi\right)  =-\text{Str}\left(  \frac{1}{2}\Psi\star\left(  \hat
{Q}_{B}\Psi\right)  +\frac{g_{0}}{3}\Psi\star\Psi\star\Psi\right)
\label{Switten}%
\end{equation}
to our new MSFT formalism by the rules developed in the previous sections for
representing QM operators. In particular, for a fermionic $A\left(
x,p\right)  $ (i.e. $\left(  -1\right)  ^{A}=-1$) recall that the BRST
operator $\hat{Q}_{B}$ is represented in iQM with an anticommutator involving
the field $Q\left(  x,p\right)  $
\begin{equation}
\hat{Q}_{B}A=\left\{  Q,A\right\}  _{\star}=Q\star A+A\star Q.
\label{QanticommA}%
\end{equation}
The fermionic field $Q\left(  x,p\right)  $ is the specific string field given
in Eq.(\ref{QisJint}) for any CFT (or Eq.(\ref{Qfield-flat}) for the flat CFT).

We digress temporarily to discuss ghost numbers of string fields before we
construct the action. The ghost number operator $\hat{N}_{gh}$ has the
following representation when applied on any string field $A\left(
x,p\right)  $
\begin{align}
\hat{N}_{gh}A  &  =\frac{1}{2}\int_{0}^{\pi}d\sigma\left(
\begin{array}
[c]{c}%
\frac{2}{\pi}\hat{g}+x^{c}\left(  \sigma\right)  \frac{\partial}{\partial
x^{c}\left(  \sigma\right)  }+p_{b}\left(  \sigma\right)  \frac{\partial
}{\partial p_{b}\left(  \sigma\right)  }\\
-x^{b}\left(  \sigma\right)  \frac{\partial}{\partial x^{b}\left(
\sigma\right)  }-p_{c}\left(  \sigma\right)  \frac{\partial}{\partial
p_{c}\left(  \sigma\right)  }%
\end{array}
\right)  A\label{Nghost}\\
&  =\left(
\begin{array}
[c]{c}%
g_{A}+x_{e}^{c}\frac{\partial}{\partial x_{e}^{c}}+p_{ob}\frac{\partial
}{\partial p_{ob}}\\
-x_{0}^{b}\frac{\partial}{\partial x_{0}^{b}}-x_{e}^{b}\frac{\partial
}{\partial x_{e}^{b}}-p_{oc}\frac{\partial}{\partial p_{oc}}%
\end{array}
\right)  A,
\end{align}
where we used $x_{0}^{b}=ib_{0}$ and there is an implied summation over the
even and odd integer modes, $e=2,4,6,\cdots$ and $o=1,3,5,\cdots.$ The
operator $\hat{g}A=g_{A}A$ gives an eigenvalue $g_{A}$ that corresponds to a
ghost charge $g_{A}$ assigned to the field $A.$ The eigenvalue $g_{A}$ may
differ for various fields $A\left(  x,p\right)  $. The star products of any
$A$ with the field $Q,$ namely the $Q\star A$ or $A\star Q$ that appear in
(\ref{QanticommA}), must have the total $\hat{N}_{gh}$ increased by $+1$
relative to $A$ since the operator $\hat{Q}_{B}$ has ghost number $+1$,
namely$\left[  \hat{N}_{gh},\hat{Q}_{B}\right]  =+\hat{Q}_{B}.$ This implies
that, for the special field $Q\left(  x,p\right)  $
\begin{equation}
\hat{N}_{gh}Q\left(  x,p\right)  =\left(  +1\right)  Q\left(  x,p\right)  \,.
\label{ghostNQ}%
\end{equation}
Taking into account the explicit form of $Q\left(  x,p\right)  $ in
(\ref{QisJint},\ref{Qfield-flat}) and the action of the derivatives in
(\ref{Nghost}) on this $Q,$ we conclude that the eigenvalue of $\hat{g}$ that
appears in (\ref{Nghost}) when applied on $Q$ is zero $g_{Q}=0.$

We now consider the kinetic term of the Witten action (\ref{Switten})
transcribed to our basis%
\begin{align}
S_{kin}\left(  A\right)   &  =-\text{Str}\left(  \frac{1}{2}A\star\left(
\hat{Q}_{B}A\right)  \right)  =-\text{Str}\left(  \frac{1}{2}A\star\left\{
Q,A\right\}  _{\star}\right) \nonumber\\
&  =-\text{Str}\left(  A\star Q\star A\right)  . \label{kinQ}%
\end{align}
In the last line we used the cyclic property of the Str to simplify the
kinetic term to its final form. In Eq.(\ref{kinQ}) the $\star$ product is our
regulated Moyal $\star$ product of Eq.(\ref{msftstar}) which has zero ghost
number, and the supertrace is the phase space integral over half of the full
string's classical phase space as in Eq.(\ref{supertrace}), which includes the
ghost zero mode $x_{0}^{b}$ or equivalently the ghost midpoint $\bar{x}%
^{b}\equiv x^{b}\left(  \pi/2\right)  $ mode. This integration
rule\footnote{The Str or integration rule $\int$ contains an equal number of
\textit{non-zero} $b,c$ ghost modes $\left(  x_{e}^{b},x_{e}^{c},p_{ob}%
,p_{oc}\right)  ,$ where $e=2,4,6,\cdots$ labels even modes and
$o=1,3,5,\cdots$ labels odd modes, while both $A$ and $\int$ depend on only
one ghost zero mode $x_{0}^{b}$. The other ghost zero mode, $\hat{c}_{0},$ is
not in the half phase space: it acts on $A\left(  x,p\right)  $ as a
derivative $\hat{c}_{0}A\left(  x,p\right)  =i\partial_{b_{0}}A\left(
x,p\right)  $. Since $b,c$ have opposite ghost numbers, the ghost numbers of
the non-zero modes cancel out in the integration rule, leaving unbalanced only
the zero mode $dx_{0}^{b}.$ Hence the total integration rule has ghost number
$+1,$ which is opposite to that of $x_{0}^{b},$ since Grassman integration is
defined as being equivalent to the derivative, $\int dx_{0}^{b}\times
x_{0}^{b}=\partial_{x_{0}^{b}}x_{0}^{b}=1.$ Another approach for the same
result is to consider the measure of integration in the $\sigma$ basis. which
contains at each $\sigma\neq\pi/2$ the ghost pairs $Dx_{+}^{b}\left(
\sigma\right)  Dp_{-b}\left(  \sigma\right)  $ and similarly $Dx_{+}%
^{c}\left(  \sigma\right)  Dp_{-c}\left(  \sigma\right)  .$ At each
$\sigma\neq\pi/2$ the measure of integration has ghost number $0$ because
$x,p$ have opposite ghost numbers. However at $\sigma=\pi/2,$ there is an
additional unpaired $Dx_{+}^{b}\left(  \pi/2\right)  $ with ghost number $+1$
(opposite to ghost number of $x^{b}$). This is because all momenta
$p^{b,c}\left(  \sigma\right)  $ vanish at $\sigma=\pi/2$ while $x_{+}%
^{c}\left(  \pi/2\right)  $ is not integrated as an additional independent
mode since $x^{c}\left(  \sigma\right)  $ has no zero mode as explained after
Eq.(\ref{PcbModes}). Thus the unpaired midpoint mode $\bar{x}^{b}\equiv
x_{+}^{b}\left(  \pi/2\right)  $ causes the integration rule $\int d\bar
{x}^{b}$ or the Str to have ghost number $+1.$} has ghost number $+1$ (because
of the $\int dx_{0}^{b}$ or $\int d\bar{x}^{b}$), therefore the integrand, or
the argument of the Str, namely $\left(  A\star Q\star A\right)  ,$ must have
ghost number $-1.$ Accordingly, taking into account (\ref{ghostNQ}), the total
ghost number of the field $A$ in Eq.(\ref{kinQ}) must be $-1.$ Then, for the
field $A$ that appears in the action we must have total ghost numbers assigned
as follows
\begin{equation}
\hat{N}_{gh}A\left(  x,p\right)  =-A\left(  x,p\right)  ,\;\hat{N}_{gh}\left(
Q\left(  x,p\right)  \star A\left(  x,p\right)  \right)  =0.
\end{equation}
This implies that in Eq.(\ref{Nghost}) a nontrivial $\hat{g}A=g_{A}A,$ is
included in the total ghost number $\hat{N}_{gh}A=\left(  -1\right)  A.$ We
will see that the perturbative vacuum, including zero modes of the ghosts,
will require $g_{A}=2$ for $A\left(  x,p\right)  .$

Next we turn to the cubic interaction term. Given $\hat{N}_{gh}A=-A$ and
$\hat{N}_{gh}$Str$=\left(  +1\right)  $Str, a straightforward term of the form
Str$\left(  A\star A\star A\right)  $ is inconsistent with total zero ghost
number for the action. Hence there has to be some midpoint insertions to
achieve the correct zero total ghost number. There is a unique answer that
gives the final form of the total action as follows
\begin{equation}
S\left(  A\right)  =-\text{Str}\left(  A\star Q\star A+\frac{g_{0}}{3}%
A\star\partial_{\bar{b}}A\star\partial_{\bar{b}}A\right)  . \label{pertS}%
\end{equation}
where we have defined the midpoint ghost derivative
\begin{equation}
\partial_{\bar{b}}A\equiv\partial_{\bar{x}^{b}}A=\frac{\partial A}{\partial
x^{b}\left(  \pi/2\right)  }, \label{midpointderiv}%
\end{equation}
which is discussed further below. The equation of motion that follows from
this action is
\begin{equation}
\left\{  Q,A\right\}  _{\star}+g_{0}\left\{  \partial_{\bar{b}}A,\partial
_{\bar{b}}A\right\}  _{\star}=0. \label{eom}%
\end{equation}
The action (\ref{pertS}) is invariant under the BRST gauge transformation
given by%
\begin{equation}
\delta_{\Lambda}A=\left[  Q,\Lambda\right]  _{\star}+g_{0}\left\{
\partial_{\bar{b}}A,\partial_{\bar{b}}\Lambda\right\}  _{\star},
\label{brstTransf}%
\end{equation}
where $\Lambda$ is an arbitrary bosonic string field with ghost number $-2$.
Then every term in this equation has total ghost number $-1.$ The proof of
this gauge symmetry, $\delta_{\Lambda}S=0,$ is given in Appendix-B.

We elaborate now on the properties of the midpoint ghost derivative
$\partial_{\bar{b}}A$. Midpoint degrees of freedom are insensitive to our star
product as explained earlier and in Appendix-A. The required midpoint
insertions in our formalism, denoted by $\partial_{\bar{b}}A$, amounts to a
derivative with respect to the midpoint ghost coordinate $x^{b}\left(
\pi/2\right)  .$ This could also be written as a star-anticommutator with the
field $p_{b}\left(  \sigma\right)  $ at the midpoint
\begin{equation}
\partial_{\bar{b}}A\equiv\frac{\partial A\left(  x,p\right)  }{\partial
x^{b}\left(  \pi/2\right)  }=-i\left\{  p_{b}\left(  \frac{\pi}{2}\right)
,A\left(  x,p\right)  \right\}  _{\star}. \label{dbarbA1}%
\end{equation}
Then the field $\partial_{\bar{b}}A$ has ghost number zero and our interaction
term in (\ref{pertS}), $A\star\partial_{\bar{b}}A\star\partial_{\bar{b}}A,$
has the desired ghost number $-1.$ More generally, after doing the integral
over the midpoint $\bar{x}^{b}$ we obtain, with any three distinct fields,
\begin{equation}
\int d\bar{x}^{b}\left(  A_{1}\star\partial_{\bar{b}}A_{2}\star\partial
_{\bar{b}}A_{3}\right)  =\int^{\prime}\partial_{\bar{b}}A_{1}\star
\partial_{\bar{b}}A_{2}\star\partial_{\bar{b}}A_{3}, \label{bbarIntegtal}%
\end{equation}
where $\int^{\prime}$ implies integration over the remaining modes (excluding
the midpoint $\bar{x}^{b}\equiv x^{b}\left(  \pi/2\right)  $). Hence, under
the integral $\int d\bar{x}^{b}$ in (\ref{pertS}) there is a symmetry in
moving the derivative $\partial_{\bar{b}}$ from one field to another (noting
$\partial_{\bar{b}}^{2}=0$), so our interaction term in the action
(\ref{pertS}) is symmetric under the cyclic interchange of the fields. Hence
in (\ref{pertS}) there is no need for parentheses that prescribe the order of
operations of the star products and/or derivatives $\partial_{\bar{b}}$.

It must be noted that, if the field is expressed in terms of independent ghost
modes that distinguish the midpoint (see section (\ref{old-new})), $\left(
\bar{x}^{b},x_{e}^{b,c},p_{\left(  b,c\right)  o}\right)  ,$ where $\bar
{x}^{b}=x^{b}\left(  \pi/2\right)  =x^{0}+\sqrt{2}\sum_{e}\cos\frac{e\pi}%
{2}~x_{e}$ is the midpoint, then the midpoint derivative $\partial_{\bar{b}}$
is equivalent to a derivative with respect to the single midpoint mode
$\bar{x}^{b}$
\begin{equation}
\partial_{\bar{b}}A=\frac{\partial A\left(  x^{b,c}\left(  \sigma\right)
,p_{b,c}\left(  \sigma\right)  ,\cdots\right)  }{\partial x^{b}\left(
\pi/2\right)  }=\partial_{\bar{x}^{b}}\tilde{A}\left(  \bar{x}^{b},x_{e}%
^{b,c},p_{\left(  b,c\right)  o},\cdots\right)  . \label{dbarbA2}%
\end{equation}
We can use any of the expressions in (\ref{dbarbA2}) for computing
$\partial_{\bar{b}}A$ depending on the basis which is used in
specific computations, i.e. the new continuous $\sigma$ basis whose
$\star$ product does not distinguish the midpoint, or the old
discrete mode basis whose $\star$ product did distinguish the
midpoint (see section (\ref{old-new}) for more details).

\subsection{Siegel gauge}

The general field in MSFT $A\left(  x^{M},p_{M}\right)  $ may be written more
explicitly in terms of the ghosts in the $b,c$ combinations given in
(\ref{bcxp}) as $A\left(  x^{\mu},p_{\mu},b,c\right)  .$ We recall that in
this basis $c\left(  \sigma,\varepsilon\right)  $ has no zero mode while
$b\left(  \sigma,\varepsilon\right)  $ has a zero mode $b_{0}=\frac{1}{\pi
}\int_{0}^{\pi}d\sigma b\left(  \sigma,\varepsilon\right)  =-ix_{0}^{b}$ (see
Eq.(\ref{XcbModes}))$.$ The field may be expanded in powers of $x_{0}^{b},$
and since this is a fermion, the general expansion has only two terms
\begin{equation}
A=x_{0}^{b}A^{\left(  0\right)  }+A^{\left(  -1\right)  } \label{Ab0}%
\end{equation}
where $A^{\left(  0\right)  }$ or $A^{\left(  -1\right)  }$ do not contain
$x_{0}^{b},$ and the labels $\left(  i\right)  $ mean ghost number $i=0,-1.$

Now we choose the Siegel gauge which satisfies $\hat{x}_{0}^{b}A_{s}=0,$ where
the subscript $s$ indicates the Siegel gauge. The zero mode ghost operator
$\hat{x}_{0}^{b}$ is given by $\hat{x}_{0}^{b}=\left(  \frac{i}{2\pi}\int
_{0}^{\pi}d\sigma\left(  \hat{B}\left(  \sigma,\varepsilon\right)  +\hat
{B}\left(  -\sigma,\varepsilon\right)  \right)  \right)  .$ From the iQM
representation of the QM operator $\hat{B}$ given in Eq.(\ref{BonA}), we note
that the operator $\hat{x}_{0}$ is diagonal on $A\left(  x,p\right)  $, so
that $\hat{x}_{0}^{b}A_{s}=x_{0}^{b}A_{s}=\left(  0A_{s}^{\left(  0\right)
}+x_{0}^{b}A_{s}^{\left(  -1\right)  }\right)  =0$, where we have used
$\left(  x_{0}^{b}\right)  ^{2}=0.$ Hence, in this gauge we must have
$A_{s}^{\left(  -1\right)  }=0,$ or
\begin{equation}
A_{s}=x_{0}^{b}A_{s}^{\left(  0\right)  }, \label{siegelGauge}%
\end{equation}
with the understanding that the zero-ghost-number field $A_{s}^{\left(
0\right)  }$ is independent of $x_{0}^{b}.$ In the further deliberations it is
sometimes convenient to consider the midpoint coordinate $\bar{x}^{b}$ instead
of the zeroth mode $x_{0}^{b}$ because the star product of the midpoint
$\bar{x}^{b}$ with any string field is trivial as discussed earlier and in
Appendix-A. That is, taking into account $x_{0}^{b}=\bar{x}^{b}+\sum_{e}%
w_{e}x_{e}^{b},$ with $w_{e}\equiv-\sqrt{2}\left(  -1\right)  ^{e/2}$ as in
(\ref{xoxbar}), any field of the form (\ref{Ab0}) can be rewritten as:%
\begin{equation}
A=x_{0}^{b}A^{\left(  0\right)  }+A^{\left(  -1\right)  }=\bar{x}%
^{b}A^{\left(  0\right)  }+\tilde{A}^{\left(  -1\right)  },\text{ with }%
\tilde{A}^{\left(  -1\right)  }=\sum_{e}w_{e}x_{e}^{b}A^{\left(  0\right)
}+A^{\left(  -1\right)  }, \label{Abbar}%
\end{equation}
where neither $A^{\left(  0\right)  }$ nor $\tilde{A}^{\left(  -1\right)  }$
contain the zero mode $x_{0}^{b}$ or the midpoint mode $\bar{x}^{b}.$ Hence in
the Siegel gauge%
\begin{equation}
A_{s}=x_{0}^{b}A_{s}^{\left(  0\right)  }=\bar{x}^{b}A_{s}^{\left(  0\right)
}+\tilde{A}_{s}^{\left(  -1\right)  },\;\text{with }\tilde{A}_{s}^{\left(
-1\right)  }\equiv w_{e}x_{e}^{b}A_{s}^{\left(  0\right)  }. \label{As}%
\end{equation}
We now see that the midpoint ghost derivative is $\partial_{\bar{b}}%
A_{s}=A_{s}^{\left(  0\right)  },$ and insert it where it occurs in the action
(\ref{pertS}) and the equation of motion (\ref{eom}) in the Siegel gauge
\begin{equation}
\partial_{\bar{b}}A_{s}\star\partial_{\bar{b}}A_{s}=A_{s}^{\left(  0\right)
}\star A_{s}^{\left(  0\right)  }.
\end{equation}

To complete the Siegel gauge we also need to identify the dependence of the
field $Q\left(  x,p\right)  $ on the zero ghost mode $x_{0}^{b}.$ Recall that
for the QM operator $\hat{Q}_{B}A$ it is well known that the $x_{0}^{b}$
dependence is isolated as follows \cite{siegel}
\begin{equation}
\hat{Q}_{B}A=\left(  \hat{L}_{0}-1\right)  \partial_{x_{0}^{b}}A+\hat{Q}%
_{1}A+x_{0}^{b}\hat{Q}_{2}A,
\end{equation}
where the operators $\left(  \hat{Q}_{1},\hat{Q}_{2}\right)  ,$ with ghost
numbers $\left(  1,2\right)  $ respectively, do not depend on $x_{0}^{b}.$
Taking into account that we have, $p_{b}\left(  \sigma\right)  \star A=\left(
p_{b}\left(  \sigma\right)  +\frac{i}{2}\partial_{x^{b}\left(  \sigma\right)
}\right)  A$ when $\sigma\leq\pi/2$, and $A\star p_{b}\left(  \sigma\right)
=A\left(  p_{b}\left(  \sigma\right)  -\frac{i}{2}\overleftarrow{\partial
}_{x^{b}\left(  \sigma\right)  }\right)  $ when $\sigma\geq\pi/2$, we deduce
that for the field $Q\left(  x,p\right)  $ we have the parallel property
\begin{equation}
\hat{Q}_{B}A=\left\{  Q,A\right\}  _{\star}=\left(  \left\{  \mathcal{L}%
_{0},\partial_{x_{0}^{b}}A\right\}  _{\star}-\partial_{x_{0}^{b}}A\right)
+\left\{  Q_{1},A\right\}  _{\star}+x_{0}^{b}\left[  Q_{2},A\right]  _{\star},
\end{equation}
where $\mathcal{L}_{0}\left(  x,p\right)  ,$ $Q_{1}\left(  x,p\right)
,Q_{2}\left(  x,p\right)  $ are fields given below (of the corresponding ghost
numbers) that do not contain any dependence on $x_{0}^{b},$ and applied on $A$
with anticommutators or commutators in the iQM as determined by their ghost numbers.

Now we examine $\left\{  Q,A_{s}\right\}  $ in the Siegel gauge $A_{s}%
=x_{0}^{b}A_{s}^{\left(  0\right)  },$ and after taking into account $\left(
x_{0}^{b}\right)  ^{2}=0,$ we find
\begin{equation}
\left\{  Q,A_{s}\right\}  =\left(  \left\{  \mathcal{L}_{0},A_{s}^{\left(
0\right)  }\right\}  _{\star}-A_{s}^{\left(  0\right)  }\right)  -x_{0}%
^{b}\left[  Q_{1},A_{s}^{\left(  0\right)  }\right]  _{\star}, \label{QAs}%
\end{equation}
where the last term is a star-commutator since $Q_{1}$ is fermionic while
$A_{s}^{\left(  0\right)  }$ is bosonic.

Using the results in Eqs.(\ref{As}-\ref{QAs}) we now evaluate the action
(\ref{pertS}) in the Siegel gauge by using the rules of Grassmann integration,
$\int dx_{0}^{b}x_{0}^{b}=1$ and $\int dx_{0}^{b}=0,$ we obtain
\begin{equation}
S_{s}=-\text{Str}^{\prime}\left(  A_{s}^{\left(  0\right)  }\star\left(
\mathcal{L}_{0}-\frac{1}{2}\right)  \star A_{s}^{\left(  0\right)  }%
+\frac{g_{0}}{3}A_{s}^{\left(  0\right)  }\star A_{s}^{\left(  0\right)
}\star A_{s}^{\left(  0\right)  }\right)  , \label{SsFixed}%
\end{equation}
where Str$^{\prime}$ no longer contains the integration over ghost zero mode
$x_{0}^{b}$. Similarly we evaluate the equation of motion (\ref{eom}) in the
Siegel gauge by identifying the coefficients for the zeroth power and the
first power of $x_{0}^{b}$
\begin{equation}
\left\{  \mathcal{L}_{0},A_{s}^{\left(  0\right)  }\right\}  _{\star}%
-A_{s}^{\left(  0\right)  }+g_{0}A_{s}^{\left(  0\right)  }\star
A_{s}^{\left(  0\right)  }=0,\;\;\left[  Q_{1},A_{s}^{\left(  0\right)
}\right]  _{\star}=0. \label{oms-s}%
\end{equation}
Note that the last equation is a constraint that supplements the equation of
motion that follows from the gauge fixed action (\ref{SsFixed}). The
constraint amounts to applying all the remaining Virasoro constraints on the
Siegel gauge field $A_{s}^{\left(  0\right)  }.$

To complete this section we give the explicit form of the fields
$\mathcal{L}_{0}\left(  x,p\right)  $ and $Q_{1}\left(  x,p\right)  $ that
correspond to the iQM representation of corresponding QM operators $\hat
{L}_{0}$ and $\hat{Q}_{1}.$ The string field $Q_{1}\left(  x,p\right)  $ is
the field $Q\left(  x,p\right)  $ after dropping all the effects of the zero
ghost mode $x_{0}^{b}.$ The string field $\mathcal{L}_{0}\left(  x,p\right)  $
is obtained from Eqs.(\ref{J_star}-\ref{QisJint}) and is given by
\begin{equation}
\mathcal{L}_{0}\left(  x,p\right)  =\frac{1}{\pi}\int_{0}^{\pi/2}d\sigma
\sum_{\pm}\left(  T_{\star\pm}^{m}+T_{\star\pm}^{gh}\right)  \left(  x_{+}%
^{M},p_{-M}\right)  ,\;
\end{equation}
where the integral is over half the string, and matter $T_{\pm}^{m}$ is for
any CFT. To be fully explicit, we give the example of the flat CFT in $d=26,$
for which the expressions for $T_{\star\pm}^{m},T_{\star\pm}^{gh}$ are similar
as given in (\ref{flatTm},\ref{TghostSp2}). Combining all terms,
$\mathcal{L}_{0}$ takes the following SO$\left(  25,1\right)  \times$
Sp$\left(  2\right)  $ symmetric form (normal ordering is implied)%
\begin{equation}
\mathcal{L}_{0}=\frac{1}{2\pi}\int_{0}^{\pi/2}d\sigma\left(
\begin{array}
[c]{c}%
\eta_{\mu\nu}\left(
\begin{array}
[c]{c}%
\pi^{2}e^{-\varepsilon\left\vert \partial_{\sigma}\right\vert }p^{\mu}\left(
\sigma\right)  \star e^{-\varepsilon\left\vert \partial_{\sigma}\right\vert
}p^{\nu}\left(  \sigma\right) \\
+e^{-\varepsilon\left\vert \partial_{\sigma}\right\vert }\partial_{\sigma
}x^{\mu}\left(  \sigma,\varepsilon\right)  \star e^{-\varepsilon\left\vert
\partial_{\sigma}\right\vert }\partial_{\sigma}x^{\nu}\left(  \sigma
,\varepsilon\right)
\end{array}
\right) \\
+i\varepsilon_{mn}\left(
\begin{array}
[c]{c}%
\pi^{2}e^{-\varepsilon\left\vert \partial_{\sigma}\right\vert }\partial
_{\sigma}p^{m}\left(  \sigma\right)  \star e^{-\varepsilon\left\vert
\partial_{\sigma}\right\vert }\partial_{\sigma}p^{n}\left(  \sigma\right) \\
+e^{-\varepsilon\left\vert \partial_{\sigma}\right\vert }x^{m}\left(
\sigma,\varepsilon\right)  \star e^{-\varepsilon\left\vert \partial_{\sigma
}\right\vert }x^{n}\left(  \sigma,\varepsilon\right)
\end{array}
\right)
\end{array}
\right)  . \label{L0m+g-field}%
\end{equation}
The only zero modes that survive in this expression are the matter zero modes
$\left(  x_{0}^{\mu},p_{0\mu}\right)  .$ Since the star product is symmetric
under OSp$\left(  d|2\right)  ,$ the cubic term in the action (\ref{SsFixed})
is supersymmetric under transformations that mix matter and ghost degrees of
freedom even when we have any CFT with curved backgrounds. However, the
supersymmetry is broken by the quadratic term because $\mathcal{L}_{0}$ is not
symmetric under OSp$\left(  26|2\right)  :$ For a non-trivial CFT the curved
background in $\mathcal{L}_{0}$ breaks even the linear SO$\left(  d\right)  ;$
for the flat CFT SO$\left(  26\right)  $ is valid in $\mathcal{L}_{0}$ but
OSp$\left(  26|2\right)  $ is broken since in Eq.(\ref{L0m+g-field}) the
$\partial_{\sigma}$ derivatives are applied on the ghost momenta $p^{m}$
rather than on ghost positions $x^{m}.$ However, we can bring the expression
(\ref{L0m+g-field}) to the expected supersymmetric form with a simple change
of the basis in the ghost sector. This will be discussed in the next section
(\ref{alpha-gen}).

We have reached the stage where we can now make direct contact in detail with
the MSFT formulation using old star $\star$ product for the flat CFT and
distinguishing the midpoint. This is important because we can then claim that
all the previous successful computations are now also a direct consequence of
the more general new formalism. The correspondence between the old and new
formulations is obtained through the Siegel gauge action (\ref{SsFixed}) and
the discussion in Appendix-A about the relation between the various bases of
the degrees of freedom, the new $\sigma$-basis, the new mode basis including
the center of mass mode $x_{0},$ and the old mode basis including the midpoint
mode,%
\begin{equation}
A\left(  x\left(  \sigma\right)  ,p\left(  \sigma\right)  \right)  =A\left(
x_{0},x_{e},p_{o}\right)  =\tilde{A}\left(  \bar{x},x_{e},p_{o}\right)  ,
\label{Abases}%
\end{equation}
assuming that the ghost zero mode $x_{0}^{b}$ or the corresponding
midpoint mode $\bar{x}^{b}$ is already integrated out. The only
remaining zero mode is the matter zero mode $x_{0}^{\mu}.$ Then the
appropriate star product in $A\left(  x_{0},x_{e},p_{o}\right)  $
basis takes the form given in Eqs.(\ref{new_star},\ref{newstar2}).
As discussed in Appendix-A this reproduces all the results obtained
in the old basis $\tilde{A}\left(  \bar {x},x_{e},p_{o}\right)  $
\cite{B}-\cite{BP}. Hence we have reproduced all of the previous
work based on the flat CFT in the Siegel gauge. We are now prepared
to tackle non-perturbative computations with a more efficient tool.

\subsection{OSp$\left(  d|2\right)  $ Supersymmetry Acting on Matter and
Ghosts \label{alpha-gen}}

In the discussion above we noted that the star product has an OSp$\left(
d|2\right)  $ matter-ghost supersymmetry for any CFT, but the stress tensors
for matter versus ghosts as well as the ghost structure of the BRST operator
break this supersymmetry. So, the kinetic term of the SFT action (\ref{pertS})
breaks the supersymmetry while the interactions are supersymmetric.

Consequently, in Feynman-like diagram computations in the Siegel gauge, in the
flat CFT case, the breaking of OSp$\left(  d|2\right)  $ is in the propagator
and not in the interactions. Since the breaking of the supersymmetry amounts
to moving the $\partial_{\sigma}$ derivative from the ghost $P^{m}$ to the
ghost $X^{m}$ in the expression for $\hat{L}_{0}$ in (\ref{LoOper}) or
$\mathcal{L}_{0}$ in (\ref{L0m+g-field}), we can construct a simple algorithm
to simplify all computations as if there is OSp$\left(  d|2\right)  $ symmetry
in the full theory, and then modify the final computation with a simple rule
that takes into account this breaking of the symmetry in propagators. This
algorithm was noted and used efficiently in past computations in MSFT
\cite{BM}\cite{BKM}\cite{BP} and will be illustrated below with an example.

However, we discovered that there is a better approach: it is possible to
rewrite the theory in a slightly modified basis of the ghost degrees of
freedom such as to display fully the OSp$\left(  d|2\right)  $ symmetry in the
Siegel gauge (but not in the general gauge) including in the kinetic term or
the propagators. We may then use supersymmetric propagators in all
computations in the Siegel gauge. The final results in computations, such as
amplitudes, are the same as before \cite{BM}\cite{BKM}\cite{BP}.

To show how this works we need to review the ghost sector and show that there
is a more general way to extract the ghost phase space operators $\left(
\hat{X}^{m},\hat{P}_{m}\right)  $ from the ghost $\left(  \hat{B},\hat
{C}\right)  $ operators. Namely, instead of Eq.(\ref{BCXP}), we can introduce
a more general formula which includes a parameter\emph{ }$\alpha$ as follows%
\begin{equation}
\text{ }%
\begin{array}
[c]{c}%
\hat{B}\left(  \pm\sigma\right)  =\left(  -i\hat{X}^{b}\left(  \sigma\right)
\pm\pi\left\vert \partial_{\sigma}\right\vert ^{\alpha-1}\partial_{\sigma}%
\hat{P}_{c}\left(  \sigma\right)  \right)  ,\\
\hat{C}\left(  \pm\sigma\right)  =\left(  \pi\hat{P}_{b}\left(  \sigma\right)
\mp i\left\vert \partial_{\sigma}\right\vert ^{-\alpha-1}\partial_{\sigma}%
\hat{X}^{c}\left(  \sigma\right)  \right)  .
\end{array}
\label{BCXP2}%
\end{equation}
Taking $\alpha=1$ reproduces (\ref{BCXP}). Results of any computations should
be independent of\emph{ }$\alpha$ since this is only a rewriting of the same
$\left(  \hat{B},\hat{C}\right)  $ operators. Indeed we have checked that
physical quantities, such as amplitudes, are independent of $\alpha.$ So we
may explore if the parameter $\alpha$ leads to some interesting consequences
resulting from this rearrangement of the degrees of freedom. The answer is
yes: as it will be pointed out below, choosing $\alpha=-1$ rather than
$\alpha=1$, will be useful to display a supersymmetry between matter and ghost
degrees of freedom in the Siegel gauge.

For the more general definition (\ref{BCXP2}), the modes in
Eqs.(\ref{XcbModes},\ref{PcbModes}) generalize to the following form including
the parameter $\alpha$
\begin{align}
\hat{X}_{0}^{b}  &  =i\hat{b}_{0},\;\hat{X}_{n\geq1}^{b}=\frac{i}{\sqrt{2}%
}\left(  \hat{b}_{n}+\hat{b}_{-n}\right)  ,\;\hat{X}_{n\geq1}^{c}%
=-\frac{n^{\alpha}}{\sqrt{2}}\left(  \hat{c}_{n}-\hat{c}_{-n}\right)  ,\\
\hat{P}_{b0}  &  =\hat{c}_{0},\;\;\hat{P}_{b,n\geq1}=\frac{1}{\sqrt{2}}\left(
\hat{c}_{n}+\hat{c}_{-n}\right)  ,\;\hat{P}_{c,n\geq1}=\frac{i}{\sqrt
{2}n^{\alpha}}\left(  \hat{b}_{n}-\hat{b}_{-n}\right)  ,
\end{align}
and similarly for the inverse relations%
\begin{align}
\hat{b}_{0}  &  =-i\hat{X}_{0}^{b},\;\hat{b}_{n\geq1}=\frac{i}{\sqrt{2}%
}\left(  -\hat{X}_{n\geq1}^{b}-n^{\alpha}\hat{P}_{c,n\geq1}\right)  ,\;\hat
{b}_{\left(  -n\leq1\right)  }=\frac{i}{\sqrt{2}}\left(  -\hat{X}_{n\geq1}%
^{b}+n^{\alpha}\hat{P}_{c,n\geq1}\right)  ,\\
\hat{c}_{0}  &  =\hat{P}_{b0},\;\hat{c}_{n\geq1}=\frac{1}{\sqrt{2}}\left(
-n^{-\alpha}\hat{X}_{n\geq1}^{c}+\hat{P}_{b,n\geq1}\right)  ,\;\hat
{c}_{\left(  -n\leq1\right)  }=\frac{1}{\sqrt{2}}\left(  n^{-\alpha}\hat
{X}_{n\geq1}^{c}+\hat{P}_{b,n\geq1}\right)  .
\end{align}
Then, the regulated expression in (\ref{BCXP-regulated}) are generalized as%
\begin{align}
\hat{B}\left(  \pm\sigma,\varepsilon\right)   &  =\left(  -ie^{-\varepsilon
\left\vert \partial_{\sigma}\right\vert }\hat{X}^{b}\left(  \sigma
,\varepsilon\right)  \pm\pi\left\vert \partial_{\sigma}\right\vert ^{\alpha
-1}\partial_{\sigma}\hat{P}_{c}\left(  \sigma\right)  \right)  ,\\
\hat{C}\left(  \pm\sigma,\varepsilon\right)   &  =\left(  \pi\hat{P}%
_{b}\left(  \sigma\right)  \mp ie^{-\varepsilon\left\vert \partial_{\sigma
}\right\vert }\left\vert \partial_{\sigma}\right\vert ^{-\alpha-1}%
\partial_{\sigma}\hat{X}^{c}\left(  \sigma,\varepsilon\right)  \right)  ,
\end{align}
The ghost stress tensor Eq.(\ref{TghostSp2}) is also generalized to include
the effects of $\alpha$%
\begin{equation}
\hat{T}_{\pm\pm}^{gh}=\frac{1}{4}\left(  i\epsilon_{mm^{\prime}}\right)
:\left(
\begin{array}
[c]{c}%
\pi\left\vert \partial_{\sigma}\right\vert ^{\frac{\alpha-1}{2}}%
\partial_{\sigma}\hat{P}^{m}\left(  \sigma\right) \\
\mp e^{-\varepsilon\left\vert \partial_{\sigma}\right\vert }\left\vert
\partial_{\sigma}\right\vert ^{\frac{1-\alpha}{2}}\hat{X}^{m}\left(
\sigma,\varepsilon\right)
\end{array}
\right)  \left(
\begin{array}
[c]{c}%
\pi\left\vert \partial_{\sigma}\right\vert ^{\frac{\alpha-1}{2}}%
\partial_{\sigma}\hat{P}^{m^{\prime}}\left(  \sigma\right) \\
\mp e^{-\varepsilon\left\vert \partial_{\sigma}\right\vert }\left\vert
\partial_{\sigma}\right\vert ^{\frac{1-\alpha}{2}}\hat{X}^{m^{\prime}}\left(
\sigma,\varepsilon\right)
\end{array}
\right)  :-\frac{1}{2}i\partial_{\pm\sigma}J_{\pm}^{gh}%
\end{equation}
\bigskip while the zero mode Virasoro operator in (\ref{LoOper}) generalizes
to the following form
\begin{equation}
\hat{L}_{0}=\frac{1}{\pi}\int_{0}^{\pi}d\sigma\left(
\begin{array}
[c]{c}%
\frac{1}{2}\eta_{\mu\nu}\left(  \pi^{2}\hat{P}^{\mu}\hat{P}^{\nu}+\hat{X}%
^{\mu}\left\vert \partial_{\sigma}\right\vert ^{2}\hat{X}^{\nu}\right) \\
+\frac{1}{2}i\epsilon_{mn}\left(  \pi^{2}\hat{P}^{m}\left\vert \partial
_{\sigma}\right\vert ^{1+\alpha}\hat{P}^{n}+\hat{X}^{m}\left\vert
\partial_{\sigma}\right\vert ^{1-\alpha}\hat{X}^{n}\right)
\end{array}
\right)  . \label{LoOper2}%
\end{equation}
Now it is evident that for $\alpha=-1$ (as opposed to $\alpha=1$ in
(\ref{LoOper})) this expression has the same form for both the matter and
ghost parts. In fact, for $\alpha=-1$ this displays an OSp($d|2$)
supersymmetry between matter and ghost degrees of freedom in the operator
$\hat{L}_{0}$ that determines the kinetic term in the Siegel gauge. Recall
that the interaction terms are already supersymmetric. Therefore computations
in the Siegel gauge can now be performed is a supersymmetric fashion at every
step of any computation provided we adopt the new expressions for $\hat{L}%
_{0}$ given above with $\alpha=-1$.

Similarly we record here the iQM version of the BRST operator for the case
$\alpha=-1$ for the flat CFT (which is different than the corresponding
expressions in (\ref{Qfield-flat}) or (\ref{L0m+g-field}))%
\begin{equation}
Q\left(  x,p\right)  =\frac{1}{2\pi}\int_{0}^{\pi/2}d\sigma\left[
\begin{array}
[c]{c}%
\pi p_{b}\left(  \pi^{2}p^{\mu}p_{\mu}+x^{\prime\mu}x_{\mu}^{\prime}+i\pi
^{2}p_{c}p_{b}+ix^{\prime b}x^{\prime c}\right) \\
-i\pi\left(  \left\vert \partial_{\sigma}\right\vert ^{-2}x^{\prime c}\right)
\left(  2x^{\prime\mu}p_{\mu}+x^{\prime c}p_{c}+x^{\prime b}p_{b}\right)
\end{array}
\right]  . \label{Qfield-flat-susy}%
\end{equation}
where $x^{\prime M}\equiv\partial_{\sigma}x^{M}.$ This $Q(x,p)$ is
not supersymmetric, and hence the kinetic term in the general gauge
is not supersymmetric; However, the kinetic term in the Siegel gauge
becomes accidentally OSp($d|2$) invariant after integrating out the
$x_{0}^{b}$ mode as described above.

\subsection{Effective non-Perturbative Purely Cubic Quantum Action
\label{npSection}}

Our discussion of the Siegel gauge so far is at the classical field theory
level, so in the action (\ref{SsFixed}) we have the zero ghost number field
$A_{s}^{\left(  0\right)  }$. However, in a quantum version of string field
theory the path integral includes Faddeev-Popov ghosts. In this case there are
also ghosts of ghosts of $b,c$ types ad infinitum \cite{thorn}. Including all
of these additional ghost fields, the quantum action can be written in a
convenient notation. The full quantum SFT effective action takes a similar
form to (\ref{pertS}), but now the string field includes all positive and
negative ghost numbers, $A=\sum A^{\left(  i\right)  }$, not only the
classical fields $A_{s}^{\left(  0\right)  },$ so the effective action in the
path integral in the Siegel gauge contains this generalized $A$, and in our
description takes the form
\begin{equation}
S_{eff}=-\text{Str}\left(  A\star Q\star A+\frac{g_{0}}{3}A\star A\star
A\right)  . \label{SeffA}%
\end{equation}
It is interesting to note that, after using $Q\star Q=0,$ this action may be
rewritten in the purely cubic form
\begin{equation}
S_{eff}\left(  \bar{A}\right)  =-\frac{1}{3g_{0}^{2}}\text{Str}\left(  \bar
{A}\star\bar{A}\star\bar{A}\right)  ,\;\text{with }\bar{A}\equiv g_{0}A+Q.
\label{nonpertS}%
\end{equation}
This action is invariant under the general gauge transformation
\begin{equation}
\delta_{\Lambda}\bar{A}=[\bar{A},\Lambda\}_{\star}=\bar{A}\star\Lambda-\left(
-1\right)  ^{A\Lambda}\Lambda\star\bar{A}. \label{generalGaugeSymm}%
\end{equation}
where $\Lambda$ includes all ghost numbers, just like $A$ does. From this we
see that (\ref{SeffA}) is also invariant by substituting $\bar{A}$ in terms of
$A$ as in (\ref{nonpertS}).

The purely cubic version has been noted before by many authors \cite{cubic} as
a formal property of SFT for the open string. But in many treatments various
anomalies, including associativity anomalies \cite{BM}\cite{Erler}, emerged
that could not be satisfactorily resolved, so as to cast doubt on the utility
of this observation. Indeed not having a satisfactory resolution of anomalies
leads to wrong conclusions \cite{BM}. In our formalism we have introduced a
reliable regulator $\varepsilon$ as part of the definition of the new MSFT.
With this regulator there are no anomalies and this allows us to use the
purely cubic form of the action reliably.

Thus we will take the fundamental \textit{non-perturbative form of
regulated MSFT} to be of the purely cubic form (\ref{nonpertS}),
including our reliable regulator $\varepsilon$ discussed throughout
this paper. This form of the action has some remarkable properties
as follows

\begin{itemize}
\item The $\star$ is background independent, it does not depend on any
specific CFT, it is defined only by an abstract phase space. Similarly, the
field $\bar{A}\left(  x,p\right)  $ is independent of any CFT.

\item This action has a huge amount of symmetry because all supercanonical
transformations of $\left(  x,p\right)  $ leave the Moyal product invariant.
The action is invariant when the field $\bar{A}$ transforms under a similarity
transformation in the iQM as follows
\begin{equation}
\bar{A}\rightarrow\bar{A}^{\prime}=U\star\bar{A}\star U^{-1},\text{ where
}U=\left(  e^{\varepsilon\left(  x,p\right)  }\right)  _{\star}\text{ with any
}\varepsilon\left(  x,p\right)  ,
\end{equation}
where $\varepsilon\left(  x,p\right)  $ is regarded as a generator of
supercanonical transformations on matter and ghosts in the iQM.

\item A subset of supercanonical transformations is a finite global subset of
super rotations OSp$\left(  d|2\right)  $ that act linearly on the
supervectors $\left(  x^{M},p_{M}\right)  ,$ namely $x^{M}\rightarrow(Sx)^{M}$
and $p_{M}\rightarrow\left(  S^{-1}p\right)  _{M},$ with $S\in$ OSp$\left(
d|2\right)  $ (more accurately OSp($\left(  d-1,1\right)  |2$)). These
supertransformations mix the matter and ghost degrees of freedom. The Moyal
$\star$ in (\ref{msftstar}) and the phase space integration measure
(\ref{supertrace}) are manifestly invariant under this OSp$\left(  d|2\right)
$. Hence when $\bar{A}\left(  x,p\right)  $ is transformed as $\bar{A}\left(
x,p\right)  \rightarrow\bar{A}\left(  Sx,S^{-1}p\right)  $ the action is
invariant. When the action is rewritten in terms of $Q$ and $A$ as in
Eq.(\ref{SeffA}), the matter-ghost symmetry is spontaneously broken (i.e.
hidden); it is still manifest in the cubic term but broken in the quadratic
term because $Q$ is not manifestly symmetric. Keeping track of the broken
symmetry is easy and it turns out to be valuable because this simplifies
computations (see below).

\item Rewriting the purely cubic theory back into the form in Eq.(\ref{SeffA})
is analogous to spontaneous breakdown. It is a rearrangement of the
non-perturbative theory into a perturbative expansion around a classical
solution of the non-perturbative equation of motion
\begin{equation}
\bar{A}\left(  x,p\right)  \star\bar{A}\left(  x,p\right)  =0. \label{npEq}%
\end{equation}
The BRST field $Q\left(  x,p\right)  $ associated to any CFT, as given in
Eqs.(\ref{QisJint}) is an exact solution of this equation $\bar{A}%
_{sol}\left(  x,p\right)  =Q\left(  x,p\right)  $. \textit{Our approach for
the construction of }$Q\left(  x,p\right)  $\textit{ for any CFT given in the
previous section provides an infinite number of solutions to }Eq.(\ref{npEq}%
)\textit{, namely one for each exact conformal CFT}. The perturbative
expansion of the field in power of $g_{0}$ around this solution is $\bar
{A}\left(  x,p\right)  =Q\left(  x,p\right)  +g_{0}A\left(  x,p\right)  .$
Inserting this in the non-perturbative action (\ref{nonpertS}) produces the
perturbative setup in Eq.(\ref{SeffA}) or Eq.(\ref{pertS}).

\item Although the non-perturbative theory is background independent, the
perturbative expansion $\bar{A}\left(  x,p\right)  =Q\left(  x,p\right)
+g_{0}A\left(  x,p\right)  $ is obviously dependent on the background field
$Q\left(  x,p\right)  $ that defines the CFT associated to the choice of the
solution $Q\left(  x,p\right)  .$ All CFTs correspond to solutions of the
non-perturbative equation. But it is not clear if all solutions of
(\ref{npEq}) are CFTs.

\item Using the observation in the previous paragraph we can now obtain a
large class of non-perturbative solutions to the standard string field theory
equation of motion $\hat{Q}A+g_{0}A\star A=0,$ namely, since this equation may
be written as $\left(  Q_{1}\left(  x,p\right)  +g_{0}A\left(  x,p\right)
\right)  ^{2}=0$ where $Q_{1}\left(  x,p\right)  $ is associated with some
CFT$_{1},$ we can give a solution for $A$ in the form%
\begin{equation}
g_{0}A\left(  x,p\right)  =Q_{2}\left(  x,p\right)  -Q_{1}\left(  x,p\right)
\label{solA}%
\end{equation}
where $Q_{2}\left(  x,p\right)  $ is another CFT$_{2}$ that is constructed
from the same degrees of freedom $\left(  x,p\right)  .$ In principle there
are an infinite number of solutions. In practice, going over pairs of exactly
conformal CFTs that we know how to handle (such as those similar to
\cite{IBcurved}-\cite{sl2rStrings}), Eq.(\ref{solA}) provides a
non-perturbative explicit solution to string field theory.
\end{itemize}

\section{Illustrations with Flat Space CFT \label{flatCFT}}

In this section we will illustrate some computations when the CFT corresponds
to the flat Minkowski background in $d=26$. Our goal here is to make our
notation transparent to the reader by showing how to proceed in explicit
simple computations using our formalism.

We have seen in Eqs.(\ref{flatTm},\ref{TghostSp2}) with $\alpha=-1$ that in
this case the total stress tensor is OSp$\left(  26|2\right)  $ invariant and
has the form
\begin{equation}
\hat{T}_{\pm\pm}^{m+gh}\left(  \sigma,\varepsilon\right)  =\frac{1}{4}\left[
\pi\hat{P}\left(  \sigma\right)  \pm e^{-\varepsilon\left\vert \partial
_{\sigma}\right\vert }\partial_{\sigma}\hat{X}\left(  \sigma,\varepsilon
\right)  \right]  ^{2}, \label{Ttotflat}%
\end{equation}
where it is implied that the indices on $\hat{P}^{M},\hat{X}^{M}$ are summed
by using the metric for OSp$\left(  26|2\right)  ,$
\begin{equation}
g_{MN}=\left(
\begin{array}
[c]{cc}%
\eta_{\mu\nu} & 0\\
0 & i\epsilon_{mn}%
\end{array}
\right)  ,
\end{equation}
where $\epsilon_{mn}$ is given in Eq.(\ref{sp2metric}). In the limit
$\varepsilon\rightarrow0$ this $\hat{T}_{\pm\pm}^{m+gh}\left(
\sigma ,\varepsilon\right)  $ reduces to the usual flat-space
Virasoro operators if rewritten in terms of free string modes
including ghosts.

\subsection{Oscillators in $\sigma$-space and Perturbative Vacuum
\label{perturbativeVac}}

It is useful to define regulated creation/annihilation operators in $\sigma$
space as follows. In $\varepsilon\rightarrow0$ limit these are equivalent to
the standard string oscillators in mode space and are given by
\begin{equation}
\hat{a}_{\pm}^{M}\left(  \sigma,\varepsilon\right)  =\frac{1}{\sqrt{2}}\left(
\pi\hat{P}_{N}\left(  \sigma\right)  \eta^{NM}\pm ie^{-\varepsilon\left\vert
\partial_{\sigma}\right\vert }\left\vert \partial_{\sigma}\right\vert \hat
{X}^{M}\left(  \sigma,\varepsilon\right)  \right)  . \label{createAnnih}%
\end{equation}
The inverse relation is%
\begin{align}
\pi\hat{P}_{M}\left(  \sigma\right)   &  =\frac{1}{\sqrt{2}}\left(  \hat
{a}_{+}^{N}\left(  \sigma,\varepsilon\right)  +\hat{a}_{-}^{N}\left(
\sigma,\varepsilon\right)  \right)  g_{NM},\label{Paa}\\
e^{-\varepsilon\left\vert \partial_{\sigma}\right\vert }\hat{X}^{M}\left(
\sigma,\varepsilon\right)   &  =\frac{-i}{\sqrt{2}\left\vert \partial_{\sigma
}\right\vert }\left(  \hat{a}_{+}^{M}\left(  \sigma,\varepsilon\right)
-\hat{a}_{-}^{M}\left(  \sigma,\varepsilon\right)  \right)  . \label{Xaa}%
\end{align}
Note that in comparing Eqs.(\ref{createAnnih},\ref{Ttotflat}) one should
distinguish $\left\vert \partial_{\sigma}\right\vert \equiv\sqrt
{-\partial_{\sigma}^{2}}$ from $\partial_{\sigma}.$ These $\hat{a}_{\pm}%
^{M}\left(  \sigma,\varepsilon\right)  $ satisfy (using $[\hat{X}^{M},\hat
{P}_{N}\}=i\delta_{N}^{M}$)%
\begin{equation}
\lbrack\hat{a}_{-}^{M}\left(  \sigma,\varepsilon\right)  ,\hat{a}_{+}%
^{N}\left(  \sigma^{\prime},\varepsilon\right)  \}=\pi e^{-\varepsilon
\left\vert \partial_{\sigma}\right\vert }\left\vert \partial_{\sigma
}\right\vert \delta_{\varepsilon}\left(  \sigma,\sigma^{\prime}\right)
g^{MN}.
\end{equation}
In terms of $\hat{a}_{\pm}^{M}\left(  \sigma,\varepsilon\right)  $ the
Virasoro operator $\hat{L}_{0}^{m+gh}$ takes the normal ordered form
\begin{equation}
\hat{L}_{0}^{m+gh}=\frac{1}{2\pi}\int_{0}^{\pi}\sum_{\pm}:\hat{T}_{\pm\pm
}^{m+gh}\left(  \sigma,\varepsilon\right)  :~=\frac{g_{MN}}{\pi}\int_{0}^{\pi
}\hat{a}_{+}^{M}\left(  \sigma,\varepsilon\right)  \hat{a}_{-}^{N}\left(
\sigma,\varepsilon\right)  .
\end{equation}
So, the vacuum state in position space, which satisfies $\hat{a}_{-}%
^{N}\left(  \sigma,\varepsilon\right)  \Psi_{0}\left(  X\left(  \sigma
,\varepsilon\right)  \right)  =0,$ is given by the Gaussian%
\begin{equation}
\Psi_{0}\left(  X\left(  \cdot,\varepsilon\right)  \right)  \sim\exp\left\{
-\frac{g_{MN}}{2\pi}\int_{0}^{\pi}d\sigma X^{M}\left(  \sigma,\varepsilon
\right)  \left\vert \partial_{\sigma}\right\vert X^{N}\left(  \sigma
,\varepsilon\right)  \right\}  . \label{vacuum_full}%
\end{equation}
In the limit $\varepsilon\rightarrow0$ this reduces to the expected familiar
vacuum state in the oscillator basis expressed in position space as a
Gaussian. By using the derivative representation of $\hat{P}_{M}\left(
\sigma\right)  $ in position space (see Eq.\ref{Pdiff}),$\;\hat{P}_{M}\left(
\sigma\right)  \Psi\left(  X\right)  =-ie^{-\varepsilon\left\vert
\partial_{\sigma}\right\vert }\left(  \partial\Psi/\partial X_{M}\left(
\sigma,\varepsilon\right)  \right)  ,$ one can verify that indeed $\hat{a}%
_{-}^{N}\left(  \sigma,\varepsilon\right)  \Psi_{0}\left(  X\left(
\sigma,\varepsilon\right)  \right)  =0$ is satisfied for both matter and
ghosts. Hence $\hat{L}_{0}^{m+gh}\Psi_{0}\left(  X\left(  \cdot,\varepsilon
\right)  \right)  =0$ so that $\Psi_{0}\left(  X\left(  \cdot,\varepsilon
\right)  \right)  $ is indeed the perturbative vacuum state in position space.

Now we turn to the field in the Moyal space $A\left(  x,p\right)  .$ The star
representation of the creation-annihilation operators above are obtained by
using the prescription in Eq.(\ref{Ofield}).%
\begin{equation}
\hat{a}_{\pm}^{M}\left(  \sigma,\varepsilon\right)  A\left(  x,p\right)
=\left\{
\begin{array}
[c]{l}%
a_{\pm}^{M}\left(  \sigma,\varepsilon\right)  \star A\left(  x,p\right)
,\text{ if }0\leq\sigma\leq\pi/2,\\
A\left(  x,p\right)  \star a_{\pm}^{M}\left(  \sigma,\varepsilon\right)
\left(  -1\right)  ^{MA},\text{ if }\pi/2\leq\sigma\leq\pi,
\end{array}
\right.
\end{equation}
where $a_{\pm}^{M}\left(  \sigma,\varepsilon\right)  $ (without the hat
$\symbol{94}$) are string fields, constructed from $\left(  x,p\right)  ,$
just like $A\left(  x,p\right)  $
\begin{equation}
a_{\pm}^{M}\left(  \sigma,\varepsilon\right)  =\frac{1}{\sqrt{2}%
}e^{-\varepsilon\left\vert \partial_{\sigma}\right\vert }\left(  \pi
p_{N}\left(  \sigma\right)  g^{NM}\pm i\left\vert \partial_{\sigma}\right\vert
x^{M}\left(  \sigma,\varepsilon\right)  \right)  . \label{createAnnihFields}%
\end{equation}
The vacuum state is identified as the field $A_{0}\left(  x,p\right)  $ that
is annihilated by $a_{-}^{M}\left(  \sigma,\varepsilon\right)  $ either from
the left or the right under star products,
\begin{align}
a_{-}^{M}\left(  \sigma,\varepsilon\right)  \star A_{0}\left(  x,p\right)   &
=0,\;\text{ if }0\leq\sigma\leq\pi/2,\\
A_{0}\left(  x,p\right)  \star a_{-}^{M}\left(  \sigma,\varepsilon\right)   &
=0,\;\;\text{if }\pi/2\leq\sigma\leq\pi.
\end{align}
The solution is
\begin{equation}
A_{0}\left(  x,p\right)  =\mathcal{N}_{0}\exp\left\{  -\frac{1}{2}\int
_{0}^{\pi}d\sigma\left(
\begin{array}
[c]{c}%
g_{MN}~x^{M}\left(  \sigma,\varepsilon\right)  \frac{\left\vert \partial
_{\sigma}\right\vert }{\pi}x^{N}\left(  \sigma,\varepsilon\right) \\
+~g^{MN}~p_{M}\left(  \sigma\right)  \frac{\pi}{\left\vert \partial_{\sigma
}\right\vert }p_{N}\left(  \sigma\right)
\end{array}
\right)  \right\}  . \label{vacuumField}%
\end{equation}
As expected, this $A_{0}\left(  x_{+},p_{-}\right)  $ is consistent with the
Fourier transform of the position space field $\Psi_{0}\left(  X\right)
=\Psi_{0}\left(  x_{+},x_{-}\right)  $ with respect to the variable
$x_{-}\left(  \sigma,\varepsilon\right)  .$ Note that the center of mass
momentum for matter vanishes on the vacuum field%
\begin{align}
\hat{P}_{\mu}^{cm}A_{0}  &  =\int_{0}^{\pi/2}d\sigma e^{-\varepsilon\left\vert
\partial_{\sigma}\right\vert }p_{\mu}\left(  \sigma\right)  \star A_{0}%
+\int_{\pi/2}^{\pi}d\sigma A_{0}\star e^{-\varepsilon\left\vert \partial
_{\sigma}\right\vert }p_{\mu}\left(  \sigma\right) \nonumber\\
&  =\int_{0}^{\pi}d\sigma e^{-\varepsilon\left\vert \partial_{\sigma
}\right\vert }\left(  \frac{1}{2}\frac{-i\partial A_{0}}{\partial x^{\mu
}\left(  \sigma,\varepsilon\right)  }\right)  +0\nonumber\\
&  =\frac{i}{2}A_{0}\int_{0}^{\pi}d\sigma e^{-\varepsilon\left\vert
\partial_{\sigma}\right\vert }\frac{\left\vert \partial_{\sigma}\right\vert
}{\pi}x_{\mu}\left(  \sigma,\varepsilon\right)  =0.
\end{align}
In the second line the \textquotedblleft0\textquotedblright\ represents the
fact that the non-derivative piece in the star product drops out because
$p_{\mu}\left(  \sigma\right)  $ is odd under reflections from $\pi/2$. In the
last line the integral vanishes since $\left\vert \partial_{\sigma}\right\vert
x_{\mu}\left(  \sigma,\varepsilon\right)  $ has no zero mode while the
integrals over the remaining even modes vanish $\int_{0}^{\pi}d\sigma\cos
e\sigma=0.$

The normalization $N_{0}$ in Eq.(\ref{vacuumField}) is determined by
demanding
\begin{equation}
1=\text{Str}\left(  A_{0}\star A_{0}\right)  =\left(  N_{0}\right)  ^{2}\int
DxDp\exp\left\{  -\int_{0}^{\pi}d\sigma\left(
\begin{array}
[c]{c}%
g_{MN}~x^{M}\left(  \sigma,\varepsilon\right)  \frac{\left\vert \partial
_{\sigma}\right\vert }{\pi}x^{N}\left(  \sigma,\varepsilon\right) \\
+~g^{MN}~p_{M}\left(  \sigma\right)  \frac{\pi}{\left\vert \partial_{\sigma
}\right\vert }p_{N}\left(  \sigma\right)
\end{array}
\right)  \right\}  .
\end{equation}
The Gaussian integral gives determinants and this fixes $N_{0}$ as follows
\begin{equation}
N_{0}=\left(  \det\left(  \left\vert \partial_{\sigma}\right\vert
^{-1/2}\right)  _{+}\det\left(  \left\vert \partial_{\sigma}\right\vert
^{1/2}\right)  _{-}\right)  ^{-\left(  d-2\right)  /2},\text{ with }d=26.
\end{equation}
The reason for $\left(  d-2\right)  $ is because the integral in the bosonic
sector contributes the $\left(  d\right)  $ and the Grasmannian integral in
the fermionic ghost sector contributes the $\left(  -2\right)  .$ Another way
of thinking about this is that we have a superdeterminant in the space of
OSp$\left(  d|2\right)  $ and this is why we get $\left(  d-2\right)  $ rather
than $d+2.$ The determinants are to be evaluated in the even and odd sectors
since $x_{+}^{M}\left(  \sigma,\varepsilon\right)  $ has only even modes and
$p_{-M}\left(  \sigma\right)  $ has only odd modes. The result is\footnote{The
computation is ambiguous because the product of the eigenvalues of $\left\vert
\partial_{\sigma}\right\vert $ can be arranged as $\lim_{N\rightarrow\infty}%
{\displaystyle\prod\limits_{n=1}^{N}}
\left(  \frac{2n}{\left(  2n-1\right)  }\right)  =\infty,$ or $\lim
_{N\rightarrow\infty}%
{\displaystyle\prod\limits_{n=1}^{N}}
\left(  \frac{\left(  2n\right)  }{\left(  2n+1\right)  }\right)  =0$. To get
a well defined result we take the product of these two and take the square
root
\begin{equation}
\lim_{N\rightarrow\infty}\left(
{\displaystyle\prod\limits_{n=1}^{N}}
\left(  \frac{2n}{\left(  2n-1\right)  }\right)  \left(  \frac{2n}{\left(
2n+1\right)  }\right)  \right)  ^{1/2}=\sqrt{\pi/2}.
\end{equation}
}
\begin{equation}
\det\left(  \left\vert \partial_{\sigma}\right\vert \right)  _{+}\det\left(
\left\vert \partial_{\sigma}\right\vert ^{-1}\right)  _{-}=\frac{2\cdot
4\cdot8\cdot\cdots\cdot2n\cdot\cdots}{1\cdot3\cdot5\cdot\cdots\cdot\left(
2n+1\right)  \cdot\cdots}=\sqrt{\pi/2}.
\end{equation}
Therefore
\begin{equation}
N_{0}=\left(  \pi/2\right)  ^{\left(  d-2\right)  /8}=\left(  \pi/2\right)
^{3},\text{ with }d=26.
\end{equation}

Often we are interested in the vacuum expectation values of the basic
operators $\hat{X}\left(  \sigma,\varepsilon\right)  ,\hat{P}\left(
\sigma\right)  $. These are computed easily by using the oscillator
expressions in Eqs.(\ref{Paa},\ref{Xaa}) and the properties of the vacuum
state. For example, for $0<\sigma_{1},\sigma_{2}<\pi/2$,%
\begin{align}
&  \langle e^{-\varepsilon\left\vert \partial_{\sigma_{1}}\right\vert }\hat
{X}^{M}\left(  \sigma_{1},\varepsilon\right)  e^{-\varepsilon\left\vert
\partial_{\sigma_{2}}\right\vert }\hat{X}^{N}\left(  \sigma_{2},\varepsilon
\right)  \rangle\nonumber\\
&  =-\frac{-i}{\sqrt{2}\left\vert \partial_{\sigma_{1}}\right\vert }\frac
{-i}{\sqrt{2}\left\vert \partial_{\sigma_{2}}\right\vert }\frac{\text{Str}%
\left(  A_{0}\star\left(  \hat{a}_{-}^{M}\left(  \sigma_{1},\varepsilon
\right)  \hat{a}_{+}^{N}\left(  \sigma_{2},\varepsilon\right)  A_{0}\right)
\right)  }{\text{Str}\left(  A_{0}\star A_{0}\right)  }\nonumber\\
&  =\frac{\pi}{2}e^{-\varepsilon\left\vert \partial_{\sigma_{1}}\right\vert
}e^{-\varepsilon\left\vert \partial_{\sigma_{2}}\right\vert }\left\vert
\partial_{\sigma_{2}}\right\vert ^{-1}\delta\left(  \sigma_{1},\sigma
_{2}\right)  g^{MN}%
\end{align}
and similarly for vacuum expectation values that involve $\hat{P}_{M}\left(
\sigma\right)  $ we find
\begin{align}
\langle e^{-\varepsilon\left\vert \partial_{\sigma_{1}}\right\vert }\hat
{X}^{M}\left(  \sigma_{1},\varepsilon\right)  \hat{P}_{N}\left(  \sigma
_{2}\right)  \rangle &  =i\frac{\pi}{2}e^{-\varepsilon\left\vert
\partial_{\sigma_{1}}\right\vert }e^{-\varepsilon\left\vert \partial
_{\sigma_{2}}\right\vert }\delta\left(  \sigma_{1},\sigma_{2}\right)
\delta_{N}^{M}\\
\langle\hat{P}_{M}\left(  \sigma_{1}\right)  \hat{P}_{N}\left(  \sigma
_{2}\right)  \rangle &  =\frac{\pi}{2}e^{-\varepsilon\left\vert \partial
_{\sigma_{1}}\right\vert }e^{-\varepsilon\left\vert \partial_{\sigma_{2}%
}\right\vert }\delta\left(  \sigma_{1},\sigma_{2}\right)  ~g_{MN}%
\end{align}
Using these expressions we can use Wick's theorem (or equivalently operator
products) to rewrite products of operators $\hat{X}$'s or $\hat{P}$'s in terms
of normal ordered products. The exact parallel steps is available in the Moyal
formulation, but these products occur as star products either on the left side
or the right side of the field $A\left(  x,p\right)  ,$ for example if both
$\sigma_{1},\sigma_{2}$ are less than $\pi/2$
\begin{equation}
\hat{O}\left(  \sigma_{1}\right)  \hat{O}\left(  \sigma_{2}\right)  A\left(
x,p\right)  =O_{\star}\left(  \sigma_{1}\right)  \star O_{\star}\left(
\sigma_{2}\right)  \star A\left(  x,p\right)  .
\end{equation}
Wick's theorem or operator products computed in familiar operator language
have the exact parallel in the Moyal formalism and therefore the c-number
coefficients are identical in either formalism, as in Eq.(\ref{opProducts}).
Therefore we can borrow well known results for quantum operator products in
the CFT and use them directly for the star products of the corresponding
fields in the induced iQM.

The expression for the matter vacuum (\ref{vacuumField}) includes the ghost
contribution
\begin{equation}
A_{0}^{ghost}\left(  x,p\right)  =N_{0}^{gh}\exp\left\{  -i\int_{0}^{\pi
}d\sigma\left(  x^{b}\left(  \sigma,\varepsilon\right)  \frac{\left\vert
\partial_{\sigma}\right\vert }{\pi}x^{c}\left(  \sigma,\varepsilon\right)
+~p_{c}\left(  \sigma\right)  \frac{\pi}{\left\vert \partial_{\sigma
}\right\vert }p_{b}\left(  \sigma\right)  \right)  \right\}  .
\label{vacuum_ghost}%
\end{equation}
which we discuss a bit more to emphasize that this is the ghost $SL\left(
2,\mathbb{R}\right)  $ vacuum. It is known that there are two ghost vacuum
states $A_{0\pm}$ that by definition satisfy the following relations:%

\begin{equation}
\hat{b}_{0}A_{0+}=A_{0-},~\hat{c}_{0}A_{0-}=A_{0+},
\end{equation}
while%
\begin{equation}
\hat{b}_{0}A_{0-}=\hat{c}_{0}A_{0+}=0.
\end{equation}
Their ghost numbers should differ by one:%
\begin{equation}
N_{gh}\left(  A_{0+}\right)  =N_{gh}\left(  A_{0-}\right)  +1,
\end{equation}
and their star product satisfies (recall Str has ghost number 1):%
\begin{align}
\text{Str}\left(  A_{0+}\star A_{0-}\right)   &  =Str\left(  A_{0+}%
\star\left(  \hat{b}_{0}A_{0+}\right)  \right)  =\text{Str}\left(  A_{0-}%
\star\left(  \hat{c}_{0}A_{0-}\right)  \right)  =1,\\
\text{Str}\left(  A_{0+}\star A_{0+}\right)   &  =\text{Str}\left(
A_{0-}\star A_{0-}\right)  =0.
\end{align}
Therefore, we can conclude that the ghost number of $A_{0-}$ is $-1$, while
$N_{gh}\left(  A_{0+}\right)  =0$. The $SL\left(  2,\mathbb{R}\right)  $
vacuum state has the ghost number $0$, therefore it can be considered as the
vacuum state $A_{0+}$.

Hence, the ghost vacua are the following%
\begin{align}
A_{0+}\left(  x,p\right)   &  =N_{0}^{gh}\exp\left\{  -i\int_{0}^{\pi}%
d\sigma\left(  x^{b}\left(  \sigma,\varepsilon\right)  \frac{\left\vert
\partial_{\sigma}\right\vert }{\pi}x^{c}\left(  \sigma,\varepsilon\right)
+~p_{c}\left(  \sigma\right)  \frac{\pi}{\left\vert \partial_{\sigma
}\right\vert }p_{b}\left(  \sigma\right)  \right)  \right\}  ,\\
A_{0-}\left(  x,p\right)   &  =-iN_{0}^{gh}x_{0}^{b}\exp\left\{  -i\int
_{0}^{\pi}d\sigma\left(  x^{b}\left(  \sigma,\varepsilon\right)
\frac{\left\vert \partial_{\sigma}\right\vert }{\pi}x^{c}\left(
\sigma,\varepsilon\right)  +~p_{c}\left(  \sigma\right)  \frac{\pi}{\left\vert
\partial_{\sigma}\right\vert }p_{b}\left(  \sigma\right)  \right)  \right\}  .
\end{align}
This works correctly with the normalization conditions above (recall
Str includes the Grassmannian integral $\int dx_{0}^{b}$). In the
Siegel gauge (\ref{siegelGauge}) we had,
$A_{s}=x_{0}^{b}A_{s}^{\left(  0\right)  },$ with $A_{s}^{\left(
0\right)  }$ not containing the ghost zero mode $x_{0}^{b}.$ Hence
$A_{s}^{\left(  0\right)  }\sim A_{0+}^{gh}$ matches all the
properties as the SL$\left(  2,R\right)  $ invariant vacuum.

\subsection{The Monoid in the $\sigma$-basis}

A very useful tool for computations in MSFT is the monoid algebra developed in
\cite{BM}\cite{BKM}. This arises as follows. We have seen above that the
perturbative vacuum state $A_{0}$ is the Gaussian string field
(\ref{vacuumField}). Excited perturbative string states are represented by the
same Gaussian field multiplied by polynomials of $\left(  x,p\right)  $. The
polynomials in perturbative states can be generated by taking derivatives with
respect to the shift $\lambda_{x}$ or $\lambda_{p}$ of a shifted gaussian
$A\sim\exp\left(  -xx-pp+\lambda_{x}x+\lambda_{p}p\right)  ,$ and then setting
the $\lambda^{\prime}s$ to zero. On the other hand, at least some
non-perturbative states are also shifted gaussian-like states, but with a
different quadratic and linear exponent than the one for the perturbative
states. This suggests that fields of the shifted-gaussian form are very common
in explicit computations. It was found in \cite{BM}\cite{BKM} that they have
nice mathematical properties that are directly useful in the computation of
amplitudes, including the Veneziano amplitude \cite{BP}.

The set of shifted gaussians of interest are of the form
\begin{equation}
A_{\mathcal{N},M,\lambda}=\mathcal{N}e^{-\xi^{i}M_{ij}\xi^{j}-\xi^{i}%
\lambda_{i}}. \label{generating}%
\end{equation}
where the $\xi^{i}$ stand for $\left(  x,p\right)  $ and the symbols
$M_{ij},\lambda_{i}$ are parameters, while $\mathcal{N}$ is a normalization.
They close under the Moyal star product as follows
\begin{equation}
A_{\mathcal{N}_{1},M_{1},\lambda_{1}}\star A_{\mathcal{N}_{2},M_{2}%
,\lambda_{2}}=A_{\mathcal{N}_{12},M_{12},\lambda_{12}},
\end{equation}
The $\xi^{i}$ are a set of non-commutative (super)variables (matter and ghost
phase spaces) which satisfy%
\begin{equation}
\lbrack\xi^{i},\xi^{j}\}_{\star}=s^{ij},
\end{equation}
with $s^{ij}$ a constant matrix which is antisymmetric when both $i$ and $j$
are bosonic and is symmetric when both $i$ and $j$ are fermionic. The $\left(
\mathcal{N}_{12},M_{12},\lambda_{12}\right)  $ are computed from $\left(
\mathcal{N}_{1},M_{1},\lambda_{1}\right)  $ and $\left(  \mathcal{N}_{2}%
,M_{2},\lambda_{2}\right)  $ as follows. Given the data for $\left(
\mathcal{N}_{1},M_{1},\lambda_{1}\right)  $ and $\left(  \mathcal{N}_{2}%
,M_{2},\lambda_{2}\right)  $ we first define the matrices $m$%
\begin{equation}
m_{1}=M_{1}s,\;m_{2}=M_{2}s,\;m_{12}=M_{12}s;
\end{equation}
then the result for $m_{12},\lambda_{12},\mathcal{N}_{12}$ takes the form
\cite{B}\cite{BM}\cite{BKM}
\begin{align}
m_{12}  &  =\left(  m_{1}+m_{2}m_{1}\right)  \left(  1+m_{2}m_{1}\right)
^{-1}+\left(  m_{2}-m_{1}m_{2}\right)  \left(  1+m_{1}m_{2}\right)
^{-1},\label{m12}\\
\lambda_{12}  &  =\left(  1-m_{1}\right)  \left(  1+m_{2}m_{1}\right)
^{-1}\lambda_{2}+\left(  1+m_{2}\right)  \left(  1+m_{1}m_{2}\right)
^{-1}\lambda_{1}\label{lambda12}\\
\mathcal{N}_{12}  &  =\frac{\mathcal{N}_{1}\mathcal{N}_{2}}{\text{sdet}\left(
1+m_{2}m_{1}\right)  ^{1/2}}e^{\frac{1}{4}\left(  \left(  \lambda_{1}%
+\lambda_{2}\right)  \left(  M_{1}+M_{2}\right)  ^{-1}\left(  \lambda
_{1}+\lambda_{2}\right)  -\bar{\lambda}_{12}\left(  M_{12}\right)
^{-1}\lambda_{12}\right)  }. \label{n12}%
\end{align}
The reader may consult \cite{B}\cite{BM}\cite{BKM} for detailed properties of
this monoid algebra and how it is used for both perturbative and
non-perturbative computations in string field theory. In our formulation here
this algebra is consistent with the OSp$\left(  d|2\right)  $ supersymmetry,
and therefore we use supertrace and superdeterminants instead of the trace and
determinant in \cite{B}\cite{BM}\cite{BKM}.

The results in (\ref{m12}-\ref{n12}) were computed for the Moyal star product
in a discrete mode space, but the same formal result applies also with our new
Moyal star product in $\sigma$-space. In the present case the non-commutative
variables are labelled by $i$ which is a combination of discrete $\left(
M\right)  $ and continuous $\left(  \sigma\right)  $ labels. The shifted
gaussian in our new formalism is
\begin{equation}
A_{\mathcal{N},M,\lambda}\equiv\mathcal{N}\exp\left(  -\left(  \int_{0}^{\pi
}\int_{0}^{\pi}d\sigma d\sigma^{\prime}\xi^{i}\left(  \sigma\right)
M_{ij}\left(  \sigma,\sigma^{\prime}\right)  \xi^{j}\left(  \sigma^{\prime
}\right)  \right)  -\int_{0}^{\pi}d\sigma\xi^{i}\left(  \sigma\right)
\lambda_{i}\left(  \sigma\right)  \right)
\end{equation}
where $\xi^{i}\left(  \sigma\right)  =\left(  x^{M}\left(  \sigma\right)
,p_{M}\left(  \sigma\right)  \right)  $ are the string half-phase degrees of
freedom, $M_{ij}\left(  \sigma,\sigma^{\prime}\right)  $ is a complex square
matrix and $\lambda_{i}\left(  \sigma\right)  $ is a complex column matrix.
Under the Moyal star product they form a closed algebra called a monoid (which
is almost a group, except for the inverse condition). Taking into account the
star commutation rules in (\ref{iQM[]}) we identify the matrix $s^{ij}$ with
the new type of labels as follows%
\begin{equation}
s^{ij}=\overset{i=}{%
\begin{array}
[c]{c}%
x^{M}\left(  \sigma\right) \\
p_{M}\left(  \sigma\right)
\end{array}
}\overset{j=%
\begin{array}
[c]{cc}%
x^{N}\left(  \sigma^{\prime}\right)  \;\;\;, & \;\;\;p_{N}\left(
\sigma^{\prime}\right)
\end{array}
}{\left(
\begin{array}
[c]{cc}%
0 & i^{1-N}\delta_{~N}^{M}\hat{\delta}_{+-}\left(  \sigma,\sigma^{\prime
}\right) \\
\left(  -i\right)  ^{1-N^{\prime}}\delta_{M}^{~N}\hat{\delta}_{-+}\left(
\sigma,\sigma^{\prime}\right)  & 0
\end{array}
\right)  }%
\end{equation}
where $\hat{\delta}_{+-}\left(  \sigma,\sigma^{\prime}\right)  $ is given in
Eqs.(\ref{deltaHat1}-\ref{deltaHat2}), while $\hat{\delta}_{-+}\left(
\sigma,\sigma^{\prime}\right)  $ is the transpose of the \textquotedblleft
matrix\textquotedblright\ $\hat{\delta}_{+-}$ and then re-labelled by
replacing $\sigma\leftrightarrow\sigma^{\prime}.$ The parameters $M_{ij}$ and
$\lambda_{i}$ of the shifted gaussian take the form%
\begin{equation}
M_{ij}=\overset{i=}{%
\begin{array}
[c]{c}%
x^{M}\left(  \sigma\right) \\
p_{M}\left(  \sigma\right)
\end{array}
}\overset{j=%
\begin{array}
[c]{cc}%
x^{N}\left(  \sigma^{\prime}\right)  \;\;\;, & \;\;\;p_{N}\left(
\sigma^{\prime}\right)  ~~~~~~
\end{array}
}{\left(
\begin{array}
[c]{cc}%
a_{MN}\left(  \sigma,\sigma^{\prime}\right)  & b_{M}^{~~N}\left(
\sigma,\sigma^{\prime}\right) \\
b_{~N}^{M}\left(  \sigma,\sigma^{\prime}\right)  & ~d^{MN}\left(
\sigma,\sigma^{\prime}\right)
\end{array}
\right)  ,\;}\lambda_{i}=\left(
\begin{array}
[c]{c}%
\lambda_{x^{M}}\left(  \sigma\right) \\
\lambda_{p_{M}}\left(  \sigma\right)
\end{array}
\right)
\end{equation}
where the diagonal entries of $M$ are (super)symmetric matrices while the off
diagonal entries are related by a (super) transposition
\begin{equation}
b_{~N}^{M}\left(  \sigma,\sigma^{\prime}\right)  =\left(  -1\right)
^{MN}b_{M}^{~~N}\left(  \sigma^{\prime},\sigma\right)  .
\end{equation}
The matrix $m=Ms$ becomes%
\begin{equation}
m_{i}^{j}=M_{ik}s^{kj}=\left(
\begin{array}
[c]{cc}%
~b_{M}^{~N}\left(  \sigma,\sigma^{\prime}\right)  \text{sign}\left(
\pi/2-\sigma^{\prime}\right)  & ~a_{MN}\left(  \sigma,\sigma^{\prime}\right)
\text{sign}\left(  \pi/2-\sigma^{\prime}\right)  \left(  -1\right)  ^{N}\\
d^{MN}\left(  \sigma,\sigma^{\prime}\right)  \text{sign}\left(  \pi
/2-\sigma^{\prime}\right)  & b_{~N}^{M}\left(  \sigma,\sigma^{\prime}\right)
\text{sign}\left(  \pi/2-\sigma^{\prime}\right)  \left(  -1\right)  ^{N}%
\end{array}
\right)
\end{equation}
Then they are combined according to the rules (\ref{m12}-\ref{n12}) to obtain
the result for the monoid algebra.

As an example, for perturbative computations, the matrix $M_{ij}\left(
\sigma,\sigma^{\prime}\right)  $ is fairly simple. It follows from the vacuum
state given in (\ref{vacuumField})%
\begin{equation}
M_{ij}^{pert}=\left(
\begin{array}
[c]{cc}%
g_{MN}\frac{\left\vert \partial_{\sigma}\right\vert }{\pi}\delta_{\varepsilon
}^{+}\left(  \sigma,\sigma^{\prime}\right)  & 0\\
0 & ~g^{MN}\frac{\pi}{\left\vert \partial_{\sigma}\right\vert }\delta
_{\varepsilon}^{-}\left(  \sigma,\sigma^{\prime}\right)
\end{array}
\right)  \label{Mpert}%
\end{equation}
In general the \textit{regulated} delta function in (\ref{Mpert}) satisfies
D-brane boundary conditions as discussed in section (\ref{deltaSection}). The
regulator $\varepsilon$ insures that all computations are well defined. The
regulator is removed after renormalization of the cubic coupling constant
$g_{0}$ as shown in \cite{BP}. For example, for the D25 brane we would use in
(\ref{Mpert}) the $\delta_{\varepsilon}^{\pm nn}\left(  \sigma,\sigma^{\prime
}\right)  $ in all directions $M.$ But if more complicated D-brane boundary
conditions are desired, then in the corresponding directions $M$ we would use
the $\delta_{\varepsilon}^{\pm nn}$ or $\delta_{\varepsilon}^{\pm dd}$ given
in section (\ref{deltaSection})$.$ In this way, by making only minimal changes
through the regulated delta functions, non-trivial D-brane boundary conditions
are implemented easily in our new MSFT formalism.

If the $M$ in (\ref{Mpert}) is used in monoid computations directly in the
form shown in (\ref{Mpert}) with $\delta^{nn}$, then this approach reproduces
all the results of the computations obtained previously in \cite{B}%
\cite{BM}\cite{BKM}\cite{BP}, including the \textit{off-shell}
4-tachyon (Veneziano) scattering amplitude.

\section{Outlook \label{outlook}}

The central structure in this paper is the new Moyal $\star$ product in the
$\sigma$-basis in Eq.(\ref{msftstar}) that implements the interactions of
strings. The string fields $A\left(  x,p\right)  $ that are multiplied with
this product are labeled by \textit{half of the phase space} of the string
$\left(  x_{+}^{M}\left(  \sigma\right)  ,p_{-M}\left(  \sigma\right)
\right)  $ as opposed to the full phase space $\left(  X^{M}\left(
\sigma\right)  ,P_{M}\left(  \sigma\right)  \right)  .$ The label $M=\left(
\mu,b,c\right)  $ includes both the spacetime \textquotedblleft matter $\mu
$\textquotedblright\ and the $\left(  b,c\right)  $ ghosts in an OSp$\left(
d|2\right)  $ covariant notation. The star product $\star$, which is
independent of the details of any conformal field theory on the worldsheet
(CFT), is background independent and is invariant under this supersymmetry for
all CFTs.

The symmetric $x_{+}^{M}\left(  \sigma\right)  =\frac{1}{2}\left(
X^{M}\left(  \sigma\right)  +X^{M}\left(  \pi-\sigma\right)  \right)  $ and
the antisymmetric $p_{-M}\left(  \sigma\right)  =\frac{1}{2}\left(
P_{M}\left(  \sigma\right)  -P_{M}\left(  \pi-\sigma\right)  \right)  $
commute with each other in the quantum mechanics (QM) of the first quantized
string, and therefore they are simultaneous observables in QM. The eigenvalues
$\left(  x_{+},p_{-}\right)  $ of these simultaneous observables provide a
complete set of labels for the first quantized string states $\langle
x_{+}^{M}\left(  \sigma\right)  ,p_{-M}\left(  \sigma\right)  |.$ The string
field $A\left(  x,p\right)  $ corresponds to the probability amplitude of a
general string state $|A\rangle$ that has the given phase space configuration
$A\left(  x,p\right)  =\langle x_{+}^{M}\left(  \sigma\right)  ,p_{-M}\left(
\sigma\right)  |A\rangle.$ Hence the Moyal product in Eq.(\ref{msftstar})
which creates a non-commutativity in the space of eigenvalues $\left(
x_{+}^{M}\left(  \sigma\right)  ,p_{-M}\left(  \sigma\right)  \right)  $ has
nothing to do with the Moyal product in QM despite the close mathematical
similarities. However this close similarity is interpreted in this paper as an
induced quantum mechanics (iQM) that governs the fundamental \textit{string
interactions}.

To be able to compute reliably in string field theory we need a regulator to
obtain unambiguous results. The essential role of the regulator is to tame
some singularities associated with the midpoint of the string at $\pi/2$. In
this paper the regulator is the dimensionless small parameter $\varepsilon.$
We regulated at first the QM quantum operators of the first quantized string
by defining $\hat{X}\left(  \sigma,\varepsilon\right)  ,$ while no
regularization is needed for the operator $\hat{P}_{M}\left(  \sigma\right)
,$ as discussed in section (\ref{Regulator_section}). Consequently, the
eigenvalues of the half phase space are also regularized, so that the string
field is regularized with the $\varepsilon$ through its labels $A\left(
x_{+}^{M}\left(  \sigma,\varepsilon\right)  ,p_{-M}\left(  \sigma\right)
\right)  .$ On this regularized basis we showed how to represent all QM
operators that belong to any CFT, by their string-field-counterparts in the
induced iQM. This representation involves only star products of the string
fields. In particular some crucial operators that are needed to construct
string field theory, i.e. the stress tensor $T_{\pm\pm}$, the BRST current
$j_{B}$ and the BRST operator, for any CFT are constructed as
\textit{regularized string fields} that operate in the induced iQM$.$ Using
them we constructed the regularized action for the new Moyal string field
theory (MSFT).

An important aspect of the new formulation is that the regularized midpoint of
the string $x_{+}^{M}\left(  \pi/2,\varepsilon\right)  $ is not isolated from
the rest of the string degrees of freedom in its treatment under the $\star$
product. Nevertheless the new product has the magical property that the
midpoint $x_{+}^{M}\left(  \pi/2,\varepsilon\right)  $ acts trivially as a
complex number (no derivatives induced on the field) when it is
star-multiplied with any string field $x_{+}^{M}\left(  \pi/2,\varepsilon
\right)  \star A\left(  x,p\right)  =A\left(  x,p\right)  \star x_{+}%
^{M}\left(  \pi/2,\varepsilon\right)  =x_{+}^{M}\left(  \pi/2,\varepsilon
\right)  A\left(  x,p\right)  .$ Remarkably, this important property of the
midpoint in string joining is automatically implemented by the new Moyal
$\star$.

We constructed an infinite number of solutions to the non-perturbative
equation of motion $\bar{A}\star\bar{A}=0$ of the purely cubic theory, in the
form $\bar{A}_{sol}\left(  x,p\right)  =Q\left(  x,p\right)  ,$ where the
\textit{BRST field} $Q\left(  x,p\right)  ~$is derived with our methods from
any CFT on the worldsheet. We turned this observation into a method for
finding an infinite number of non-perturbative solutions to the string field
equation, $\hat{Q}A+g_{0}A\star A=0,$ in the form, $g_{0}A=Q_{2}-Q_{1},$ where
$Q_{i}\left(  x,p\right)  $ correspond to the \textit{BRST fields} (as
obtained with our methods) of a pair of conformal field theories $CFT_{i}.$

We have shown that all successful computations previously accomplished in MSFT
using the old regularized Moyal star product (which was tied to the flat CFT
in $d=26,$ and treated the midpoint as special), are also reproduced by the
new formalism when the same flat CFT is used. However the advantage of the new
approach is that it also applies to any curved CFT and can easily include the
effects of D-brane boundary conditions. The applications of the new features
will be explored in future work in several directions as follows.

It would be a very interesting exercise to apply our formalism to some simple
cases of exact CFTs. Sometime ago some of the earliest examples of exact
conformal field theories that describe strings in curved spaces with
\textit{one time coordinate} were suggested \cite{IBcurved} and studied at the
classical and first quantized levels \cite{BSf}-\cite{sl2rStrings}. The BRST
operator $Q\left(  x,p\right)  $ associated with such models is constructed in
terms of a current algebra (or Kac-Moody algebra) basis which replaces the
half phase space $\left(  x\left(  \sigma\right)  ,p\left(  \sigma\right)
\right)  $. Using the properties of the current algebra should be an important
tool to compute in these special curved spaces using our MSFT approach.

We are eager to aim our new MSFT approach to investigate the physical
circumstances in which string theory should play its most important physical
role. Noting that string field theory (SFT) is a complete approach that
incorporates generally both the curved background and the interactions in the
non-perturbative description of the theory, it is very important to pursue the
SFT avenue despite the fact that computations may be difficult. The areas that
we think are important to investigate with our MSFT formalism includes very
early cosmology in the vicinity of cosmological singularities as well as black
hole or black D-brane type singularities. In particular, there has been some
new developments in identifying uniquely cosmological backgrounds that are
geodesically complete across cosmological singularities \cite{BSTstring} to
which we plan to apply our formalism.

An understanding of the very early cosmology of the universe through string
theory has been tradionally a hope that it would eventually yield an
explanation of why we live in four dimensions and provide the ingredients of
the Standard Model of particle physics, such as the number of generations and
their symmetry structures. It is believed that string physics is unavoidable
in the deeply small and highly curved quantum mechanical regions of
space-time. In this paper we have developed sharper tools to address such
issues in the context of the new MSFT, including string-string interactions
both perturbatively and non-perturbatively, and hope to make further progress
in the pursuit of these goals.

We wish to conclude with a speculation on the origin of quantum mechanics. A
by-product of our approach is an astonishing suggestion of the formalism: the
roots of ordinary quantum mechanics may originate from the non-commutative
interactions in string theory. Indeed, the string joining Moyal star induces
non-commutativity between otherwise commutative string degrees of freedom
$\left(  \hat{x}_{+}\left(  \sigma,\varepsilon\right)  ,\hat{p}_{-}\left(
\sigma\right)  \right)  $. We draw again the attention of the reader to the
remarkable property that even though the operators $\left(  \hat{x}_{+}\left(
\sigma,\varepsilon\right)  ,\hat{p}_{-}\left(  \sigma\right)  \right)  $
commute in QM, their eigenvalues $\left(  x_{+}\left(  \sigma,\varepsilon
\right)  ,p_{-}\left(  \sigma\right)  \right)  $ do not commute with each
other under the induced iQM as seen in Eq.(\ref{iQM[]}). This
non-commutativity under the string-joining star product is what led to the
representation of the QM operators in the basis of iQM as in Eqs.(\ref{iQMX2}%
,\ref{iQMP2}) in the language of only the string joining star
product. The reader is invited to read section (\ref{QM=iQM}) in
reverse by starting from Eqs.(\ref{iQMX2},\ref{iQMP2}) and
interpreting its contents as the emergence of QM from iQM rather
than the other way around. Then it seems astonishing that the
half-phase-space $\left(  x_{+}\left(  \sigma,\varepsilon\right)
,p_{-}\left(  \sigma\right)  \right)  $ in iQM under the string
joining $\star$ generates the conventional QM commutation rules for
the full phase space operators $\left(  \hat{X}\left(
\sigma,\varepsilon\right)  ,\hat {P}\left(  \sigma\right)  \right)
$. Assuming that string theory is right that it underlies all
physics, it is then very tempting to speculate that the source of QM
rules in all physics may simply be the rules of interactions in
string theory as seen explicitly in our paper. This could be the
long sought explanation of where QM comes from. This exciting point
is very important in its own right, and it will be pursued further.

\begin{acknowledgments}
This research was partially supported by the U.S. Department of Energy. IB
thanks CERN for hospitality while part of this research was performed.
\end{acknowledgments}

\section{Appendix}

\subsection{Old Discrete Basis Versus New $\sigma$-Basis \label{old-new}}

In this section we are going to show that the new $\sigma$-basis formalism
that does not distinguish the midpoint is equivalent to the old discrete basis
that distinguished the midpoint. The old formalism was developed in
\cite{B}\cite{BM}\cite{BKM}\cite{BP}. By not distinguishing the midpoint we
have a cleaner and more efficient approach in practical computations.

The unregulated position and momentum degrees of freedom $x_{+}\left(
\sigma\right)  $ and $p_{-}\left(  \sigma\right)  $ in the new $\sigma$ basis
can be written in terms of modes with even and odd \textit{cosine} expansions
respectively,%
\begin{equation}
x_{+}\left(  \sigma\right)  =x_{0}+\sqrt{2}\sum_{e}x_{e}\cos e\sigma
;\;\;\;p_{-}\left(  \sigma\right)  =\frac{\sqrt{2}}{\pi}\sum_{o}p_{o}\cos
o\sigma, \label{xpmodes}%
\end{equation}
where $e=2,4,6,\cdots$ and $o=1,3,5,\cdots.$ Their Moyal $\star$ commutator,
using the new $\star$ product (\ref{msftstar}), is the delta-function that has
two forms (\ref{iQM[]})%
\begin{equation}
\left[  x_{+}\left(  \sigma\right)  ,p_{-}\left(  \sigma^{\prime}\right)
\right]  _{\star new}=i\delta^{+nn}\left(  \sigma,\sigma^{\prime}\right)
\varepsilon\left(  \pi/2-\sigma^{\prime}\right)  =i\varepsilon\left(
\pi/2-\sigma\right)  \delta^{-nn}\left(  \sigma,\sigma^{\prime}\right)  .
\label{x+p-localcomm}%
\end{equation}
Using the mode expansion for these delta functions we can compare the two
results
\begin{equation}
\left(  \frac{2}{\pi}+\frac{4}{\pi}\sum_{e}\cos e\sigma\cos e\sigma^{\prime
}\right)  \varepsilon\left(  \pi/2-\sigma^{\prime}\right)  =\varepsilon\left(
\pi/2-\sigma\right)  \frac{4}{\pi}\sum_{o}\cos o\sigma\cos o\sigma^{\prime}.
\end{equation}
It can be verified that this is an identity \cite{BM}. The commutator contains
two terms, one with and one without the zero mode:%
\begin{equation}
\left[  x_{+}\left(  \sigma\right)  ,p_{-}\left(  \sigma^{\prime}\right)
\right]  _{\star new}=\frac{\sqrt{2}}{\pi}\sum_{o}\left[  x_{0},p_{o}\right]
_{\star new}\cos o\sigma^{\prime}+\frac{2}{\pi}\sum_{e,o}\left[  x_{e}%
,p_{o}\right]  _{\star new}\cos e\sigma\cos o\sigma^{\prime}, \label{xp*new}%
\end{equation}
By comparing the Fourier modes to the answer (\ref{x+p-localcomm}) we extract
the mode commutators and find that they are expressed in terms of the matrix
elements of the special matrix $T$ introduced in \cite{B}%
\begin{equation}
\left[  x_{0},p_{o}\right]  _{\star new}=2iT_{0o},\;\;\left[  x_{e}%
,p_{o}\right]  _{\star new}=2iT_{eo}, \label{old1}%
\end{equation}
where
\begin{align}
T_{eo}  &  =\frac{4}{\pi}\int_{0}^{\pi/2}d\sigma\cos e\sigma\cos o\sigma
=\frac{4}{\pi}\frac{o\left(  -1\right)  ^{\left(  e-o-1\right)  /2}}%
{e^{2}-o^{2}},\label{old2}\\
T_{0o}  &  =\frac{2\sqrt{2}}{\pi}\int_{0}^{\pi/2}d\sigma\cos o\sigma
=\frac{2\sqrt{2}}{\pi o}\sin\frac{o\pi}{2}=\frac{2\sqrt{2}}{\pi}\frac{\left(
-1\right)  ^{\left(  o-1\right)  /2}}{o}. \label{old3}%
\end{align}
A useful identity is \cite{BM} (a sum over $e$ is implied)
\begin{equation}
\cos o\sigma=\varepsilon\left(  \frac{\pi}{2}-\sigma\right)  \left(  \cos
e\sigma-\cos e\frac{\pi}{2}\right)  T_{eo}. \label{old4}%
\end{equation}
These star commutators for the modes (\ref{old1}-\ref{old3}) that were
obtained with the new Moyal product (\ref{msftstar}) in the $\sigma$ basis are
identical to those given by the old Moyal product in the discrete basis
\textit{for the flat CFT case}, once we write $p_{o}=p_{e}T_{eo}$ where
$p_{e}$ is introduced \cite{B} as a convenient change of basis (it does
\textit{not} mean the even momentum modes in $p_{+}\left(  \sigma\right)  $).
Hence the old and new Moyal products are equivalent for the flat CFT. But the
new $\sigma$ basis $\star$ product (\ref{msftstar}) is more powerful since it
is valid for all CFTs due to the fact that the phase space notation in the
$\sigma$ basis is independent of the background fields.

There is one more subtle point in this equivalence which is related to the
midpoint. In the old approach the midpoint,
\begin{equation}
\bar{x}\equiv x\left(  \pi/2\right)  =x_{0}+\sqrt{2}\sum_{e\geq2}x_{e}%
\cos\frac{e\pi}{2},
\end{equation}
was distinguished as an independent mode (instead of $x_{0}$) and did not
appear in the old star product which was expressed only in terms of
$x_{e},p_{e}$ (where $p_{e}$ is related to $p_{o}=p_{e}T_{eo}$). Then the
midpoint $\bar{x}$ had a trivial star product (i.e. it acted on fields with
the ordinary product for complex numbers) and in particular commuted with the
momentum modes under the string joining $\star.$ We want to verify that the
new $\star$ product also has the same properties for the midpoint even though
the new $\star$ (\ref{msftstar}) does not distinguish the midpoint. It is
useful to note the following notation for $w_{e},v_{o}$ and the relations
among these symbols and the midpoint
\begin{align}
x_{0}  &  =\bar{x}+w_{e}x_{e},\;\text{with }w_{e}=-\sqrt{2}\cos\frac{e\pi}%
{2}=-\sqrt{2}\left(  -1\right)  ^{e/2},\label{xoxbar}\\
v_{o}  &  \equiv T_{0o}=w_{e}T_{eo},
\end{align}
where a summation over $e$ is implied. The last equation follows from
integrating both sides of (\ref{old4}). Note that $v_{o}$ or equivalently
$T_{0o}$ is really related to the midpoint at $\pi/2$ since
\begin{equation}
\left[  \bar{x},p_{o}\right]  _{\star new}=\left[  \left(  x_{0}-w_{e}%
x_{e}\right)  ,p_{o}\right]  _{\star}=2i\left(  T_{0o}-w_{e}T_{eo}\right)  =0.
\label{midpointcommutator}%
\end{equation}
This verifies the last crucial point in the equivalence of the old and new
star products: even though the midpoint $\bar{x}$ was not distinguished in the
new star product, $\bar{x}$ behaves as if it is insensitive to the derivatives
implicit in the new Moyal product and therefore it acts just like a complex
number under the new $\star,$ which is the same as under the old $\star.$

It is also instructive to verify the equivalence in reverse. That is, suppose
we are given the commutators (\ref{old1}-\ref{old3}) and
(\ref{midpointcommutator}) under the old product $\star_{old}$ with the same
result, and let us derive the local commutators in the continuous $\sigma
$-basis (\ref{x+p-localcomm}) by using only the old star product. By writing
out $x_{+}\left(  \sigma\right)  ,p_{-}\left(  \sigma^{\prime}\right)  $ in
terms of modes (\ref{xpmodes}) and using (\ref{old1}-\ref{old3}) with
$\star_{old}$ we get the same form as (\ref{xp*new})%

\begin{equation}
\left[  x_{+}\left(  \sigma\right)  ,p_{-}\left(  \sigma^{\prime}\right)
\right]  _{\star old}=2i\frac{\sqrt{2}}{\pi}\sum_{o}T_{0o}\cos o\sigma
^{\prime}+2i\frac{2}{\pi}\sum_{e,o}T_{eo}\cos e\sigma\cos o\sigma^{\prime}.
\end{equation}
Now insert the results (\ref{old2}-\ref{old4}) for $T_{0o},T_{eo}$ given in
the old literature \cite{B} and compute the right hand side of this equation.
We find \emph{ }%
\begin{align}
&  =\left[
\begin{array}
[c]{c}%
2i\frac{\sqrt{2}}{\pi}\sum_{o}\left(  \frac{2\sqrt{2}}{\pi}\int_{0}^{\pi
/2}d\sigma_{1}\cos o\sigma_{1}\right)  \cos o\sigma^{\prime}\\
+2i\frac{2}{\pi}\sum_{e,o}\left(  \frac{4}{\pi}\int_{0}^{\pi/2}d\sigma_{1}\cos
e\sigma_{1}\cos o\sigma_{1}\right)  \cos e\sigma\cos o\sigma^{\prime}%
\end{array}
\right] \nonumber\\
&  =i\frac{2}{\pi}\int_{0}^{\pi/2}d\sigma_{1}\delta^{-}\left(  \sigma
_{1},\sigma^{\prime}\right)  +i\int_{0}^{\pi/2}d\sigma_{1}\left(  \delta
^{+}\left(  \sigma_{1},\sigma\right)  -\frac{2}{\pi}\right)  \delta^{-}\left(
\sigma_{1},\sigma^{\prime}\right) \nonumber\\
&  =i\delta^{+nn}\left(  \sigma,\sigma^{\prime}\right)  \varepsilon\left(
\pi/2-\sigma^{\prime}\right)  .
\end{align}
Therefore, the old product $\star_{old}$ reproduces the new product
$\star_{new}$ in the $\sigma$ basis.

Hence the same string field can be rewritten in the different bases%
\begin{equation}
A\left(  x_{+}\left(  \sigma\right)  ,p_{-}\left(  \sigma\right)  \right)
=A\left(  x_{0},x_{e},p_{o}\right)  =\left.  A\left(  x_{0},x_{e}%
,p_{o}\right)  \right\vert _{x_{0}=\bar{x}+w_{e}x_{e}}=\tilde{A}\left(
\bar{x},x_{e},p_{o}\right)  . \label{AAtilde}%
\end{equation}
and the joining of strings can be expressed equivalently in either the old or
the new star products if the CFT is flat. To show some subtleties of how this
works, we begin with the relation between the old and new bases. The old star
product is (using (\ref{orderedMoyal1}) with fixed $\bar{x}$)
\begin{equation}
\tilde{A}_{1}\left(  \bar{x},x_{e},p_{o}\right)  \star_{old}\tilde{A}%
_{2}\left(  \bar{x},x_{e},p_{o}\right)  =\tilde{A}_{1}\left(  \bar{x},\left(
x_{e}^{\prime}+iT_{eo}\partial_{p_{o}}\right)  ,\left(  p_{o}^{\prime}%
-iT_{eo}\partial_{x_{e}}\right)  \right)  \star\tilde{A}_{2}\left(  \bar
{x},x_{e},p_{o}\right)
\end{equation}
where in the last formula we used Eq.(\ref{orderedMoyal1}). Rewrite this in
terms of the new basis $A_{1,2}\left(  x_{0},x_{e},p_{o}\right)  $ making
explicit that $x_{0}$ is a function of $\bar{x}$ and $x_{e}$ when the old star
product is used:
\begin{align}
&  \tilde{A}_{1}\left(  \bar{x},x_{e},p_{o}\right)  \star_{old}\tilde{A}%
_{2}\left(  \bar{x},x_{e},p_{o}\right) \label{oldnew1}\\
&  =A_{1}\left(  x_{0}\left(  \bar{x},x_{e}\right)  ,x_{e},p_{o}\right)
\star_{old}A_{2}\left(  x_{0}\left(  \bar{x},x_{e}\right)  ,x_{e},p_{o}\right)
\\
&  =A_{1}\left(  \left(  \bar{x}^{\prime}+w_{e}\left(  x_{e}^{\prime}%
+iT_{eo}\partial_{p_{o}}\right)  \right)  ,\left(  x_{e}^{\prime}%
+iT_{eo}\partial_{p_{o}}\right)  ,\left(  p_{o}^{\prime}-iT_{eo}%
\partial_{x_{e}}\right)  \right)  ~A_{2}\left(  x_{0}\left(  \bar{x}%
,x_{e}\right)  ,x_{e},p_{o}\right) \\
&  =A_{1}\left(  \left(  x_{0}^{\prime}+\frac{i}{2}\partial_{\bar{p}}\right)
,\left(  x_{e}^{\prime}+iT_{eo}\partial_{p_{o}}\right)  ,\left(  p_{o}%
^{\prime}-iT_{eo}\left(  \partial_{x_{e}}+w_{e}\partial_{x_{0}}\right)
\right)  \right)  A_{2}\left(  x_{0},x_{e},p_{o}\right)  .\label{A1A2old2}\\
&  =A_{1}\left(  x_{0},x_{e},p_{o}\right)  \star_{new}A_{2}\left(  x_{0}%
,x_{e},p_{o}\right)  \label{oldnew2}%
\end{align}
where the last step is proven below. The following manipulations are used in
obtaining (\ref{A1A2old2}) from the previous line: for the first factor,
$\left(  x_{0}^{\prime}+\frac{i}{2}\partial_{\bar{p}}\right)  ,$ note
\begin{equation}%
\begin{array}
[c]{c}%
\bar{x}^{\prime}+w_{e}\left(  x_{e}^{\prime}+iT_{eo}\partial_{p_{o}}\right)
=x_{0}^{\prime}+iw_{e}T_{eo}\partial_{p_{o}}=x_{0}^{\prime}+iT_{0o}%
\partial_{p_{o}}=x_{0}^{\prime}+\frac{i}{2}\partial_{\bar{p}},\\
\text{with~ }\partial_{\bar{p}}\equiv2T_{0o}\partial_{p_{o}}=2v_{o}%
\partial_{p_{o}}=\frac{1}{\pi}\int_{0}^{\pi}d\sigma\varepsilon\left(
\frac{\pi}{2}-\sigma\right)  \partial_{p_{-}\left(  \sigma\right)  },
\end{array}
\label{dpbar}%
\end{equation}
and for the last factor note,
\begin{equation}
\left(  p_{o}^{\prime}-iT_{eo}\partial_{x_{e}}\right)  A_{2}\left(
x_{0}\left(  \bar{x},x_{e}\right)  ,x_{e},p_{o}\right)  =\left(  p_{o}%
^{\prime}-iT_{eo}\left(  \frac{\partial x^{0}}{\partial x_{e}}\partial_{x_{0}%
}+\partial_{x_{e}}\right)  \right)  A_{2}, \label{chain}%
\end{equation}
where the $\partial_{x_{0}}$ term in (\ref{chain}) takes care of applying the
derivative $T_{eo}\partial_{x_{e}}$ on the $x_{e}$ in $x_{0}\left(  \bar
{x},x_{e}\right)  $ by using the chain rule, with
\begin{equation}
\frac{\partial x^{0}}{\partial x_{e}}=w_{e},\text{ and }\partial_{x_{0}}%
=\frac{1}{2}\int_{0}^{\pi}d\sigma\partial_{x_{+}\left(  \sigma\right)  },
\end{equation}
while the last $\partial_{x_{e}}$ in (\ref{chain}) is applied to $x_{e}$ which
is not inside $x_{0}\left(  \bar{x},x_{e}\right)  $ in $A_{2}.$ In this way we
have established in general the equivalence of the old and new star products
(\ref{oldnew1}) and (\ref{oldnew2}). Any computation can be performed by
switching between the old/new versions of the star as long as sufficient care
is used as demonstrated above.

On the way to prove the equivalence of (\ref{A1A2old2}) and the new $\star$
product in the $\sigma$-basis via (\ref{oldnew2}), we first note that the
expression in (\ref{A1A2old2}) is reproduced from (\ref{oldnew2}) by defining
a new star product in the $\left(  x_{0},x_{e},p_{o}\right)  $ mode basis that
includes the center of mass mode $x_{0}$ that is explicitly shown as follows
\begin{equation}
\star_{new}=\left(  \star_{x_{e},p_{o}}\right)  \exp\left(  \frac{i}{2}\left(
\overleftarrow{\partial}_{x_{0}^{M}}\overrightarrow{\partial}_{\bar{p}_{M}%
}-\overleftarrow{\partial}_{\bar{p}_{M}}\overrightarrow{\partial}_{x_{0}^{M}%
}\right)  \right)  ,\text{ } \label{new_star}%
\end{equation}
where $\partial_{\bar{p}_{M}}=2v_{o}\partial_{p_{Mo}}$ is defined as in
(\ref{dpbar}), and $\left(  \star_{x_{e},p_{o}}\right)  $ is the non-zero mode
contribution given by
\begin{equation}
\left(  \star_{x_{e},p_{o}}\right)  =\exp\left(  iT_{eo}\left(  \overleftarrow
{\partial}_{x_{e}^{M}}\overrightarrow{\partial}_{p_{Mo}}-\overleftarrow
{\partial}_{p_{Mo}}\overrightarrow{\partial}_{x_{e}^{M}}\right)  \right)  .
\label{newstar2}%
\end{equation}
In particular, as applications of this mode-version of the new star, note that
for products with the center of mass mode, $x_{0}^{\mu}$ or $x_{0}^{b},$ we
get
\begin{equation}
x_{0}^{M}\star_{new}A\left(  x_{0},x_{e},p_{o}\right)  =\left(  x_{0}%
^{M}+\frac{i}{2}\partial_{\bar{p}_{M}}\right)  A\left(  x_{0},x_{e}%
,p_{o}\right)  ,
\end{equation}
however, for products with the midpoint we get (as in
(\ref{midpointcommutator}))%
\begin{equation}
\bar{x}^{M}\star_{new}A\left(  x_{0},x_{e},p_{o}\right)  =\bar{x}^{M}A\left(
x_{0},x_{e},p_{o}\right)  ,
\end{equation}
where there are no derivatives on the right hand side, showing again that the
midpoint $\bar{x}$ is insensitive to the derivatives in the new star product,
and acts just like a complex number.

Finally, to prove the equality between the star products in the new $\sigma
$-basis and the new mode basis (\ref{new_star}) we use the mode expansions of
the position and momentum (\ref{xpmodes}) and compute the mode expansion of
their derivatives by using the chain rule as follows:%
\begin{align}
&
\begin{array}
[c]{l}%
x_{+}\left(  \sigma\right)  =x_{0}+\sqrt{2}\sum_{e}x_{e}\cos e\sigma,~~\\
\partial_{x_{+}\left(  \sigma\right)  }=\frac{2}{\pi}\left(  \partial_{x_{0}%
}+\sqrt{2}\sum_{e}\cos e\sigma~\partial_{x_{e}}\right) \\
\partial_{x_{+}\left(  \sigma\right)  }x_{+}\left(  \sigma^{\prime}\right)
=\delta^{+nn}\left(  \sigma,\sigma^{\prime}\right)  ,
\end{array}
\\
&
\begin{array}
[c]{l}%
p_{-}\left(  \sigma\right)  =\frac{\sqrt{2}}{\pi}\sum_{o}p_{o}\cos
o\sigma,~~\\
\partial_{p_{-}\left(  \sigma\right)  }=2\sqrt{2}\sum_{o}\cos o\sigma
~\partial_{p_{o}},\\
\partial_{p_{-}\left(  \sigma\right)  }p_{-}\left(  \sigma^{\prime}\right)
=\delta^{-nn}\left(  \sigma,\sigma^{\prime}\right)  .
\end{array}
\end{align}
The first term (and similarly the second term) in the exponential of the new
star product in (\ref{msftstar}) can be evaluated in terms of modes%
\begin{equation}%
\begin{array}
[c]{l}%
\frac{i}{4}\int_{0}^{\pi}d\sigma\overrightarrow{\partial}_{p_{-}\left(
\sigma\right)  }\cdot\overleftarrow{\partial}_{x\left(  \sigma\right)
}\varepsilon\left(  \frac{\pi}{2}-\sigma\right) \\
=\frac{i}{4}\int_{0}^{\pi}d\sigma\varepsilon\left(  \frac{\pi}{2}%
-\sigma\right)  \left(  2\sqrt{2}\sum_{o}\cos o\sigma~\overrightarrow
{\partial}_{p_{o}}\right)  \cdot\left(  \frac{2}{\pi}\left(  \overleftarrow
{\partial}_{x_{0}}+\sqrt{2}\sum_{e}\cos e\sigma~\overleftarrow{\partial
}_{x_{e}}\right)  \right) \\
=\left(  \frac{i}{4}\frac{2}{\pi}2\sqrt{2}\int_{0}^{\pi}d\sigma\varepsilon
\left(  \pi/2-\sigma\right)  \cos o\sigma\right)  \overrightarrow{\partial
}_{p_{o}}\cdot\overleftarrow{\partial}_{x_{0}}~+~"\overrightarrow{\partial
}_{p_{o}}\overleftarrow{\partial}_{x_{e}}"\\
=\left(  \frac{i}{4}\frac{2}{\pi}2\sqrt{2}2\int_{0}^{\pi/2}d\sigma\cos
o\sigma\right)  \overrightarrow{\partial}_{p_{o}}\cdot\overleftarrow{\partial
}_{x_{0}}~+~"\overrightarrow{\partial}_{p_{o}}\overleftarrow{\partial}_{x_{e}%
}"\\
=iT_{0o}\overrightarrow{\partial}_{p_{o}}\cdot\overleftarrow{\partial}_{x_{0}%
}~+iT_{eo}\overrightarrow{\partial}_{p_{o}}\cdot\overleftarrow{\partial
}_{x_{e}}.
\end{array}
\end{equation}
where we used $"\overrightarrow{\partial}_{p_{o}}\overleftarrow{\partial
}_{x_{e}}"$ as a short notation to include the integrals that are not shown
explicitly to save space. Putting together all the terms in the exponential of
(\ref{msftstar}), leads exactly to Eq. (\ref{new_star}), thus proving that the
$\star$ in the continuous $\sigma$-basis (\ref{msftstar}) and the $\star$ in
the discreet basis (\ref{new_star}) that includes the center of mass mode are identical.

Therefore, we have proven (noting the relation (\ref{AAtilde}) between $A$ and
$\tilde{A}$) that\emph{ }%
\begin{align}
&  \tilde{A}_{1}\left(  \bar{x},x_{e},p_{o}\right)  \star_{old}\tilde{A}%
_{2}\left(  \bar{x},x_{e},p_{o}\right)  ,\nonumber\\
&  =A_{1}\left(  x_{0},x_{e},p_{o}\right)  \star_{new}A_{2}\left(  x_{0}%
,x_{e},p_{o}\right)  ,\\
&  =A_{1}\left(  x_{+}\left(  \sigma\right)  ,p_{-}\left(  \sigma\right)
\right)  \star_{new}A_{2}\left(  x_{+}\left(  \sigma\right)  ,p_{-}\left(
\sigma\right)  \right)  .\nonumber
\end{align}
In the $\star_{new}$the center of mass mode $x_{0}$ is active in the string
joining as an independent degree of freedom; this is in contrast to
$\star_{old}$ where the midpoint $\bar{x}$ is passive as an independent degree
of freedom. We have shown that $\star_{old}$ and $\star_{new}$ are completely
equivalent, but they must be used consistently in their own basis. As we saw
above in Eq.(\ref{A1A2old2}), $\partial_{x_{e}}$ does not mean the same thing
in the various bases because partial derivatives imply that some variables are
fixed while evaluating derivatives, but the quantities held fixed are
different in the various bases: $\bar{x}$ fixed in the old basis, while
$x_{0}$ fixed in the new basis. Hence one must be careful in such
computations. Clearly, as long as we stick consistently to only one basis, or
be careful in the translation as in the steps in Eqs.(\ref{oldnew1}%
-\ref{oldnew2}), there will be no errors.

The advantage of the new star is that there is a lot of
simplification in the $\sigma$ formalism because the midpoint does
not need to be distinguished. Furthermore, the $\sigma$ basis is
clearly background independent and applies to all CFT backgrounds
that may be used in the constructions of the BRST field $Q\left(
x,p\right)  .$ Moreover, quantum operators products that are well
known in the QM of any CFT apply directly also in the parallel
induced iQM of the MSFT formalism, thus rendering the computations
in the new MSFT much easier.

\subsection{The BRST Gauge Transformations for an Invariant Action}

The MSFT action (\ref{pertS}) has the following form when the midpoint
integration is made explicit%
\begin{align}
S  &  =-Str^{\prime}\int d\bar{x}^{b}\left(  A\star Q\star A+\frac{g_{0}}%
{3}A\star\partial_{\bar{x}^{b}}A\star\partial_{\bar{x}^{b}}A\right)
\nonumber\\
&  =-Str^{\prime}\left(  \partial_{\bar{x}^{b}}\left(  A\star Q\star A\right)
+\frac{g_{0}}{3}\partial_{\bar{x}^{b}}A\star\partial_{\bar{x}^{b}}%
A\star\partial_{\bar{x}^{b}}A\right)  ,
\end{align}
$Str^{\prime}$ is the remainder of the phase space integration that does not
include the midpoint ghost modes. The gauge transformation that leaves this
action invariant is the following%
\begin{equation}
\delta_{\Lambda}A=\left[  Q,\Lambda\right]  _{\star}+g_{0}\{\partial_{\bar
{x}^{b}}A,\partial_{\bar{x}^{b}}\Lambda\}_{\star}~. \label{gauge_transform}%
\end{equation}
Analyzing the ghost numbers in Eq. (\ref{gauge_transform}) we conclude
that\ the gauge parameter $\Lambda$ is bosonic with ghost number $-2$.

Let us check the invariance under the transformation (\ref{gauge_transform}).
\begin{align}
\delta_{\Lambda}S  &  =-Str^{\prime}\partial_{\bar{x}^{b}}\left[
\begin{array}
[c]{c}%
\left(  \delta_{\Lambda}A\star Q\star A+A\star Q\star\delta_{\Lambda}A\right)
\\
+g_{0}\delta_{\Lambda}A\star\left(  \partial_{\bar{x}^{b}}A\right)  _{\star
}^{2}%
\end{array}
\right] \\
&  =-Str^{\prime}\partial_{\bar{x}^{b}}\left[
\begin{array}
[c]{c}%
\left(
\begin{array}
[c]{c}%
\left(  Q\star\Lambda-\Lambda\star Q+g_{0}\left\{  \partial_{\bar{x}^{b}%
}A,\partial_{\bar{x}^{b}}\Lambda\right\}  _{\star}\right)  \star Q\star A\\
+A\star Q\star\left(  Q\star\Lambda-\Lambda\star Q+g_{0}\left\{
\partial_{\bar{x}^{b}}A,\partial_{\bar{x}^{b}}\Lambda\right\}  _{\star
}\right)
\end{array}
\right) \\
+g_{0}\left(  Q\star\Lambda-\Lambda\star Q+g_{0}\left\{  \partial_{\bar{x}%
^{b}}A,\partial_{\bar{x}^{b}}\Lambda\right\}  _{\star}\right)  \star\left(
\partial_{\bar{x}^{b}}A\right)  _{\star}^{2}%
\end{array}
\right]
\end{align}
Next we use the cyclic property of the supertrace, and the star nilpotency of
the BRST field $Q\star Q=0,$ to cancel kinetic terms and reorganize the
interaction terms%
\begin{equation}
\delta_{\Lambda}S=Str^{\prime}\left[
\begin{array}
[c]{c}%
\partial_{\bar{x}^{b}}\left(
\begin{array}
[c]{c}%
\left(  Q\star\Lambda\star Q\star A-Q\star\Lambda\star Q\star A\right) \\
+\left(  -\Lambda\star Q_{\star}^{2}\star A+A\star Q_{\star}^{2}\star
\Lambda\right)
\end{array}
\right) \\
+g_{0}\partial_{\bar{x}^{b}}\left(  Q\star\left[
\begin{array}
[c]{c}%
-\partial_{\bar{x}^{b}}\left(  A\star\partial_{\bar{x}^{b}}A\star
\Lambda\right) \\
+\partial_{\bar{x}^{b}}\left(  A\star\partial_{\bar{x}^{b}}\Lambda\star
A\right) \\
+\partial_{\bar{x}^{b}}\left(  \Lambda\star\partial_{\bar{x}^{b}}A\star
A\right)
\end{array}
\right]  \right)
\end{array}
\right]
\end{equation}
The first term vanishes explicitly while the second term takes the form%
\begin{equation}
\delta_{\Lambda}S=g_{0}Str^{\prime}\left[  \partial_{\bar{x}^{b}}%
Q\star\partial_{\bar{x}^{b}}\left(  \Lambda\star\partial_{\bar{x}^{b}}A\star
A+A\star\partial_{\bar{x}^{b}}\Lambda\star A-A\star\partial_{\bar{x}^{b}%
}A\star\Lambda\right)  \right]  .
\end{equation}
In the next step we expand the string field $A$ and the gauge parameter
$\Lambda$ in the powers of$~\bar{x}^{b}$:%
\begin{equation}
A=\bar{x}^{b}A^{\left(  0\right)  }+A^{\left(  -1\right)  },~~\Lambda=\bar
{x}^{b}\Lambda^{\left(  -1\right)  }+\Lambda^{\left(  -2\right)  }.
\end{equation}
The BRST charge contributes%
\begin{equation}
\partial_{\bar{x}^{b}}Q\equiv Q_{++}=\int_{0}^{\pi/2}d\sigma p_{-b}\left(
\sigma\right)  \star x_{+}^{c}\left(  \sigma\right)  .
\end{equation}
where $Q_{++}$ has ghost number $+2.$ Therefore, $\delta_{\Lambda}S$ can be
reorganized into the form
\[
\delta_{\Lambda}S=g_{0}Str^{\prime}\left\{  Q_{++}\Psi\right\}  ,\text{ }%
\]
with
\begin{equation}
\Psi=\left[  \Lambda^{\left(  -2\right)  },\left(  A^{\left(  0\right)
}\right)  ^{2}\right]  _{\star}+\left\{  A^{\left(  -1\right)  },\left\{
\Lambda^{\left(  -1\right)  },A^{\left(  0\right)  }\right\}  \right\}
_{\star}. \label{str_commut}%
\end{equation}
We dropped the star product in $Str^{\prime}\left(  Q_{++}\star\Psi\right)  $
because it is allowed under the supertrace. Next we examine this $Str^{\prime
}.$ Making the Grassmann integrals explicit as being equivalent to
derivatives, we can write
\begin{equation}%
\begin{array}
[c]{l}%
Str^{\prime}\left(  Q_{++}\Psi\right)  =\\
=Tr\left\{
{\displaystyle\prod\limits_{\sigma^{\prime}}}
\left[  \frac{\partial}{\partial p_{-b}\left(  \sigma^{\prime}\right)  }%
\frac{\partial}{\partial x_{+}^{b}\left(  \sigma^{\prime}\right)  }%
\frac{\partial}{\partial p_{-c}\left(  \sigma^{\prime}\right)  }\frac
{\partial}{\partial x_{+}^{c}\left(  \sigma^{\prime}\right)  }\left(  \int
_{0}^{\pi/2}d\sigma p_{-b}\left(  \sigma\right)  x_{+}^{c}\left(
\sigma\right)  \right)  \Psi\left(  p_{-b},x_{+}^{c},p_{-c},x_{+}^{b}\right)
\right]  \right\} \\
\sim Tr\left(  \int_{0}^{\pi/2}d\sigma\left[  \frac{\partial}{\partial
p_{-c}\left(  \sigma\right)  }\frac{\partial}{\partial x_{+}^{b}\left(
\sigma\right)  }\Psi\left(  p_{-b},x_{+}^{c},p_{-c},x_{+}^{b}\right)  \right]
_{p_{-b}=x_{+}^{c},=p_{-c},=x_{+}^{b}=0}\right)
\end{array}
\nonumber
\end{equation}
where $Tr$ is the remaining bosonic integrals. Now, substituting $\Psi$ from
(\ref{str_commut}) the two derivatives in the last line produce several
(anti)commutators. To see this we write out the bose/fermi components of $A$
and $\Lambda$
\begin{equation}
A=\bar{x}^{b}A^{\left(  0\right)  }+A^{\left(  -1\right)  },~~\Lambda=\bar
{x}^{b}\Lambda^{\left(  -1\right)  }+\Lambda^{\left(  -2\right)  }.
\end{equation}
and then find that $Str^{\prime}\left(  Q_{++}\Psi\right)  $ takes the form
\begin{equation}
Tr\int_{0}^{\pi/2}d\sigma\left(
\begin{array}
[c]{c}%
\begin{array}
[c]{c}%
\left[  \frac{\partial}{\partial p_{-c}\left(  \sigma\right)  }\frac{\partial
}{\partial x_{+}^{b}\left(  \sigma\right)  }\Lambda^{\left(  -2\right)
},\left(  A^{\left(  0\right)  }\right)  ^{2}\right]  _{\star}\\
+\left[  \Lambda^{\left(  -2\right)  },\frac{\partial}{\partial p_{-c}\left(
\sigma\right)  }\frac{\partial}{\partial x_{+}^{b}\left(  \sigma\right)
}\left(  A^{\left(  0\right)  }\right)  ^{2}\right]  _{\star}\\
-\left\{  \frac{\partial}{\partial x_{+}^{b}\left(  \sigma\right)  }%
\Lambda^{\left(  -2\right)  },\frac{\partial}{\partial p_{-c}\left(
\sigma\right)  }\left(  A^{\left(  0\right)  }\right)  ^{2}\right\}  _{\star
}\\
+\left\{  \frac{\partial}{\partial p_{-c}\left(  \sigma\right)  }%
\Lambda^{\left(  -2\right)  },\frac{\partial}{\partial x_{+}^{b}\left(
\sigma\right)  }\left(  A^{\left(  0\right)  }\right)  ^{2}\right\}  _{\star}%
\end{array}
\\%
\begin{array}
[c]{c}%
-\left\{  \frac{\partial}{\partial p_{-c}\left(  \sigma\right)  }%
\frac{\partial}{\partial x_{+}^{b}\left(  \sigma\right)  }A^{\left(
-1\right)  },\left\{  \Lambda^{\left(  -1\right)  },A^{\left(  0\right)
}\right\}  \right\}  _{\star}\\
-\left\{  A^{\left(  -1\right)  },\frac{\partial}{\partial p_{-c}\left(
\sigma\right)  }\frac{\partial}{\partial x_{+}^{b}\left(  \sigma\right)
}\left\{  \Lambda^{\left(  -1\right)  },A^{\left(  0\right)  }\right\}
_{\star}\right\}  _{\star}\\
-\left[  \frac{\partial}{\partial x_{+}^{b}\left(  \sigma\right)  }\left\{
\Lambda_{-}^{\left(  -1\right)  },A^{\left(  0\right)  }\right\}  _{\star
},\frac{\partial}{\partial p_{-c}\left(  \sigma\right)  }A^{\left(  -1\right)
}\right]  _{\star}\\
-\left[  \frac{\partial}{\partial x_{+}^{b}\left(  \sigma\right)  }A^{\left(
-1\right)  },\frac{\partial}{\partial p_{-c}\left(  \sigma\right)  }\left\{
\Lambda^{\left(  -1\right)  },A^{\left(  0\right)  }\right\}  _{\star}\right]
_{\star}%
\end{array}
_{\star}%
\end{array}
\right)
\end{equation}
It is understood that we should set $p_{-b}=x_{+}^{c},=p_{-c},=x_{+}^{b}=0$
after taking the derivatives. All these terms vanish under the trace on the
basis of their bose/fermi properties, $Tr\left[  a,b\right]  =0$ when they are
both bosons and $Tr\left\{  \alpha,\beta\right\}  =0$ when they are both fermions.

Hence, we proved that the gauge transformation that gives $\delta_{\Lambda
}S=0$ has the following form%
\begin{equation}
\delta_{\Lambda}A=\left[  Q,\Lambda\right]  _{\star}+g_{0}\left\{
\partial_{\bar{x}^{b}}A,\partial_{\bar{x}^{b}}\Lambda\right\}  _{\star}\ ,
\end{equation}
for $\Lambda\left(  x,p\right)  $ a general field of ghost number
$-2$, and $A$ a general field of ghost number $-1$.

\end{document}